\documentclass[10pt,a4paper]{dmjthesis}
\pdfoutput=1 
\usepackage[american]{babel}
\usepackage{layout}     
\usepackage{psfrag}     
\usepackage{graphicx}   
\usepackage{makeidx}    
\usepackage{amssymb}    
\usepackage{amsbsy}     
\usepackage{amstext}    

\makeatletter
   \newcommand{\chapapp}{\MakeLowercase\@chapapp}
   \newcommand{\Chapapp}{\@chapapp}
\makeatother
\makeindex     

\newcommand{\clearemptydoublepage}{%
	\newpage{\pagestyle{empty}\cleardoublepage}} 

\newcommand{\be}{\begin{equation}}
\newcommand{\ee}{\end{equation}}
\newcommand{\bea}{\begin{eqnarray}}
\newcommand{\eea}{\end{eqnarray}}
\newcommand{\mrm}{\mathrm}
\newcommand{\mbf}{\mathbf}
\newcommand{\msf}{\mathsf}
\newcommand{\mcal}{\mathcal}

\newcommand{\bsa}{\boldsymbol{\alpha}}
\newcommand{\bsb}{\boldsymbol{\beta}}
\newcommand{\bsg}{\boldsymbol{\gamma}}
\newcommand{\Z}{\ensuremath{\mathcal{Z}}}  
\newcommand{\Ztil}{\ensuremath{\tilde{\mathcal{Z}}}}
\newcommand{\Ham}{\ensuremath{\mathcal{H}}}  
\newcommand{\Tr}{\mathop{\mathrm{Tr}}}  
\newcommand{\tr}{\mathop{\mathrm{tr}}}  
\newcommand{\singlet}{\ensuremath{\boldsymbol{1}}}
\newcommand{\triplet}{\ensuremath{\boldsymbol{3}}}
\newcommand{\antitriplet}{\ensuremath{\bar{\boldsymbol{3}}}}
\newcommand{\octet}{\ensuremath{\boldsymbol{8}}}
\newcommand{\dom}{\partial\Omega}

\newcommand{\Index}[1]{#1\index{#1}}	

\newcommand{\Casimir}{Casimir\index{Casimir operator}}
\newcommand{\Cartan}{Cartan\index{Cartan operator}}

\newcommand{\Eq}[1]{Eq.~(\ref{#1})}
\newcommand{\Fig}[1]{Fig.~\ref{#1}}

\newcommand{\viz}{\textit{viz.}}
\newcommand{\ie}{\textit{i.e.}}
\newcommand{\eg}{\textit{e.g.}}

\newcommand{\comment}[1]{} 

\begin{document}


\hyphenation{stran-ge-let
             mad-sen
	     jen-sen
	     chris-ti-an-sen
	     brook-ha-ven
}

\index{quark star|see{strange star}}
\index{propagator|see{Green's function}}
\index{Gaussian approximation|see{saddle-point approximation}}

\pagenumbering{roman}

\newfont{\bigtitlefont}{cmss17 scaled 3000}
\newfont{\smalltitlefont}{cmss12 scaled 1350}
\title{\vspace{-2cm}
       \rule{\linewidth}{1mm}\\[10pt]
       \vspace{0.3cm}
       {\bigtitlefont STRANGELETS}\\
       {\smalltitlefont
	Effects of Finite Size and Exact Color Singletness}
       \rule{\linewidth}{1mm}\\[1.5cm]
	} 
\author{
	{\Large Dan M\o nster Jensen}\\[2.5cm]
        Institute of Physics and Astronomy\\
        University of Aarhus\\[0.5cm]
        \includegraphics[width=4cm]{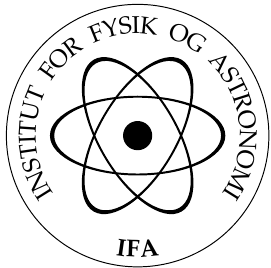} }
\date{July 1996}
\maketitle

\clearemptydoublepage
\chapter*{Preface}
\addcontentsline{toc}{chapter}{Preface}

The present volume has been submitted to the Faculty of Natural Science
at the University of Aarhus, in order to fulfill the requirements for the
Ph.D.\ degree in physics.

The work presented here has been done in collaboration with my
supervisor, Jes Madsen, during my time as a Ph.D student at the
Institute of Physics and Astronomy, University of Aarhus, and during a
stay at the Physics Department, Brookhaven National Laboratory.
I have also had the pleasure of working with another Ph.D.\ student,
Michael B.~Christiansen, here at the Institute of Physics and Astronomy.

\section*{List of Publications}
\addcontentsline{toc}{section}{List of Publications}

The work presented here has been described in part in the following
publications:
\begin{enumerate}
   \item\label{paper1}
         Dan~M{\o}nster Jensen and Jes Madsen,  \textit{Strangelets at Non-Zero
         Temperature}, in {\em Strangeness and Quark
         Matter}, edited by G. Vassiliadis, A.~D. Panagiotou, S. Kumar,
         and J. Madsen (World Scientific, Singapore, 1995), pp.\ 220--229.
   \item\label{paper2}
         Jes Madsen, Dan~M. Jensen, and Michael B.\ Christiansen, \textit{Color
         Singlet Suppression of Quark-Gluon Plasma Formation}, Phys. Rev. C
         {\bf 53},  1883 (1996).
   \item\label{paper3}
         Dan~M. Jensen and Jes Madsen, \textit{Strangelets with Finite
         Entropy}, Phys. Rev. D {\bf 53},  R4719 (1996).
   \item\label{paper4}
         Dan M{\o}nster Jensen, Jes Madsen, and Michael B. Christiansen,
		 \textit{Color Singlet Strangelets}, to appear in Heavy Ion Phys. 
		 (1996).
\end{enumerate}

The contents of Paper~\ref{paper1} has been described in
Chapter~\ref{chap:temperature} and Chapter~\ref{chap:equilibrium}, with
some additional details not included in the paper.

The subject of Paper~\ref{paper2} is described in
Chapter~\ref{chap:suppression}, but some new material has been added, as
discussed in that chapter.

Chapter~\ref{chap:color} discusses the same problem that is treated in
Paper~\ref{paper3}, but again some additional aspects are discussed,
some of them also presented in Paper~\ref{paper4}.

\section*{Notation}
\addcontentsline{toc}{section}{Notation}
I use the Einstein summation convention, where implicit summation is
understood whenever an index is repeated, as in
$$
	\mu_q N_q \equiv \sum_q \mu_q N_q ,
$$
but not if one of the indexed quantities appear as an argument
$$
	N_q(\mu_q) \not= \sum_q N_q(\mu_q).
$$

Relativistic convention:
$$
	\begin{array}{r@{\,=\,}l@{\qquad}r@{\,=\,}l}
	x^\mu & (t,\mbf{x}) & x_\mu & (t,-\mbf{x}) \\[10pt]
	\partial_\mu & (\frac{\partial}{\partial t},\nabla) &
	\partial^\mu & (\frac{\partial}{\partial t},-\nabla)
	\end{array}
$$
\Index{Minkowski metric}:
$$
	\eta^{\mu\nu} = \eta_{\mu\nu} =  \left(
	\begin{array}{rrrr}
		1 & 0 & 0 & 0\\
		0 & -1 & 0 & 0\\
		0 & 0 & -1 & 0\\
		0 & 0 & 0 & -1
	\end{array} \right)
$$

Unless otherwise noted I have used so-called natural units, in which
the Planck constant, the Boltzmann constant, and the speed of light are unity
$$
	\hbar = k_B = c = 1.
$$

\section*{Acknowledgements}
\addcontentsline{toc}{section}{Acknowledgements}

First and foremost I would like to thank my supervisor, Jes Madsen.
During all of my Ph.D.~study he has provided me with constant inspiration,
and guidance.

Many thanks to Carl Dover for making my visit at the Physics
Department, Brookhaven National Laboratory possible. I was sad to learn
that Prof.~Dover passed away recently.

I would also like to thank my office mates Michael B.~Christiansen and
Georg Bergeton Larsen for many
interesting discussions, and for being great company. At Brookhaven I
shared office, a single computer terminal, and many good times 
with Raffaele Mattiello.

My Ph.D.~study has been financed by a scholarship from the University of
Aarhus. I also appreciate the generosity and hospitality of Brookhaven
National Laboratory who funded part of my stay.

A very special thanks goes to my girlfriend Rikke F\"ursterling, for
putting up with me for all these years, especially during the writing
of this thesis. Thanks for all your love and support!

Last, but not least, I wish to thank my parents, Inge M{\o}nster Jensen
and Carl M.~Jensen, who have always backed me up in what I did. Thanks!

\tableofcontents
\clearemptydoublepage
\pagenumbering{arabic}

\chapter{Introduction\label{chap:introduction}}
This thesis deals with quark matter, in particular, strange quark matter
with the focus being on very small systems, the so-called
\emph{strangelets}. Quarks are fundamental particles, just like the
electron, but they are a sort of ghosts, who never show themselves. A
free quark has thus never been seen, and it is believed to be a law of
nature that quarks may not be observed directly. They always come 
either as three quarks, making up what we call baryons, of which the
familiar proton and neutron are the most well-known, or as a pair of
quark and antiquark, called a meson, of which the $\pi$-meson is an
example. The complicated rules stating what combinations of quarks are
allowed (the so-called color singlet states) do not limit the spectrum
to the baryons and the mesons, but also allow for the existence of
conglomerates of an
integer multiple of three quarks. The lightest of these are
the baryons, whereas a six quark state is sometimes called a dibaryon.
All of these heavier states are collectively called quark matter, and
unlike nuclei which cease to exist at a baryon number of $\sim$240,
strange quark matter, if stable, could span a range in baryon number
all the way from light nuclei to stars, as heavy as or heavier than our own
sun.

The purpose of this chapter is to give an introduction to strange quark
matter for the benefit of the non-specialist reader, as well as to give
an overview of the properties of this exotic phase of matter, and where
it might play a role.

\section{Basic Ideas}

\begin{figure}
	\centering
	\includegraphics{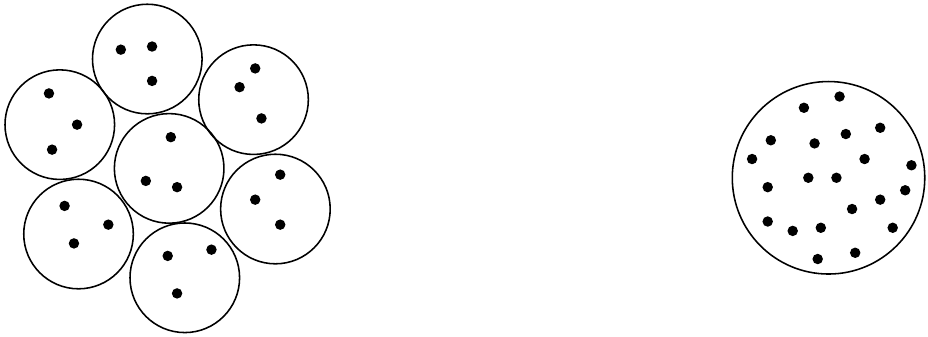}
	\caption{\label{fig:nucleus}A nucleus (left) consisting of 7
		nucleons (\textit{e.g.}, ${}^{7}$Li), with each nucleon
		made up of three quarks; and a strangelet
		(right) composed of $3\times 7 = 21$ quarks.}
\end{figure}

Quark matter is best explained by contrasting it with ordinary nuclear
matter. Whereas the quarks in a nucleus are confined within the nucleons
(protons, neutrons) the quarks in quark matter are free to roam the
entire phase (see \Fig{fig:nucleus}). Thus one can imagine quark matter
as a nucleus made up of
quarks that are not correlated in such a way as to form nucleons. From
nuclear physics we know that in fact quarks \emph{do} lump together to form
nucleons which in turn make up the nucleus.

So what makes contemplations
on quark matter more than a purely academic exercise? The answer is
\emph{strangeness}. Nature has supplied us with
three so-called families of the fermionic constituents of matter. 
Each family consists of two quarks and two leptons, each with a
corresponding antiparticle.
The particle members of a family are grouped together in
columns below
$$
\begin{array}{lll}
		u  \qquad &  c \qquad & t \\
 		d  &  s  & b \\
 		\nu_e  &  \nu_\mu & \nu_\tau \\
 		e^-  &  \mu^- & \tau^- 
\end{array} .
$$
Corresponding members in different families are alike in all respects
except for their masses. The first
family consisting of the $u$ and $d$ quarks and the electron and its
neutrino (along with their respective antiparticles) are lighter
than their counterparts in the two other families.
What makes the strange quark particularly interesting is that it is the
lightest of the heavy quarks\footnote{The current-quark mass of the
strange quark is estimated to be 100--300 MeV \cite{PDG96}.}
(or the heaviest of the light quarks), and
thus the one most easily produced. 

\begin{figure}
	\centering
	\setlength{\unitlength}{0.5cm}
\begin{picture}(22,13)(-1,0)
	\thicklines
	\put(0,1){\line(1,0){8}}
	\put(0,1){\vector(0,1){12}}
	\thinlines
	\put(2,0){\makebox(0,0)[b]{$u$}}
	\put(6,0){\makebox(0,0)[b]{$d$}}
	\put(0,10){\line(-1,0){0.2}}
	\put(-0.2,10){\makebox(0,0)[r]{$\epsilon_F$}}
	\put(1,1){\line(0,1){9}}
	\put(1,10){\line(1,0){2}}
	\put(3,1){\line(0,1){9}}
	\put(5,1){\line(0,1){9}}
	\put(5,10){\line(1,0){2}}
	\put(7,1){\line(0,1){9}}
	\thicklines
	\put(10,1){\line(1,0){12}}
	\put(10,1){\vector(0,1){12}}
	\thinlines
	\put(12,0){\makebox(0,0)[b]{$u$}}
	\put(16,0){\makebox(0,0)[b]{$d$}}
	\put(20,0){\makebox(0,0)[b]{$s$}}
	\put(10,8){\line(-1,0){0.2}}
	\put(9.8,8){\makebox(0,0)[r]{$\epsilon_F$}}
	\put(10,5){\line(-1,0){0.2}}
	\put(9.8,5){\makebox(0,0)[r]{$m_s$}}
	\put(11,1){\line(0,1){7}}
	\put(11,8){\line(1,0){2}}
	\put(13,1){\line(0,1){7}}
	\put(15,1){\line(0,1){7}}
	\put(15,8){\line(1,0){2}}
	\put(17,1){\line(0,1){7}}
	\put(19,5){\line(0,1){3}}
	\put(19,8){\line(1,0){2}}
	\put(19,5){\line(1,0){2}}
	\put(21,5){\line(0,1){3}}
\end{picture}
	\caption{\label{fig:fermi_well}
		The distribution of quarks in 2 and 3 flavor quark matter. The
		addition of an extra flavor lowers the Fermi energy, and thus
		the total energy of the system.}
\end{figure}
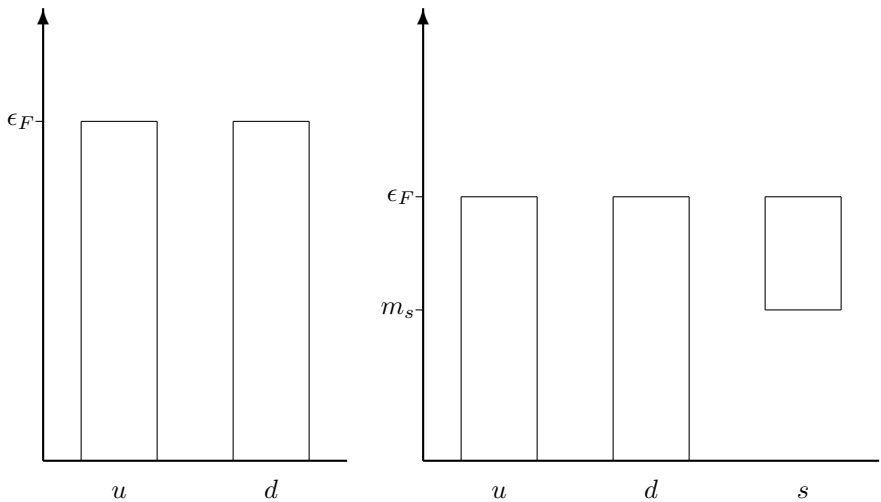

If we think of quark matter in terms of a simple Fermi-gas model, the
quarks are filled into single particle levels up to a maximum energy,
the Fermi energy, $\epsilon_F$. As long as the strange quark mass is less
than the Fermi energy, it is favorable to have strange quarks in the
system in addition to up and down quarks. This is illustrated
schematically in \Fig{fig:fermi_well}. Thus it is possible that the
energy per baryon of 3 flavor quark matter, in addition to being smaller
than that of 2 flavor quark matter, is also smaller than the energy per
baryon of nuclear matter, making it the true ground state of strongly
interacting matter.

We therefore have the following situation: 2 flavor quark matter must 
have a higher energy per baryon than nuclear matter, which would
otherwise decay into $u$, $d$ quark matter. Strange quark matter, on the
other hand, has a lower energy per baryon than non-strange quark matter,
and it could possibly be lower than that of nuclear matter. The latter
case is obviously the more exciting since it should then be possible to
produce, and observe, strange quark matter.

But wouldn't nuclei immediately decay into small lumps of strange quark
matter if that were to be the state of
lower energy? For such a decay to happen to a nucleus with baryon
number $A$, roughly one $u$ or $d$ quark per nucleon would need to be converted
into $s$ quarks \emph{simultaneously}. This means that $A$ simultaneous
$\beta$-decays must take place, which is a highly unlikely event as soon
as $A$ exceeds a few. An estimate of the probability of such a process
can be made using the Fermi theory of weak interactions, where 
each $\beta$-decay has an amplitude proportional to the Fermi
constant $G_F$, giving a total amplitude 
\be
	\mcal{A}\left( {}^AX \to {}^A\left\{uds\right\} \right)
	\propto G_F^A .
\ee
Since $G_F$ is very small\footnote{$G_F$ is approximately
$10^{-5}m_p^{-2}$, where $m_p$ is the proton mass.} the amplitude (and
thus the probability) is \emph{very} small.

So even if strange quark matter turns out to be absolutely stable
relative to nuclear matter, we need not worry that our surroundings will
suddenly turn strange.

The first to entertain the idea that strange quark matter might be
stable, was Bodmer \cite{Bodmer71a}, who in 1971 pointed out that the
existence of ``\Index{collapsed nuclei}''\footnote{Bodmer used the phrase
collapsed nuclei to denote what we now call strange quark matter.} with
lower energy than normal nuclei does not conflict with the fact that we
observe normal nuclei, but not their collapsed counterparts, since
nuclei would simply be very long-lived with respect to ``collapse''.
Bodmer also suggested that strange quark matter might have been
produced in the early hot stages of the universe, and might still be
found in cosmic rays or even in the interior of stars or planets,
possibly slowly converting ordinary matter into strange quark matter.

In 1979 Chin and Kerman \cite{Chin79a} discussed the possible
metastability of hyperstrange multiquark droplets, and suggested that
such states could be produced in the collision of heavy ions at
relativistic speed. Similar considerations were put forward
by Terazawa \cite{Terazawa79a} at about the same time.

The absolute stability of strange quark matter, as first suggested by
Bodmer, was not considered again until 1984, when Witten \cite{Witten84a}
rehabilitated the idea, giving it new impetus that resulted in a 
series of articles by many authors trying to refine and explore the many
ideas put forward by Witten (see Ref.~\cite{sqm91} for an overview).
Like Bodmer before him, Witten considered the production
of strange quark matter in the very early universe, where lumps of
strange quark matter---so-called quark nuggets---would emerge as survivors
of a first order quark-hadron phase transition.

A thorough and comprehensive study of the properties of strange quark
matter, following Witten's conjecture, was performed by Farhi and Jaffe
\cite{Farhi84a}. They considered both bulk systems, which contain
electrons in addition to quarks, and small systems where the electrons
required for overall charge neutrality form a cloud around the strange
matter nucleus. For the latter systems, which correspond to a baryon
number less than roughly 10\footnote[7]{This is not a footnote, it's an
exponent. $10^7=10 000 000$.}, they coined the term
\emph{\Index{strangelet}}.

\index{stablility!of strange quark matter|(}
It was found by Witten, and confirmed by Farhi and Jaffe's more detailed
calculations, that there is a range of model parameters (see Chapter
\ref{chap:mit} for details) for which strange quark matter is stable in
the bulk, and for the most optimistic choices also at even very low
baryon number. The condition of stability of strange quark matter
is that its energy per baryon should be lower than the proton (or neutron)
mass which is approximately 938 MeV. Absolute stability would also
require that the energy per baryon should be lower than that of the most
stable nucleus with an energy per baryon around 930 MeV, but just as a
nucleus would not readily decay into strange quark matter, so would
\comment{Isn't this correct?}
the decay of strange quark matter into a nucleus be greatly inhibited.

For an energy per baryon greater than the nucleon mass ($\sim$938 MeV)
but lower than the $\Lambda$ mass ($\sim$1116 MeV), strange quark
matter would be metastable, decaying via the emission of primarily
neutrons, followed by the conversion ($\beta$-decay) of strange quarks
into up and down quarks, to restore the most favorable flavor
composition. If the energy per nucleon exceeds the $\Lambda$ mass,
strange quark matter will be unstable; immediately dissolving into a gas
of $\Lambda$'s.
\index{stablility!of strange quark matter|)}

\section{Phase Diagram of Strongly Interacting Matter}
\index{phase diagram|(}
If ordinary nuclear matter is compressed to very high densities, say
2--3 times normal nuclear matter densities, nucleons would overlap and
loose their separate identities. This would bring about a transition
into quark matter. Likewise at very high temperature, the energy density
will be very high and the asymptotic freedom of quantum chromodynamics
will allow quarks to move freely about, in a hot \Index{quark-gluon plasma}.
The temperature at which this occurs in a baryon free environment has
been estimated from lattice calculations to be around 150--200~MeV.
\index{lattice gauge calculations}

\begin{figure}
	\includegraphics[width=\linewidth]{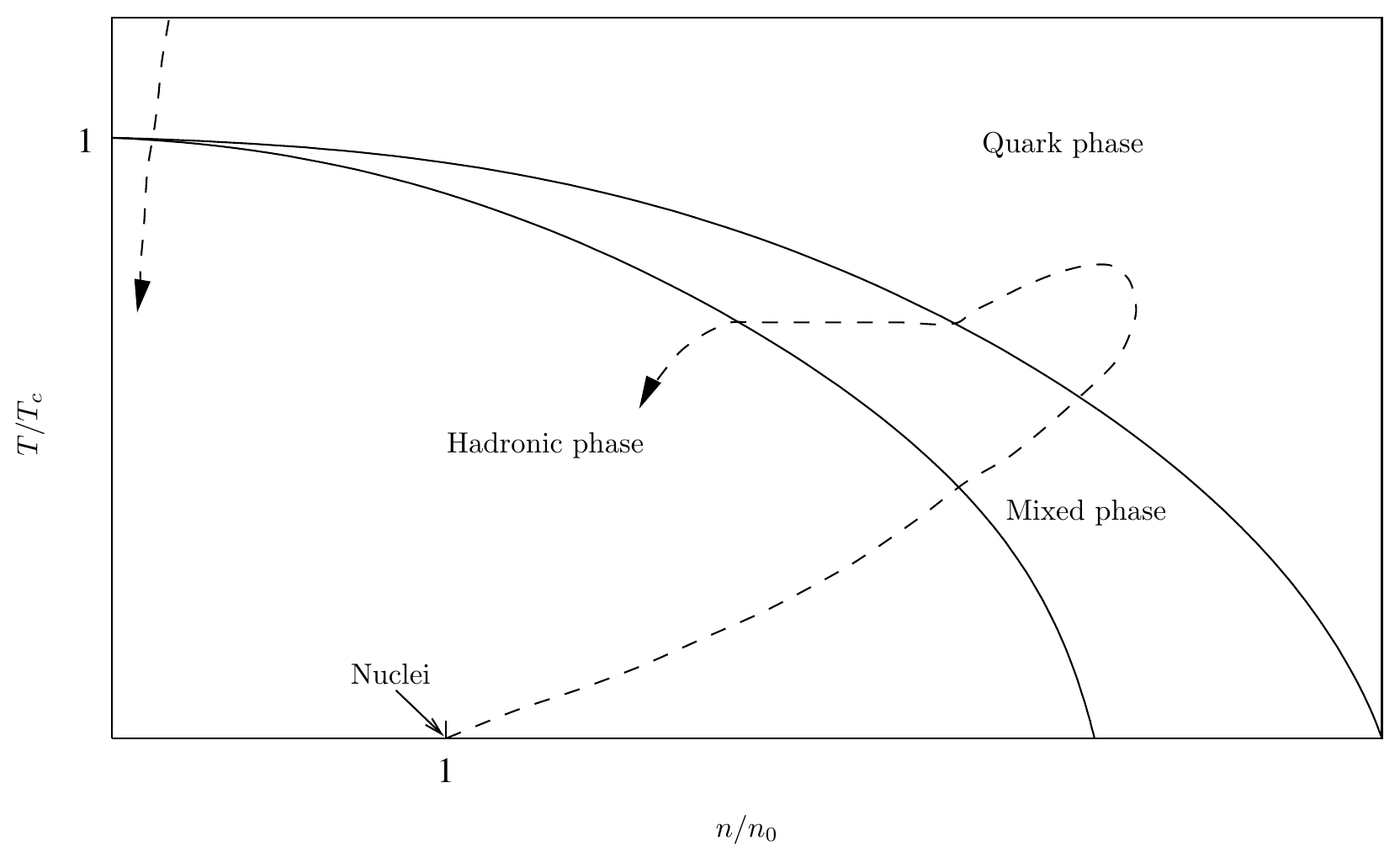}
	\caption{\label{fig:phase} Schematic phase diagram of strongly interacting
	         matter.}
\end{figure}

A schematic view of how the phase diagram of baryonic matter may look is
shown in \Fig{fig:phase}. The baryon number density $n_B$ is in units of
normal nuclear matter density $n_0$, and the temperature $T$  is in units of
the critical temperature $T_c$. At low temperature and baryon number
density the hadronic phase resides, while at high $T$  or high
$n_B$, matter is in the quark phase. There is also an area where both
phases coexist. At fixed $T$ these coexisting phases do not have the
same baryon number density, so a point in the mixed phase is
characterized by the mean baryon number density of the whole system,
even though none of the two phases actually has this density.
There are other transitions than the one shown here, \eg , there is a
transition from a nuclear liquid to a hadron gas, but here focus is
entirely on the quark-hadron phase transition.
\index{phase diagram|)}

I have tacitly assumed that the phase transition is of first order,
since otherwise there would be no mixed phase, as the energy density and
baryon number density are continuous during a second order phase
transition. The order of the phase transition is unknown, but lattice
calculations \cite{Brown90a} for pure gauge fields show a first order
phase transition. Of course the quark fields cannot be neglected, but
they are notoriously hard to incorporate in lattice gauge
calculations. The order of the phase transition depends on the quark
masses, where the pure gauge case corresponds to all quark masses being
infinite. Two massless quark flavors seem to indicate a second order
phase transition, whereas three massless flavors give a first order
phase transition. The real world with two approximately massless flavors
($u$ and $d$) and one massive flavor ($s$) lies somewhere in between.

\section{The Cosmological Quark-Hadron Phase Transition
         \label{sec:earlyuniverse}}
\index{quark-hadron phase transition!cosmological|(}
It is very difficult indeed to probe this
phase diagram, since the conditions needed to approach the phase
boundary are so extreme. In astrophysical settings, however, we find
conditions unattainable in any earthly laboratory. The hottest
environment ever to exist is to be found in the early universe, where
during some epoch matter was in the quark phase, until the universe
expanded and cooled enough to reach the quark-hadron phase transition
temperature\footnote{In fact the very pure plasma phase most likely
experienced a significant amount of supercooling.}, where bubbles of
hadronic matter started to nucleate. The path followed by matter in the
early universe is shown as the arrow crossing the phase border from
above at low baryon number density. The reason that the baryon number
density was low in the early universe is that almost as much antimatter
as matter was formed in the \Index{big bang}. The small asymmetry between
baryons and antibaryons is what provided us with the matter now present
in the universe---matter of which we ourselves are made.

It was Wittens \cite{Witten84a} original idea that quark matter might
have survived the cosmological quark-hadron phase transition, still
being around at present as a form of \Index{dark matter}.  The physical
mechanism allowing strange quark matter to survive the phase transition
is the concentration of strangeness in the quark phase. 
Notwithstanding their formation in the  phase transition, quark nuggets
may not survive the hostile environment following their creation, as
they are prone to suffer from surface evaporation of neutrons and
protons \cite{alcfar85,Madsen86a,sumkaj91}, making it unlikely that
nuggets with an initial  baryon number less than $10^{30}$--$10^{40}$
have survived. The upper limit on the baryon number is set by the baryon
number inside the particle horizon, which was around $10^{48}$ at the
time of the quark-hadron phase transition.

\comment{References?}
But even
if no nuggets survived, we may still be able to observe their
``fossilized remains,'' in that they would have produced inhomogeneities
in the baryon number distribution, owing to their higher density than
the surrounding phase, and their propensity to absorb neutrons.
This baryon number inhomogeneity would in turn influence the primordial
\Index{nucleosynthesis}, \eg , by causing less ${}^4$He to be
produced.
\index{quark-hadron phase transition!cosmological|)}

\section{Ultrarelativistic Heavy-Ion Collisions}
\index{heavy-ion collision|(}
If strange quark matter is to be produced in a laboratory, the most
likely source is the collision of two heavy nuclei at very high energy.
Such ultrarelativistic heavy-ion collisions are taking place both at
CERN and at Brookhaven National Lab., at present using fixed targets.
The Brookhaven Alternating Gradient Synchrotron (AGS) is capable of
accelerating nuclei as heavy as Au to a momentum of 14.5 GeV$/c$
per nucleon \cite{Beavis95a} ,
while the Super Proton Synchrotron (SPS) at CERN can achieve a 
momentum of 158 GeV$/c$ per nucleon for a Pb beam \cite{Appelquist96a}.
The Relativistic Heavy Ion Collider (RHIC), which is under construction
at Brookhaven, will ultimately accelerate beams of heavy ions to
energies as high as 100 GeV per nucleon, using the AGS as an injector.
Colliding the two beams of the RHIC will hopefully allow
experimentalists to observe the creation of a quark-gluon plasma, the
Holy Grail of heavy-ion physics. A possible path that the central region
of the colliding nuclei could follow to enter the quark phase, and then
cool and expand (much like the early universe scenario) ending up in the
hadronic phase again, is shown in \Fig{fig:phase}. 

\index{quark-gluon plasma|(}
So far the creation of the quark-gluon plasma has not been established,
mainly because most of the signals that would indicate the creation 
of a quark-gluon plasma phase in the early stages of a heavy-ion
collision, can be mimicked by a hot and dense hadronic gas. The creation
of a strangelet \index{strangelet!creation in heavy-ion coll.} in a
heavy-ion collision is one of the cleanest imaginable signals of the
creation of a quark-gluon plasma.
\index{quark-gluon plasma|)}

A  scenario for the production of strangelets in high-energy
heavy-ion collisions, was proposed by Greiner \textit{et al.}~%
\cite{Greiner87a,greal88,Greiner91a}. The proposed scenario builds on
the mechanism of separation of strangeness and antistrangeness, where
most of the strangeness is concentrated in the quark phase, while
antistrangeness is carried by the antistrange mesons, mainly
$K^0$ and $K^+$.

\index{heavy-ion collision|)}

\section{Strange Stars}
\begin{figure}
	\centering
	\psfrag{-}{${}^-$}
	\includegraphics[width=.50\linewidth]{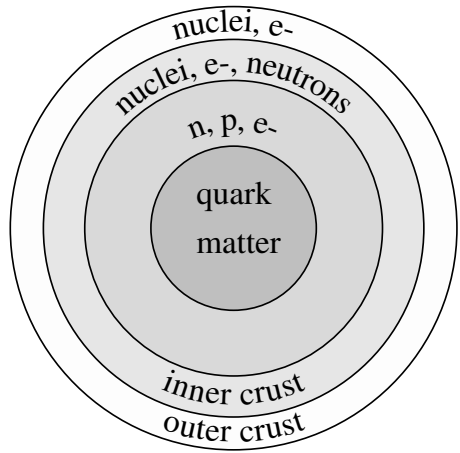}
	\caption{\label{fig:star}
		Cross section of a hybrid star. An inner core of quark matter
		is surrounded by a neutron liquid, also containing some protons
		and electrons. Further out nuclei are also present, along with
		neutrons that literally drip out of the nuclei, and $e^-$ to
		keep charge neutrality. The outermost layer is a nuclear crust.}
\end{figure}

\index{strange star|(}
\comment{References for this section?}
If strange quark matter is absolutely stable, all known neutron stars
are not neutron stars at all, but most likely \emph{strange stars},
\ie\ stars composed of strange
quark matter. In the event of metastable strange quark matter, a core
of strange quark matter could still be present, in what would then be
denoted a \emph{hybrid star}. Both strange stars and hybrid stars would
probably be surrounded by a nuclear crust composed of mainly ${}^{56}$Fe
nuclei, arranged in a Coulomb lattice, as well as  electrons.
Figure \ref{fig:star} shows a schematic cross section of a hybrid
star (see Ref.~\cite{ShapiroTeukolsky} for details).
The outermost layer is the nuclear crust, just mentioned, while
deeper in the star neutron rich nuclei, and electrons coexist with
superfluid neutrons. At greater depths the density becomes so high that
electrons and protons are converted to neutrons and neutrinos, in the
inverse $\beta$-decay reaction $p+e^- \to n +\nu_e$, and the predominant
component is a superfluid neutron liquid. Finally, if the pressure
becomes high enough, even neutrons dissolve into their constituent
quarks, giving way to a core of strange quark matter.

A strange star is what results if strange quark matter is absolutely
stable, in which case, the quark matter core would extend all the way
out to the outer crust.
For a review of strange stars see Alcock \cite{Alcock91a} or Weigel
and Weber \cite{Weigel94a}.

Unfortunately it is rather difficult to distinguish between a strange
star and a neutron star observationally. A strong indication of the
existence strange stars, and thus the absolute stability of strange
quark matter, would be the observation of a pulsar with a rotation
period less than 1 ms. The reason is that a neutron star pulsar has an
upper limit to its rotational speed of about 1000 rotations per second. 
If it were to rotate faster than this the centrifugal force at the
surface of the star would exceed the gravitational force, resulting in
mass loss, and slowing down of the stars rotation. This is the so-called
\Index{Kepler limit}. Another upper limit to a neutron stars rotation
frequency is set by Einstein's theory of general relativity, which
predicts that a neutron star spinning faster than around 1000 revolutions
per second will radiate energy in the form of \Index{gravitational
radiation} due to its deviation from spherical symmetry.

A strange star, on the other hand, is not subject to this restriction
since it is bound not by the gravitational force, but by the strong
force, allowing it to rotate about twice as fast. So far no pulsars
exceeding the limits for neutron stars have been observed---the fastest
known pulsar rotates at 641 revolutions per second \cite{Weigel94a}.

\index{strange star|)}

\index{cosmic rays|(}
Strange quark matter may be present in the cosmos, arriving at Earth as
a component of cosmic rays. Such a cosmic presence could be remains of
nuggets formed in the early universe, as discussed in section
\ref{sec:earlyuniverse}, or it could originate from the collision of two
compact objects,  one of them being a strange star.

Very strict flux limits have been set on such a cosmic presence of
strange quark matter, by Madsen \cite{Madsen88a,Madsen94b}, using stars
as detectors. The argument is that if a \Index{supernova} progenitor is
hit by a quark nugget that makes its way to the interior of the star, it
would still be around after the \Index{supernova} explosion, in the
collapsed core. Once there, it could initiate the conversion of a
\Index{neutron star} into a \Index{strange star}. A neutron star could
also absorb a quark nugget after its birth. By this argument all neutron
stars should in reality be strange stars if strange quark matter is
stable (and an initial cosmic abundance of nuggets was present).
On the other hand if a single neutron star is identified as a neutron
star, strange quark matter is likely not stable, or there are no nuggets
left over from the \Index{big bang}.
\index{cosmic rays|)}

\section{Thesis Outline}
Concluding this introductory chapter, let me give a brief outline of the
remainder of this thesis.

Chapter \ref{chap:mit} gives an introduction to the MIT bag model, which
I have used as the basis of all my work. Following this introduction, is
a description of two alternative ways to treat strange quark matter in
the MIT bag model: the shell model, in which quarks are filled into
single particle states, and the liquid drop model, which is a more
statistical approach. Both of these models are inspired by their
well-known namesakes from nuclear physics. The main advantage of the
liquid drop model, adopted for most of the remaining chapters, is that
many quantities can be evaluated analytically, albeit not in the most
general cases. In contrast, the shell model must be explored using
numerical techniques, but is able to reveal features such as shell
closures, which are not accessible in the liquid drop model.

Chapter \ref{chap:temperature} deals with strange quark matter at
non-zero temperature. The basic methods developed in this chapter is
used in several of the later chapters, where additional complications
are added to the rather simple picture presented in Chapter
\ref{chap:temperature}.

In Chapter \ref{chap:equilibrium}, I treat the equilibrium between strange
quark matter and hadronic matter, at finite temperature---a situation of
importance for the possible creation of strangelets in heavy-ion
collisions. The emphasis is on the effect of strangeness on the
equilibrium between the two phases, and the effect of finite size on the
quark matter phase.

Chapter \ref{chap:projection} develops the methods used in Chapter
\ref{chap:color} and Chapter \ref{chap:suppression} to investigate the
effect of demanding exact color singletness. Most of the material in
Chapter \ref{chap:projection} has appeared in the literature, but some
of the more detailed calculations have been added, hopefully making it
easier to read than the original papers.

In Chapter \ref{chap:color} I apply the color singlet projection,
developed in Chapter \ref{chap:projection}, to describe color singlet
strangelets at finite temperature. Most of the results are for massless
quarks, but massive quarks are also treated.

Chapter \ref{chap:suppression} deals with the suppression of the rate of
quark matter droplet formation, due to the color singlet constraint.
This has important consequences for quark-gluon plasma formation in
heavy-ion collisions, as well as the nucleation of quark matter in
neutron stars.

Finally, I have included three appendices. Appendix \ref{chap:mre} is an
introduction to the multiple reflection expansion, on which the liquid
drop model is based. Apart from reproducing some results from the
literature in a slightly generalized form, I also show how to obtain
the density of states, which is the quantity of interest in the liquid
drop model.

Appendix \ref{app:integrals} gives results of special cases of the
Fermi-Dirac and Bose-Einstein integrals, which are used throughout the
thesis, but which do not seem to be well known, and are 
somewhat inaccessible through the standard literature.

Since, as mentioned above, analytical solution of the shell model is
not possible, a description of how to solve it is equivalent to providing
an algorithm.  Therefore I have given, in Appendix \ref{app:shell}, 
an implementation of a program
to calculate strangelet masses in the shell model.

\clearemptydoublepage
\chapter{Quark Matter in the MIT Bag\label{chap:mit}}
\index{MIT bag model|(}

The standard model of elementary particle physics describes the
strong interaction in terms of a Yang-Mills
\index{Yang-Mills theory@Yang--Mills theory} gauge theory based on the
internal symmetry of color $SU(3)$, \index{SU(3)@$SU(3)$} in which the
dynamical degrees of freedom are the quark and gluon fields. In principle
the properties of quark matter should be derived on the basis of this
theoretical foundation. However, the only analytical scheme of calculation
available is \Index{perturbation theory}, which is only applicable if the strong
fine structure constant is sufficiently small, \ie\ at high energy
or equivalently at small scales.  Hence all
low-energy phenomena are inaccessible to analytical calculations, and at
present the only alternative is to do lattice calculations, which is a
numerical solution to the field equations of QCD in a discretized
space-time.

Because of our inability to perform calculations, many so-called
phenomenological models have been proposed. These are models which are
believed to contain a large part of the relevant physics, and which are
simpler to solve than the original equations. The MIT bag model---developed
at the Massachusetts Institute of Technology, whence the name---is one
such model. In the following a short introduction to this model will be
given.

\section{The MIT Bag Model\label{sec:bagmodel}}
The basic idea of the MIT bag model \cite{Chodos74a,Johnson75a} is that
quarks and gluons should be confined within a finite region of space.
This region of space therefore acts as a container for the fields, and
is somewhat colloquially referred to as a `bag'.  By construction, the
MIT bag model therefore reproduces one well known property of QCD,
\viz\ \Index{confinement}.

The region of space $\Omega$ inside the bag is endowed with a constant
energy density which, as will be seen, acts as a source of confinement.
This energy density $B$ is an extra parameter of the model, which
otherwise depends only on the usual parameters such as quark masses and
coupling constants. All the non-perturbative effects are put into the
structure of the so-called \Index{perturbative vacuum} inside the bag, in which
the quark and gluon fields propagate. This means that 
\Index{perturbation theory} is applicable to fields inside the bag
as opposed to
the field free region---the true vacuum---outside the bag. The constant
energy density $B$ of the perturbative vacuum is called the
\textit{bag constant}\index{bag constant}.

Since fields are only allowed inside the bag, we may summarize the model
by saying that it is described by the Lagrangian\index{Lagrangian!for
MIT bag model}\index{MIT bag model!Lagrangian}
\be
	L = \int d^3\mbf{r} \left( \mcal{L}_\mrm{QCD} - B
		\right) \theta_\Omega(\mbf{r}) ,
\ee
where $\theta_\Omega(\mbf{r})$ is a function which is unity for
$\mbf{r} \in \Omega$ and zero otherwise, and $\mcal{L}_\mrm{QCD}$ is the
ordinary QCD Lagrangian density. Applying the variational principle to
this Lagrangian will give a set of field equations, valid in $\Omega$,
as well as accompanying boundary conditions to be satisfied on the bag surface
\index{boundary conditions!of MIT bag model|(}
$\partial\Omega$. The field equations are the ordinary field equations
of QCD, and a careful treatment (see \cite{Chodos74a}) yields the
appropriate boundary conditions.

I will only consider non-interacting fields, but perturbative
corrections are possible, \eg\ in the framework developed by Hansson and
Jaffe \cite{Hansson83a}.
Note that I do not use the terms `non-interacting' and `free'
interchangeably. The latter term is taken to mean a quark (or gluon)
which is not confined and non-interacting, whereas the former term is
used for a non-interacting but confined particle.
For non-interacting  quarks the equation of motion and the boundary
conditions are given by
\comment{The bag surface is static}
\be \label{eq:quarkinbag}
    \begin{array}{r@{\;=\;}lr@{\;\in\;}l}
	\left( i\gamma^\mu\partial_\mu - m\right)\Psi(x) & 0 & \mbf{x}
	& \Omega \\[1ex]
	in_\mu\gamma^\mu\Psi(x)& \Psi(x) &\mbf{x}& \partial\Omega \\[1ex]
	n_\mu\partial^\mu \bar{\Psi}(x)\Psi(x) & -2B \qquad\qquad &\mbf{x}&
	\partial\Omega .
    \end{array}
\ee
Here $n^\mu = (n_0, \mbf{n})$, where $\mbf{n}$ is the outward unit
normal to the bag surface $\partial\Omega$, and $n_0=0$ in the
instantaneous rest frame, giving the normalisation $n_\mu n^\mu = -1$. 
The first equation is the Dirac equation \index{Dirac equation!for quark
in MIT bag} for a non-interacting quark. The remaining two equations are
the boundary conditions of the bag model.

The second equation is a
boundary condition saying that there is no quark current across the bag
surface. The 4-vector probability current of a Dirac field is $j^\mu =
\bar{\Psi}\gamma^\mu\Psi$. To see that the first boundary condition is
equivalent to zero quark current crossing the boundary consider the
equivalent equation for the conjugate spinor $\bar{\Psi}= 
\Psi^\dagger\gamma^0$, which is easily obtained  using the
\Index{Clifford algebra}
$\{\gamma^\mu, \gamma^\nu\}=2\eta^{\mu\nu}$, where $\eta^{\mu\nu}
=\mrm{diag}(1,-1,-1,-1)$ is the Minkowski metric,\index{Minkowski metric}
giving
\be \label{eq:antiquarkboundary}
	\bar{\Psi}(x)in_\mu\gamma^\mu = -\bar{\Psi}(x) .
\ee
Next consider the contraction of $n_\mu$ with the probability current
\be
	in_\mu j^\mu(x) = \bar{\Psi}(x) in_\mu\gamma^\mu \Psi(x) .
\ee
Using the first boundary condition in Eq.~(\ref{eq:quarkinbag}) to evaluate
the right hand side gives $\bar{\Psi}\Psi$, whereas using
Eq.~(\ref{eq:antiquarkboundary}) gives $-\bar{\Psi}\Psi$. This shows
that $\bar{\Psi}\Psi=0$ and
\be
	n_\mu j^\mu(x) = 0 \qquad\qquad \mbf{x}\in \partial\Omega ,
\ee
which means that the component of the \Index{quark current} perpendicular
to the bag surface vanishes at the surface.

The second boundary condition in Eq.~(\ref{eq:quarkinbag}) expresses
equality between the pressure exerted by the quarks on the surface and
the bag constant which acts as an external pressure. Without the quark
fields to counter this pressure the bag would not be stable. To see that
the second boundary condition is equivalent to pressure equilibrium
look at the \Index{energy momentum tensor} 
\be
	T^{\mu\nu}(x) = T_q^{\mu\nu}(x) + \eta^{\mu\nu}B
\ee
for a quark field in a bag. The energy momentum tensor for the quark
field alone is \cite{ItzyksonZuber}
\be
	T_q^{\mu\nu}(x) = -\frac{i}{2} \left\{
		\bar{\Psi}(x)\gamma^\mu\partial^\nu\Psi(x)
		-\left[\partial^\nu\bar{\Psi}(x)\right]
		\gamma^\mu\Psi(x) \right\} .
\ee
The momentum flow through the surface, as given by $n_\mu T^{\mu\nu}$,
should be zero. The contraction of $n_\mu$ with the energy momentum
tensor is easily evaluated at the bag surface by using the first
boundary condition in Eq.~(\ref{eq:quarkinbag}) and
Eq.~(\ref{eq:antiquarkboundary}) to give
\be
	n_\mu T^{\mu\nu}(x) = -\frac{1}{2}\partial^\nu
		\bar{\Psi}(x)\Psi(x) + n^\nu B  \qquad
		\mbf{x}\in\partial\Omega .
\ee
Contracting with $n_\nu$ and setting the result equal to zero, using
$n_\nu n^\nu=-1$ then gives
\be
	0 = n_\nu n_\mu T^{\mu\nu}(x) = -\frac{1}{2}n_\nu\partial^\nu
		\bar{\Psi}(x)\Psi(x) - B \qquad
		\mbf{x}\in\partial\Omega ,
\ee
which is recognized as the second boundary condition in
Eq.~(\ref{eq:quarkinbag}). The interpretation is that the quark pressure
$-\frac{1}{2}n_\nu\partial^\nu\bar{\Psi}(x)\Psi(x)$ balances the bag
pressure $B$.

The ground state of an MIT bag only contains quarks, but
if gluons are present (at finite temperature) they also contribute
to the pressure through a term \cite{ItzyksonZuber}
\be
	 T_g^{\mu\nu} = -\frac{1}{4}\eta^{\mu\nu}
			F^{\alpha\beta}F_{\alpha\beta}
			+F^{\mu\lambda}F_\lambda^{\phantom{\lambda}\nu} ,
\ee
in the energy momentum tensor, which, since the gluon field tensor 
$F^{\mu\nu}$ has to fulfill the boundary condition
$n_\mu F^{\mu\nu}=0$, gives a contribution to the pressure which is
$\frac{1}{4}F^{\alpha\beta}F_{\alpha\beta}$.

This short discussion of the boundary conditions of the MIT bag shows
that the idea of confining fields to a region of space naturally leads
to boundary conditions for the quark and gluon fields which correspond
to the fact that there is no probability current across the bag surface. A
second boundary condition is interpreted as pressure equilibrium, between
the dynamical pressure generated by quarks and gluons inside the bag on
the one side and the pressure exerted by the physical vacuum outside the
bag on the perturbative vacuum inside the bag on the other side.
\index{boundary conditions!of MIT bag model|)}
\index{MIT bag model|)}

\section{Shell Model}
\index{shell model|(}
The description of quark matter (or ordinary hadrons) in
the framework of the MIT bag model is a question of solving the Dirac
equation with the appropriate boundary conditions, as given by
Eq.~(\ref{eq:quarkinbag}). Given the single quark energy levels, it is a
matter of filling quarks into these levels using the \textit{Aufbau
principle}.
\comment{Elaborate a bit here}


For a static spherical bag with radius $R$ the Dirac equation can be
split into an angular part and a radial part. The radial equation
is \cite{Flamm82a,Madsen95a} \index{Dirac equation!radial}
\be \label{eq:diracshell}
	f_{\kappa}(x_{\nu\kappa}) = \frac{-x_{\nu\kappa}}{(x_{\nu\kappa}^2 + 
	m^2R^2)^{1/2} + mR} f_{\kappa-1}(x_{\nu\kappa}),
\ee
where $x_\kappa = k_\kappa R$ is the momentum $k_\kappa$ in units of the
inverse radius, and $m$ is the quark mass. The function $f_\kappa(x)$ is
given by
\be \label{eq:fkx}
	f_{\kappa}(x) = \left\{ \begin{array}{ll}
              j_{\kappa} (x) & \kappa\geq 0 \\
              y_{\kappa} (x)=(-1)^{\kappa +1} j_{-\kappa -1}(x)\quad
	      & \kappa <0
              \end{array}
                \right. ,
\ee
where $j_{\kappa} (x)$ and $y_{\kappa} (x)$ are spherical Bessel
functions \cite{Abramowitz} \index{Bessel function}  of order $\kappa$.
The quantum number $\kappa$ is the eigenvalue of the operator $K = \beta
(\boldsymbol{\sigma}\cdot\boldsymbol{\ell} + 1)$ where $\beta$ is the
Dirac $\beta$-matrix, $\boldsymbol{\sigma}$ the Pauli
matrices, \index{Pauli matrices} and $\boldsymbol{\ell}$ the angular
momentum operator. The possible values of $\kappa$ are
\cite{BjorkenDrellQM}
\be
	\kappa = \pm \left( j+\frac{1}{2} \right) ,
		\quad\mbox{for}\quad j = l \mp \frac{1}{2} ,
\ee
where $j(j+1)$ is the eigenvalue of the square of the total angular
momentum $\mbf{j} = \boldsymbol{\ell} + \mbf{s}$.

\begin{table}
 \begin{center}
  \begin{tabular}{ccr@{}lrr}\hline
	$l$ & $j=l\mp 1/2$ & $\kappa = \pm$&$ (j + 1/2 )$ &
	$d_\kappa$ & \multicolumn{1}{c}{$x_{1,\kappa}$}  \\\hline\hline
	0 & 1/2	& $-$&1 & 6 & 2.04279\\
	1 & 1/2 & &1    & 6 & 3.81135\\
	1 & 3/2 & $-$&2 & 12 & 3.20394\\
	2 & 3/2 & &2    & 12 & 5.12302\\
	2 & 5/2 & $-$&3 & 18 & 4.32726\\
	3 & 5/2 & &3    & 18 & 6.37010\\
  \end{tabular}
 \end{center}
 \caption{ \label{table:shell} The quantum number systematics and
            degeneracies of the solutions to Eq.~(\ref{eq:diracshell}).
			The solutions listed in the last column are for a massless
			quark species.
	  }
\end{table}

The quantum number $\nu=1,2,\ldots$ in Eq.~(\ref{eq:diracshell}) is a principal 
quantum number which labels the different solutions to the equation for
fixed $\kappa$. The degeneracy of a multiplet $|\nu\kappa\rangle$ is
$d_\kappa = 3(2j+1)$, where the factor 3 is due to the color degree of
freedom. The solutions to Eq.~(\ref{eq:diracshell}) for a massless
quarks species (in which case $x_{\nu\kappa}$ is independent of $R$) is
shown along with the degeneracies for the first few values of $\kappa$, $l$, and
$j$ in Table~\ref{table:shell}.

\begin{figure}
	\includegraphics[width=\textwidth]{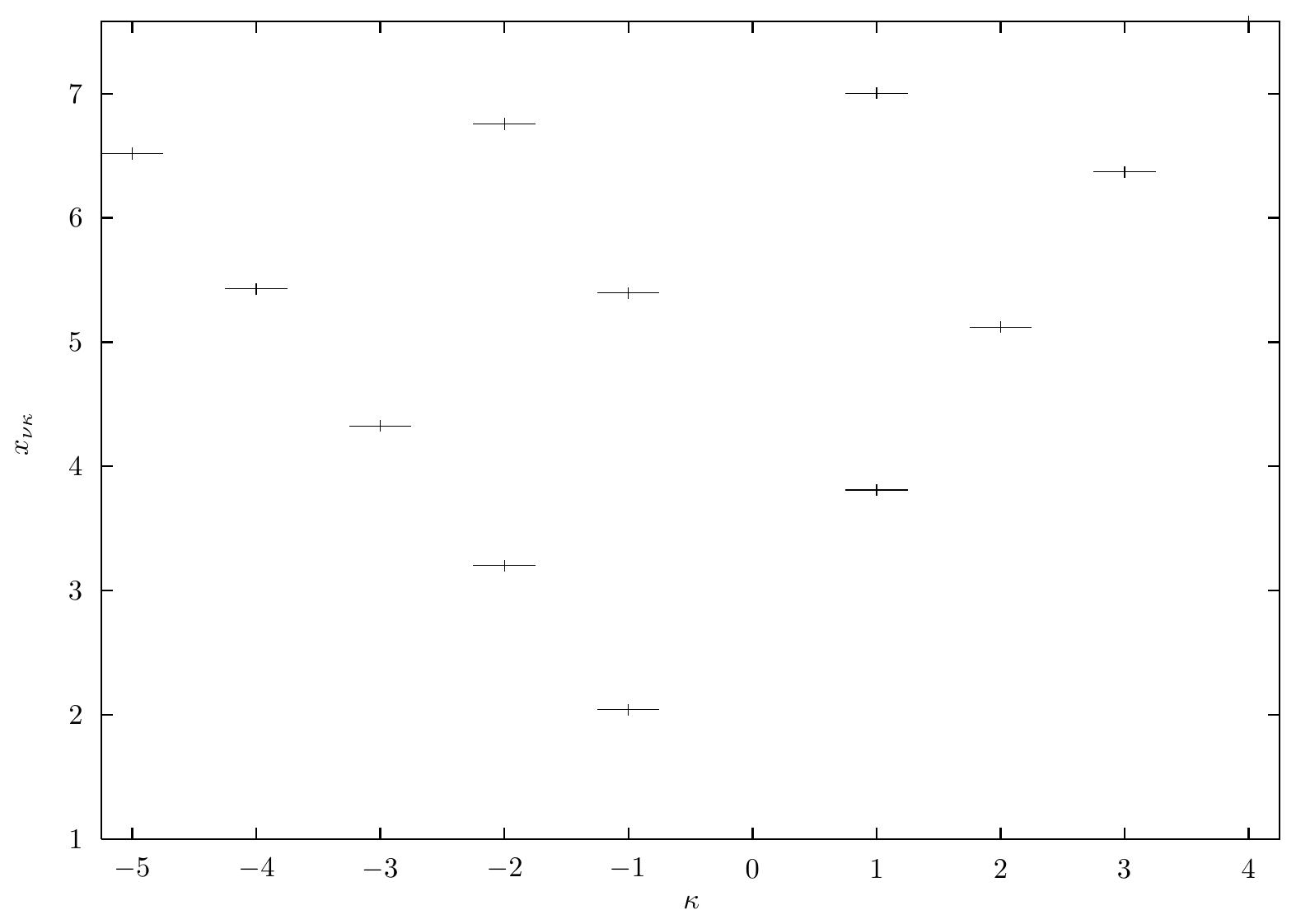}
	\caption{\label{fig:spectrum}
		The lowest lying solutions $x_{\nu\kappa}$ to
		Eq.~(\ref{eq:diracshell}), for a massless quark species.
		Some of these levels are tabulated in Table \ref{table:shell}.
		 }
\end{figure}

The first few solutions are also shown in Fig.~(\ref{fig:spectrum}),
where the order of the levels is more clearly seen. Since this is
again for a massless quark species, the ordinate is readily converted to
an energy (momentum) by dividing by the radius (in units of MeV$^{-1}$)
of the spherical cavity. For a given value of $j$ the level with
negative $\kappa$ is observed to lie below the level with positive
$\kappa$. The lowest lying $\nu =2$ level is $\nu=2$, $\kappa=-1$, which
is lower in energy than the $\nu=1$, $\kappa=3$ level listed in Table
\ref{table:shell}.

With the solutions $x_{\nu\kappa}$ to the radial equation
(\ref{eq:diracshell}) given, the total energy for a lump of quark matter
in a bag with radius $R$ is 
\be \label{eq:shellenergy}
	E = \sum_{q, \nu, \kappa, j} \epsilon_{q, \nu, \kappa}(R)
		+ \frac{4}{3}\pi R^3 B ,
\ee
where $q=u,d,s$ is a flavor index, and the single particle energies are
given by
\be
	\epsilon_{q, \nu, \kappa}(R) = \sqrt{x_{\nu\kappa}^2R^{-2} + m_q^2} .
\ee
The energy arising from the non-zero vacuum energy density, as given by
the bag constant $B$, has been added to the energy of the quarks
to obtain the total energy.
To construct a strangelet with a fixed baryon number $A$ the constraint
\be
	3A =  \sum_{q} N_q ,
\ee
must be satisfied, where $N_q$ is the number of quarks of species $q$.
The lowest energy for a given baryon number is then
obtained by minimizing the energy with respect to radius, flavor
composition, and the distribution among the different levels.

\begin{figure}
	\includegraphics[width=\textwidth]{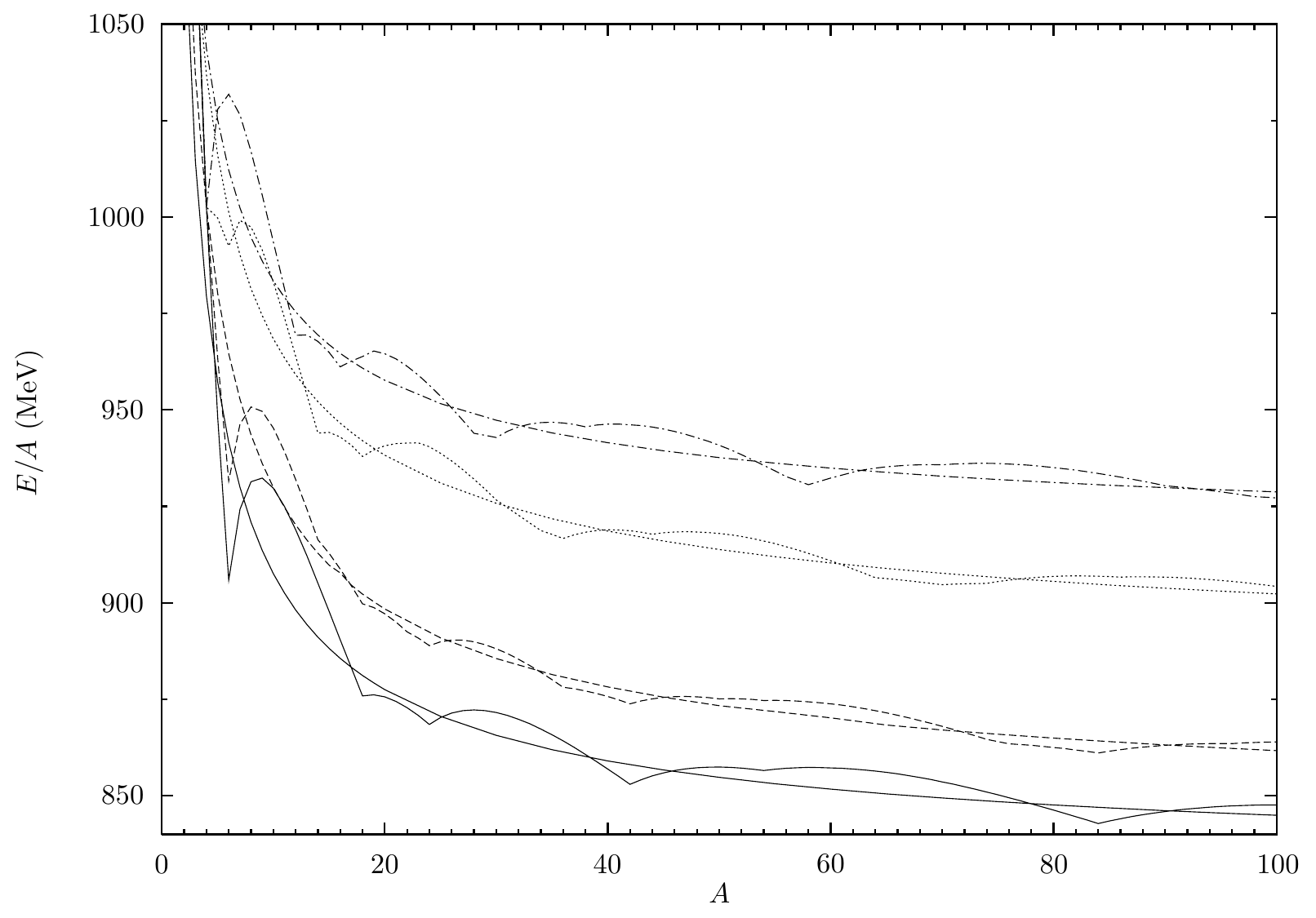}
	\caption{\label{fig:ld_and_shell}
		 The energy per baryon as a function of baryon
		 number for three flavor quark matter, with $B^{1/4} =
		 145$ MeV and $m_u=m_d=0$. Full curves are for $m_s =
		 0$, dashed curves for $m_s=50$ MeV, dotted curves for
		 $m_s=150$ MeV, and dot-dashed curves for $m_s=250$ MeV
		 For each value of $m_s$ both liquid drop and shell
		 model results are shown.}
\end{figure}

Such calculations have been performed by Vasak, Greiner, and Neise 
\cite{Vasak86a} for two flavor quark matter, and by 
Greiner, Rischke, St\"ocker, and Koch \cite{greal88}, Gilson and Jaffe
\cite{Gilson93a} and Madsen \cite{Madsen94a} for strangelets. In
Appendix \ref{app:shell} I have provided a program which implements the
solution of Eq.~(\ref{eq:diracshell}). The resulting plot of the energy
per baryon as a function of baryon number for 3 quark flavors can be seen in
Fig.~(\ref{fig:ld_and_shell}), where results of liquid drop model
calculations are also shown. Conspicuous shell effects are seen---most
\index{shell effects}
notably the shell closure that occurs at $A=6$ (except for the highest
value of $m_s$) which corresponds to 6
quarks of each flavor in the lowest energy level $\nu=1$, $\kappa=-1$.
The next three quarks added at $A=7$ must go into the level with the
next to lowest energy resulting in a sharp increase in the energy per
baryon. Shell closures also occur at higher baryon numbers, but the
locations of these ``magic numbers'' \index{magic numbers}
are seen to depend on the strange
quark mass ($u$ and $d$ are assumed massless). For moderate strange quark
masses the next shell closure occurs at $A=18$ which corresponds to
\comment{The $\kappa=-2$ level is lower than the $\kappa=1$ level}
three flavors filling the level $\nu=1$, $\kappa=-2$. For $m_s > m_u,
m_d$ (which
is certainly the case in reality) the lighter $u$ and $d$ quarks fill
the $\nu=1$, $\kappa=-1$ level before the $s$ quarks, with the result that
the shell model calculations for different values of $m_s$ yield identical
results for $A\leq 4$, since these systems consist entirely of $u$ and
$d$ quarks.

\index{shell model|)}

\section{Liquid Drop Model\label{sec:drop}}
\index{liquid drop model|(}

The Liquid drop model of strange quark matter in the MIT bag is
essentially a Fermi gas model, where each quark flavor $q=u,d,s$ is
ascribed a chemical potential $\mu_q$, and the energy levels are filled
according to the Fermi-Dirac distribution. At finite temperature there
are also antiquarks and gluons.

\subsection{Density of States\label{sec:DOS}}
\index{density of states|(}
In the case of bulk quark matter the
boundary conditions have no effect---one may as well choose periodic
\index{boundary conditions!periodic}
boundary conditions as is often done when quantizing fields in an
large volume. The allowed momenta are then 
\be
	\mbf{k}= \frac{2\pi\mbf{n}}{L}
 		\qquad\qquad n_i\in\mathbb{Z} ,
\ee
where $L$ is the linear dimension of the box. Sums over $\mbf{k}$
can be replaced by integrals in the large volume limit
\be \label{eq:sumtoint}
	\sum_{\mbf{k}} \to \frac{V}{(2\pi)^3}\int d^3\mbf{k} ,
\ee
where $V=L^3$ is the volume of the box, and $V/(2\pi)^3$ is the
number of states per volume in momentum space.
The density of states $\rho(k)$ is introduced  as
\be \label{eq:volumedos}
	\frac{V}{(2\pi)^3}\int d^3\mbf{k}\,
	= \int_0^\infty dk\, \rho(k)
	\quad\Longrightarrow\quad  \rho(k) = \frac{Vk^2}{2\pi^2} ,
\ee
For small system size there are important corrections to this density of
states, which depend on the boundary conditions imposed, as well as on
\index{boundary conditions!effect on density of states}
the detailed geometry of the confining surface. These finite size
corrections to the density of states can be calculated by using the
\index{finite size corrections|(}
\textit{multiple reflection expansion} \index{multiple reflection
expansion} method developed by Balian and
Bloch \cite{Balian70a,Balian70b,Balian74a}. I have given a detailed
description of the multiple reflection expansion in Appendix
\ref{chap:mre}, with emphasis on the case of a spinor field satisfying
the Dirac equation with MIT bag boundary conditions.
The multiple reflection expansion density of states has the form
\index{multiple reflection expansion!of the density of states}
\be \label{eq:DOS}
	\rho(k) =
		\frac{Vk^2}{2\pi^2} + f_S\left(\frac{m}{k}\right)kS
		+ f_C\left(\frac{m}{k}\right) C + \dots ,
\ee
where  $S$ is the surface area of the
confining boundary, and $C$ is the
extrinsic curvature given by the surface integral
\index{curvature!extrinsic}
\be
	C = \oint_{\partial\Omega} d^2\boldsymbol{\sigma}
		\left(\frac{1}{R_1(\boldsymbol{\sigma})}
		+\frac{1}{R_2(\boldsymbol{\sigma})} \right) .
\ee
$R_1$ and $R_2$ are the principal radii of curvature. For a spherical
system $V=4/3\pi R^3$, $S=4\pi R^2$, and $C=8\pi R$, and
Eq.~(\ref{eq:DOS}) is seen to be an expansion in powers of $(kR)^{-1}$.

The multiple reflection expansion is valid whenever the surface is
\index{multiple reflection expansion!validity}
smooth (on the scale set by the momentum), which is why the surface
$\partial\Omega$ can be represented by just two invariants, \viz\ 
the surface area
and the extrinsic curvature. Another well known invariant quantity, the
Gaussian curvature
\index{curvature!Gaussian}
\be
	K = \oint_{\partial\Omega} d^2\boldsymbol{\sigma}
	   \frac{1}{R_1(\boldsymbol{\sigma})R_2(\boldsymbol{\sigma})} ,
\ee
is of higher order in the inverse of the principal radii of curvature,
and is therefore
among the left out terms, indicated by `\dots ' in Eq.~(\ref{eq:DOS}).

The volume term
is seen to be universal in the sense that it is independent of the
nature of the field (\ie\ the field equation) and the boundary
condition.  The coefficient functions $f_S$ and $f_C$ depend on the
momentum and the mass as indicated, but also on the type of field
(scalar, spinor, vector,\dots) as well as on  the boundary condition.

\subsubsection{Quarks and Antiquarks}
\index{density of states!for quarks|(}
The density of states is the same for antiquarks as for quarks, so
all the results in this section, although stated for quarks, are also
valid for antiquarks.

The detailed calculation of the surface and curvature terms in the
multiple reflection expansion of the density of states for quarks has never
appeared in the literature, but Berger and Jaffe reported the result of
a calculation of the surface term in Ref.~\cite{Berger87a,Berger91a},
which was later
pointed out by Mardor and Svetitsky \cite{Mardor91a} to be wrong by a
factor of 2. The (corrected) result obtained by Berger and Jaffe is
\be \label{eq:quarksurfaceterm}
	f_S\left(\frac{m}{k}\right) = -\frac{1}{8\pi} \left( 1
	-\frac{2}{\pi}\tan^{-1} \frac{k}{m} \right) .
\ee
It is interesting to note that in the limit of zero mass this expression
vanishes. This has the important consequence that for massless quarks
the leading order correction to the bulk approximation to the density of
states is the curvature term.

The curvature coefficient $f_C$ has not been calculated using the
multiple reflection expansion in the general case of massive quarks,
but for massless quarks with MIT bag boundary conditions the result is
\cite{Farhi84a,Mardor91a} $f_C = -1/24\pi^2$. In the limit of infinite mass the
quark can be considered non-relativistic and the field equation reduces
to the wave equation. In this case the boundary condition of the MIT
bag is the well known Dirichlet boundary condition \cite{Chodos74a}
studied by Balian and Bloch \cite{Balian70a} who derived the result
$f_C = 1/12\pi^2$. The limits $m\to 0$ and $m \to \infty$ are thus
known, and it was shown by Madsen \cite{Madsen94a} that the expression
\be \label{eq:quarkcurvatureterm}
	f_C\left(\frac{m}{k}\right) = \frac{1}{12\pi^2} \left[ 1 -
        \frac{3k}{2m} \left(\frac{\pi}{2} - \tan^{-1} \frac{k}{m}\right)
      \right] ,
\ee
has the right limiting values and furthermore is in exceedingly good
agreement with shell model calculation (see also
Fig.~\ref{fig:ld_and_shell} and Fig.~\ref{fig:numofstates}).
In view of the very good agreement both
with the known analytic limits and with shell model calculations it
seems more than reasonable to adopt Eq.~(\ref{eq:quarkcurvatureterm})
as the curvature term for quarks.
\index{density of states!for quarks|)}

\subsubsection{Gluons}
\index{density of states!for gluons|(}

The surface term for gluons is zero, as it is for massless quarks. 
The MIT bag boundary condition $n_\mu F^{\mu\nu} = 0$ is equivalent to
the boundary condition studied by Balian and Bloch\cite{Balian70b} for the
electromagnetic field, and it gives the curvature term \cite{Mardor91a}
\be
	f_C = -\frac{1}{6\pi^2}.
\ee
\index{density of states!for gluons|)}
\index{finite size corrections|)}

\subsubsection{Comparison to Shell Model}
\begin{figure}
	\includegraphics[width=0.95\textwidth]{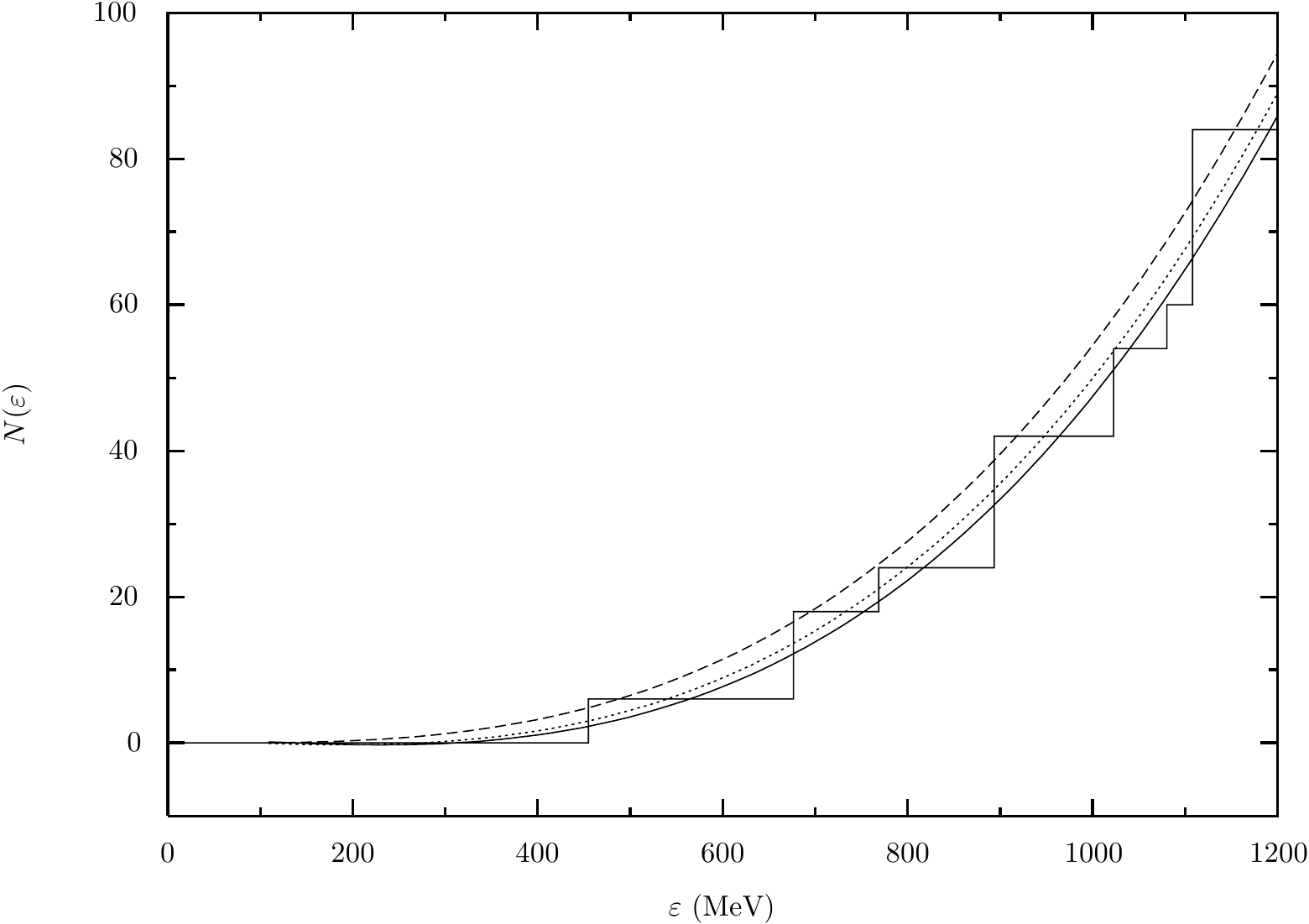}
	\caption{\label{fig:numofstates}
		 Plot of the number of states $N(\varepsilon)$
		 with energy less than $\varepsilon$ for shell model
		 and liquid drop model. Parameters chosen are: $m = 100$
		 MeV and $R=1$~fm. The smooth solid line
		 includes both volume, surface, and curvature terms, the
		 dotted line is for volume and surface terms, and the
		 dashed curve is the volume term only.}
\end{figure}

\index{liquid drop model!compared to shell model|(}
\index{shell model!compared to liquid drop model|(}
The expression for the density of states of quarks as given by
Eqs.~(\ref{eq:DOS}), (\ref{eq:quarksurfaceterm}), and
(\ref{eq:quarkcurvatureterm}) can be compared to the spectrum obtained
by solving the Dirac equation (\ref{eq:diracshell}) numerically. Since
this exact density of states is a sum of delta functions
\be
	\rho(k) =
		\sum_{\nu,\kappa} d_\kappa \delta(k - k_{\nu\kappa}) ,
\ee
it is more convenient to compare the integrated density of states
\be
	N(k) = \int_0^k dk' \rho(k') ,
\ee
which is the number of states with momentum less than $k$. This will be
a step function in the shell model. In the liquid drop model the volume,
surface, and curvature terms can immediately be integrated to give
\begin{eqnarray}
	N(k) &=& VN_V(k) + SN_S(k) + CN_C(k) \\
	N_V(k) &=& \frac{k^3}{\pi^2} \\
	N_S(k) &=& -\frac{3m^2}{4\pi^2}  \left\{ \frac{k}{m}
		+ \frac{\pi}{2}\left(\frac{k}{m}\right)^2
		-\left[ 1 + \left(\frac{k}{m}\right)^2\right]
		\tan^{-1}\frac{k}{m} \right\} \\
	N_C(k) &=& \frac{m}{8\pi^2}\left\{ \frac{k}{m} -\frac{3\pi}{2}
		\left(\frac{k}{m}\right)^2 + 3\left[ 1+\left(
		\frac{k}{m}\right)^2\right] \tan^{-1}\frac{k}{m} \right\} .
\end{eqnarray}
In Fig.~\ref{fig:numofstates} a plot of the number of states
$N(\varepsilon)$ with energy less than $\varepsilon$, has been plotted
using the above expressions for the liquid drop model, and a numerical
calculation for the shell model. The plot shown is for a single 
quark species, and it is seen that the inclusion of the surface and
curvature terms in the liquid drop calculation give a much better
agreement with the shell model result.

\index{density of states|)}
\index{liquid drop model!compared to shell model|)}
\index{shell model!compared to liquid drop model|)}

\subsection{Partition Function\label{sec:partitionfunction}}
\index{partition function|(}
In order to calculate the energy of a strangelet in the liquid drop
model to get a result that parallels Eq.~(\ref{eq:shellenergy}) in the
shell model it is convenient to go through the grand canonical partition
\index{partition function!grand canonical}
function, since it at once generalizes the liquid drop model to finite
temperature.

The grand canonical partition function \Z\ is defined as the trace over
all states
\be
	\Z = \Tr e^{-\beta\left( \Ham - \mu_i\mcal{N}_i\right) } ,
\ee
where $\beta$ is the inverse temperature, \Ham\ is the Hamiltonian, and
$\mcal{N}_i$ is the number operator for particles of species $i$, with
$\mu_i$ being the corresponding chemical potential. For a system of
non-interacting particles, as is the case here, the Hamiltonian splits
up into a sum of single particle Hamiltonians, and the partition
function factorizes. So for a collection of quarks, antiquarks, and
gluons in an MIT bag, the logarithm of the partition function may be
written as the sum
\be
	\ln \Z = \sum_i \ln \Z_i -\beta BV ,
\ee
in which the summation index $i$ runs over all species present, and the
last term is the contribution from the bag constant. The contribution
from each species is of the form
\be	
	\ln \Z_i =  \pm g_i \int_0^\infty dk\, \rho_i(k)
		\ln\left[ 1 \pm e^{ -\beta\left( \epsilon_i(k) - \mu_i
		\right) } \right] ,
\ee
where the upper sign is for fermions, and the lower for bosons. Here the
sum over discrete momentum eigenvalues has been supplanted by an
integral over $k$ weighted by the density of states $\rho_i(k)$.
$g_i$ is a statistical weight, taking into account the degeneracy due to
\eg\ spin and color.
The energy $\epsilon_i(k)$ for a particle with mass $m_i$ is given by the
relativistic dispersion relation $\epsilon_i(k) = \sqrt{k^2 + m_i^2}$.
$\mu_i$ is the chemical potential of the species, which is zero for
gluons, since the number of gluons is not a conserved quantity.

To obtain the total energy, baryon number, and other characteristics of
the system one goes through the \Index{thermodynamic potential}
$\Omega$ defined by
\be
	\Omega = -T \ln \Z ,
\ee
with the differential
\be \label{eq:diffomega}
	d\Omega = -\mcal{S}dT - N_id\mu_i -pdV +\sigma dS + \gamma dC ,
\ee
where $\mcal{S}$ is the entropy, $N_i$ is the number of particles of
species $i$ (where a sum over $i$ is implied), $\sigma$ is the surface
tension, and
$\gamma$ is the generalized force conjugate to the curvature, which is
an external parameter on the same footing as temperature, volume and
surface area. The thermodynamic potential introduced above should not be
confused with the set of points comprising the bag, which was denoted by
the same symbol in section \ref{sec:bagmodel}, but the distinction
should always be clear.

Because of the simple dependence of the density of states Eq.~(\ref{eq:DOS})
on the geometry, the thermodynamic potential for a single species
has the form
\be
	\Omega_i(T,\mu_i,V,S,C) = \Omega_{i,V}(T,\mu_i)V +
		\Omega_{i,S}(T,\mu_i)S + \Omega_{i,C}(T,\mu_i)C ,
\ee
which makes it possible to identify the pressure, surface tension, and
curvature coefficient in Eq.~(\ref{eq:diffomega}), as
\begin{eqnarray}
	p&=&-\left({\partial \Omega \over \partial V}
	\right)_{T,\mu,S,C}=
	-\Omega_V(\mu,T) \label{eq:presdef}\\
	\sigma&=&\left({\partial \Omega \over \partial S}
	\right)_{T,\mu,V,C}=
	\Omega_S(\mu,T) \label{eq:sigmadef}\\
	\gamma&=&\left({\partial \Omega \over \partial C}
	\right)_{T,\mu,V,S}=
	\Omega_C(\mu,T). \label{eq:gammadef}
\end{eqnarray}

The \Index{free energy} $F$ is obtained from the $\Omega$ potential, by the
Legendre transformation
$
	F = \Omega +\mu_i N_i ,
$
which gives the differential
\be
	dF = -\mcal{S}dT  + \mu_i dN_i - pdV + \sigma dS + \gamma dC,
\ee
exchanging the natural variable $\mu_i$ of $\Omega_i$ with $N_i$,
where $N_i$ is obtained from $\Omega$ by differentiation
\begin{eqnarray}
	N_i &=&
	 -\left(\frac{\partial\Omega_i}{\partial\mu_i}\right)_{T,V,S,C}
	\nonumber \\
	&=& N_{i,V}(T,\mu_i)V + N_{i,S}(T,\mu_i)S + N_{i,C}(T,\mu_i)C,
\end{eqnarray}
so that $F$ is also a sum of volume, surface, and curvature terms. This
is a feature shared by all those quantities, which in the bulk
limit are extensive thermodynamic variables. With the inclusion of
\Index{finite size effects} these quantities receive contributions from
the surface and curvature terms in addition to the volume term. 

Finally the energy is given by $E=F+T\mcal{S}$, with the differential
\index{energy!differential of}
\be
	dE = -Td\mcal{S} + \mu_i dN_i - pdV + \sigma dS + \gamma dC ,
\ee
which allows us calculate the mass of a strangelet.

\subsubsection{Gluons}
\index{thermodynamic potential!for gluons}
\index{thermodynamic potential|(}
Gluons have a
statistical weight of $g=16$, since they can be in either of 2 helicity
states, and form an octet in color space.

The thermodynamic potential $\Omega_g = -T\ln\Z_g$ for gluons can be
expressed
in terms of the integrals given in Appendix \ref{app:integrals}, after
performing a partial integration to get rid of the logarithm.
Using the results \Eq{eq:sm1Dingle} and \Eq{eq:sm3Dingle} immediately
gives
\be \label{eq:gluonvolume}
	\Omega_{g,V} = -\frac{8}{45} \pi^2 T^4 ,
\ee
for the volume term, and
\be \label{eq:gluoncurvature}
	\Omega_{g,C} = \frac{4}{9} T^2,
\ee
for the curvature term. Both of these terms vanish in the limit of zero
temperature, since gluons are only present as `thermal radiation'.

\subsubsection{Quarks}
\index{thermodynamic potential!for quarks|(}
The thermodynamic potential for quarks $\Omega_q = -T\ln\Z_q$
can be evaluated analytically in the limit
of zero temperature. In the limit of zero quark mass the sum of
thermodynamic potentials for a quark and its corresponding antiquark can
be evaluated provided that $\mu_{\bar{q}} = -\mu_q$, which is the case
if quarks and antiquarks are in equilibrium through processes such as
$q\bar{q} \leftrightarrow gg$ and $q\bar{q} \leftrightarrow
\gamma\gamma$.

The statistical weight $g$ for quarks and
antiquarks receives contributions from spin (a factor 2) and color (a
factor 3), totalling $g=6$.

\paragraph{Massless Quarks}
\index{thermodynamic potential!for massless quarks}
The thermodynamic potential for massless quarks can be written as an
infinite sum in powers of $\mu_q/T$
\cite{Rhodes51a,Dingle57a}, but the combination with positive and
negative chemical potential treated in Appendix~\ref{app:integrals}
gives only a finite number of terms. This is exactly the combination
relevant when a quark and its antiquark are both present as noted
above. At zero temperature the antiquark contributions disappear and
only the quark terms are left. Using the results \Eq{eq:sp1} and
\Eq{eq:sp3} gives the quark-antiquark volume term
\be \label{eq:quarkantiquarkvolume}
	\Omega_{q\bar{q},V} \equiv \Omega_{q,V}+\Omega_{\bar{q},V} =
	- \frac{7\pi^2}{60} T^4 - \frac{1}{2} \mu_q^2 T^2
	-\frac{1}{4\pi^2}\mu_q^4 ,
\ee
and the quark-antiquark curvature term
\be \label{eq:quarkantiquarkcurvature}
	\Omega_{q\bar{q},C} \equiv \Omega_{q,C}+\Omega_{\bar{q},C} =
	 \frac{1}{24} T^2 +\frac{1}{8 \pi^2} \mu_q^2 .
\ee
The \emph{net number} of quarks of species $q$, \ie , the number of
quarks \emph{minus} the number of antiquarks is then given by
\be
	N_q = -\left(\frac{\partial\Omega_{q\bar{q}}}{\partial\mu_q}
		\right)_{T,V,C} ,
\ee
with the volume and curvature contributions given by
\bea    \label{eq:nummasslessV}
	N_{q,V} &=&
	 \mu_q T^2 + \frac{1}{\pi^2} \mu_q^3	\\
	\label{eq:nummasslessC}
	N_{q,C} &=&
	\frac{1}{4\pi^2} \mu_q
\eea

\begin{figure}
	\centering
		\includegraphics[width=\textwidth]{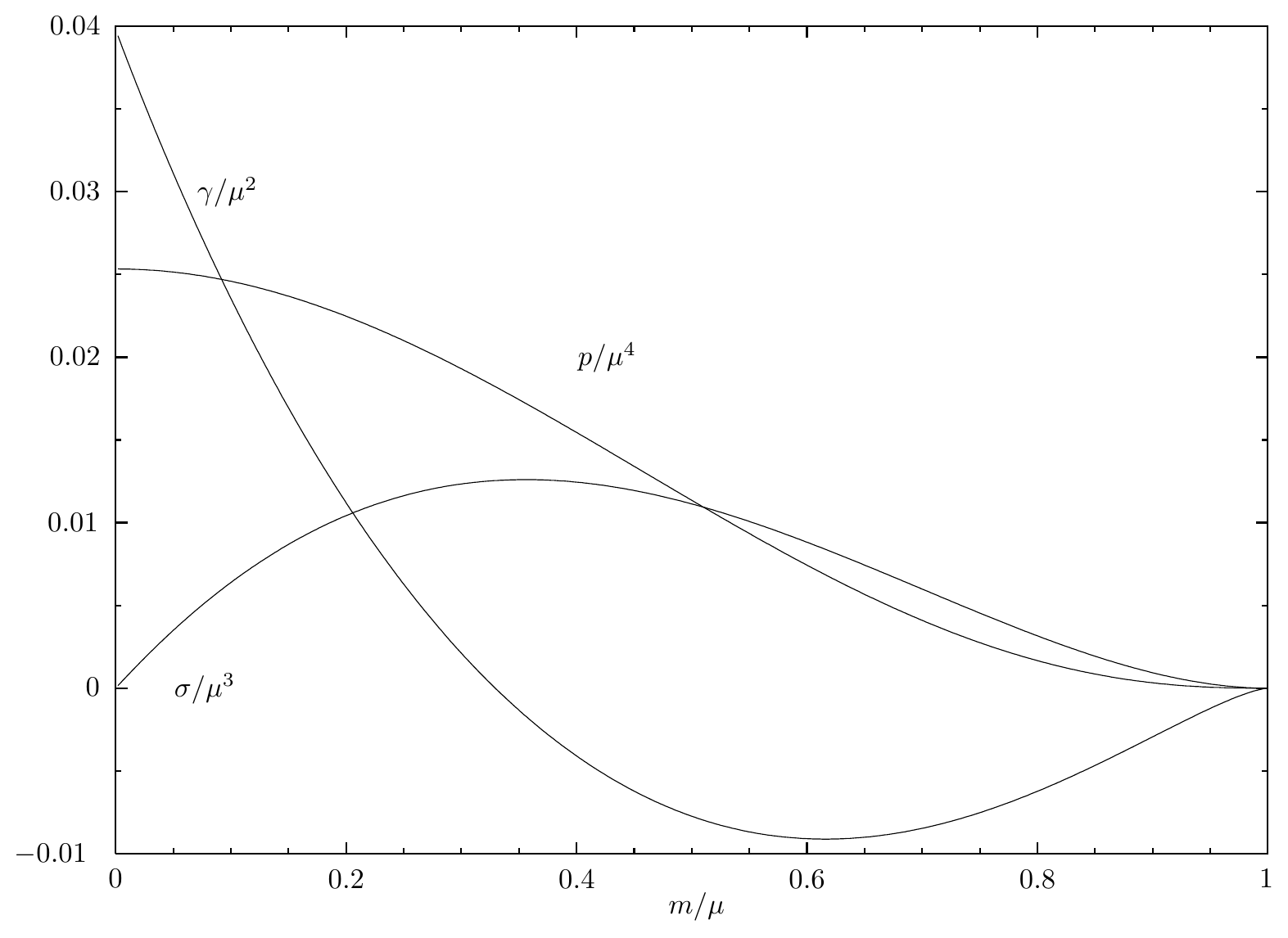}
	\caption{\label{fig:tension}
		Plot of the pressure $p$, surface tension $\sigma$, and the
		curvature coefficient $\gamma$ for a single quark species at
		zero temperature, as a function of mass $m$ in units of the chemical
		potential $\mu$.}
\end{figure}

\paragraph{Zero Temperature} At $T=0$ the thermodynamic potential is
\index{thermodynamic potential!zero temperature}
given in terms of known integrals \cite{Gradshteyn}, and the volume and
surface terms were first calculated by Berger and Jaffe
\cite{Berger87a,Berger91a}, while the curvature term was calculated by
Madsen \cite{Madsen94a}. Introducing $x=\mu_q/m_q$ as the chemical potential
in units of the mass, the relevant expressions are
\be \label{eq:omegaVT0}
	\Omega_{q,V} = 
	\frac{1}{4\pi^2}m^4 \left[ x(x^2-1)^{1/2}(x^2-5/2)
	+\frac{3}{2}\ln(x+(x^2-1)^{1/2}) \right]
\ee
\begin{eqnarray} \label{eq:omegaST0}
	\Omega_{q,S}&=&
	\frac{3}{4\pi} m^3 \bigg\{ \frac{1}{6}x(x^2-1)-\frac{1}{3}(x-1)
	-\frac{1}{3\pi} \bigg[ x^3\tan^{-1} (x^2-1)^{1/2}
	\nonumber \\
	&\phantom{=}&\phantom{\frac{3}{4\pi}m^3\bigg\{ }
	-2x(x^2-1)^{1/2}
	+\ln (x+(x^2 -1)^{1/2} ) \bigg] \bigg\}
\end{eqnarray}
\begin{eqnarray} \label{eq:omegaCT0}
	\Omega_{q,C}&=&
	\frac{1}{8\pi^2}m^2 \bigg\{ \frac{\pi}{2} x(x^2-1)-\pi (x-1)
	+\ln(x+(x^2-1)^{1/2})
	\nonumber \\
	&\phantom{=}& \phantom{\frac{1}{8\pi^2}m^2 \bigg\{}
	-x^3\tan^{-1} (x^2-1)^{1/2} \bigg\}
\end{eqnarray}
The pressure $p=-\Omega_{q,V}$, surface tension $\sigma=\Omega_{q,S}$,
and the curvature coefficient $\gamma=\Omega_{q,C}$ for a single quarks
species, as given by the above expressions, are plotted in dimensionless
units as a function of $x^{-1}$ in \Fig{fig:tension}.
Note that whereas the pressure and surface tension are always positive,
the curvature term is negative for $1/3\lesssim m/\mu < 1$. As $m$
approaches $\mu$ all three quantities vanish; reflecting the
disappearance of quarks of the given species.

\index{thermodynamic potential|)}
\index{thermodynamic potential!for quarks|)}
The quark number densities are again found by differentiation of the
thermodynamic potential with respect to chemical potential, and the
result is
\be \label{eq:numvolume}
	N_{q,V} = \frac{1}{\pi^2} m^3(x^2-1)^{3/2}
\ee
\be \label{eq:numsurface}
	N_{q,S} = -\frac{3}{4\pi}\left\{ \frac{1}{2}(x^2-1)-\frac{1}{\pi}
	\left[ x^2 \tan^{-1} (x^2-1)^{1/2} - (x^2-1)^{1/2} \right] \right\}
\ee
\be \label{eq:numcurvature}
	N_{q,C} = \frac{3}{8\pi^2} m \left[ -\frac{\pi}{2}(x^2-1)
	+\frac{1}{3}(x^2-1)^{1/2} + x^2\tan^{-1} (x^2-1)^{1/2} \right]
\ee
\index{partition function|)}

\subsection{Mass Formula\index{mass formula|(}}
At zero temperature the liquid drop model can be used to construct a
mass formula for strange quark matter, analogous to the well known
liquid drop mass formula of nuclear physics. Such mass formulae have
been studied extensively by Madsen
\cite{Madsen93a,Madsen93b,Madsen94a,Madsen95a,Madsen95b}.

The energy for a strangelet at zero temperature is given by the
expression
\be
	E = \sum_q \left(\Omega_q + \mu_qN_q\right) + BV .
\ee
Considering only spherical systems, we seek to minimize this expression
with respect to $V$ ($S$ and $C$ are now given in terms of $V$), and
with respect to each of the chemical potentials $\mu_q$, which govern
the flavor composition of the strangelet. The minimization has to be
done under the constraint that the number of quarks of all flavors add
up to give a total number of quarks equal to three times the baryon
number
\be
	A = \frac{1}{3}\sum_q N_q .
\ee

Flavor equilibrium is maintained by the \Index{weak interaction},
through processes such as
\bea \label{eq:weakprocesses}
	d &\longleftrightarrow& u + e^- + \bar{\nu}_e \nonumber \\
	s &\longleftrightarrow& u + e^- + \bar{\nu}_e \\
	u + d &\longleftrightarrow& s + u \nonumber
\eea	
Since the condition of optimal composition is
\be
	\mu_i dN_i = 0 ,
\ee
the chemical potentials are connected through
\be
	\mu_d = \mu_s = \mu_u + \mu_e + \mu_{\bar{\nu}_e} ,
\ee
but neutrinos are never present\footnote{Except in supernova
explosions, where matter densities are so high that neutrinos do not
immediately escape.} in the system so $\mu_{\bar{\nu}_e}=0$.
For bulk systems electrons would be present, but for the very small
systems considered here, electrons, which are not confined, will escape,
perhaps to form a surrounding electron cloud if the strangelet is
positively charged. Thus for strangelets we may set $\mu_e=0$, and we
get a common quark chemical potential
\be
	\mu \equiv \mu_u = \mu_d = \mu_s .
\ee

Finding the minimum energy of a strangelet with baryon number $A$ is
thus reduced to the problem of simultaneously solving the equations
\be \label{eq:T0mech}
	\sum_q \left( \Omega_{q,V}(\mu) + \frac{2}{R}\Omega_{q,S}(\mu)
		+ \frac{2}{R^2} \Omega_{q,C}(\mu) \right) + B = 0 ,
\ee
and
\be
	A = \frac{1}{3} \sum_q N_q(\mu) ,
\ee
expressing respectively mechanical equilibrium, and baryon number
conservation. In \Eq{eq:T0mech}, I have used the fact that $dS/dV =
2R^{-1}$, and $dC/dV=2R^{-2}$ for a sphere of radius $R$.
The equation \Eq{eq:T0mech} without the surface and curvature terms
would  simply be pressure equilibrium, so it is equivalent to
the second boundary condition of the MIT bag, given in \Eq{eq:quarkinbag}.
\index{boundary conditions!of MIT bag model}

These equations are easily solved on a computer, and the result of such
a numerical solution can be seen in \Fig{fig:ld_and_shell}.
The very good agreement with shell model calculations
\index{shell model!compared to liquid drop model}
shows the success of the multiple reflection expansion, and the
viability of the \textit{Ansatz} \Eq{eq:quarkcurvatureterm}, of Madsen,
for the quark curvature term.

\index{liquid drop model|)}
\index{mass formula|)}

\section{Summary and Discussion}
This chapter has introduced the MIT bag model which \emph{all} the
results in the rest of this thesis build on. It must always be remembered
that the bag model is exactly that---a model. If we know when a model
can be applied we are not doing something new; hence there is always the
chance that the model used is for some as yet unknown reason not suited
to the problem to which it is applied. Even if the model is suited it
can not be trusted to give exact results. In a way every theory should
be considered a model eventually to be superseeded by a better
model. What we today call fundamental theories may turn out tomorrow to
be approximations to some more elaborate theory. Theories are merely
models that enjoy a rare longevity. There are many indications that the
so-called standard model of physics is not \emph{the} fundamental
theory, and high-energy physicists are constantly inventing new
ones.

Setting aside these more philosophical considerations, the MIT bag model
is certainly known to have very specific shortcomings. 
The idea of a static bag surface is in conflict with basic quantum
mechanics, since according to Heisenberg's uncertainty principle there
should be a certain fuzzyness to the bag surface. The bag surface should
in reality be a dynamic object whose degrees of freedom is governed by
the fields inside the bag. Another disease of the bag model is that
the boundary conditions of the MIT bag violate chiral symmetry, which is
\index{chiral symmetry!breaking of}
a symmetry of the QCD Lagrangian. 

Notwithstanding these drawbacks, the bag model has been very successful
in reproducing many hadronic properties. Most important in this context
is the successful reproduction of the hadronic mass spectrum
\cite{Chodos74b,Johnson75a}
which gives a certain confidence when applying the model to the
calculation of strangelet masses.

The MIT bag model is not the only phenomenological model that has been
applied to the problem of hadron structure or quark matter. Some other
contenders are the solitonic bag 
\cite{Friedberg77a}, the Budapest bag model \cite{Gnadig76}, the
Skyrmion \cite{Skyrme62a,Skyrme91a}, and the chiral bag \cite{Brown79a}. In
addition to these relativistic models there have also been attempts to
use non-relativistic potential models, with constituent quarks
\cite{Sathpathy94a}.

The liquid drop model and the shell model both assume that the
ground-state configuration is that of a degenerate Fermi gas. The
possiblility of a superconducting ground-state has been investigated by
Iwasaki and Iwado \cite{Iwasaki95a,Iwasaki95b}. This could be relevant
for quark matter in neutron stars. Superconductivity in quark matter was
\index{superconductivity}
also discussed by Bailin and Love \cite{Bailin81a,Bailin82a}.

The liquid drop model of strangelets introduced in this chapter builds
on the multiple reflection expansion explained in Appendix
\ref{chap:mre}. A very good agreement between liquid drop model results
and shell model calculations performed using the program in Appendix
\ref{app:shell} has been demonstrated in this chapter. The first
comprehensive comparison including the curvature term for massive quarks
was performed by Madsen \cite{Madsen94a} at zero temperature.
It is reassuring to note the very different approach by De~Francia
\cite{DeFrancia94a} gives results identical to those obtained using the
multiple reflection expansion, in the case of massless quarks and gluons
at finite temperature. In the case of pure glue Kajantie, K\"arkk\"ainen,
and Rummukainen \cite{Kajantie92a} find that
\Index{lattice gauge calculations}
are in good agreement with the results obtained with the MIT bag
model and the multiple reflection expansion, as regards the vanishing of
the surface term and the magnitude of the curvature term. A similar
study by Huang, Potvin, and Rebbi \cite{Huang92b} finds that the curvature
term agrees with the bag model results for hadroninc bubbles, but
not for purely gluonic bubbles for which they obtain the opposite sign
relative to both Kajantie \textit{et al.}\ \cite{Kajantie92a} and the
bag model results. There is a clear sign that the curvature term is the
dominating contribution for massless species, but apparently there is
no consensus among lattice studies regarding the exact value.

When the bag model is applied to
hadrons one usually includes a term $-Z_0/R$ in the energy, where $R$ is
the radius of the bag which is normally considered spherical. $Z_0$ is a
purely phenomenological parameter. The physical basis for such a term is
twofold. One effect which gives such a dependence is a spurious
\Index{center of mass energy}, which the above term serves to subtract. Another
effect which contributes such a term is the \Index{zero-point energy}
of the fields in the bag. A fit to the hadron spectrum
\cite{Johnson75a} gives a value\footnote{The values of all parameters
in the fit mentioned are: $B^{1/4}=146$ MeV, $Z_0=1.84$,
$\alpha_c=0.55$, $m_s=279$ MeV, $m_u=m_d=0$.} of $Z_0=1.84$.

We have seen that the multiple
reflection expansion gives rise to surface and curvature terms in the
energy. For a spherical bag of radius $R$ these terms are proportional
to $R^2$ and $R$ respectively. The $R^{-1}$ term discussed above will only be
significant for very small systems, where the liquid drop model should
not be applied anyway, so for strangelets such a term is not relevant.

I mentioned in section \ref{sec:bagmodel} that it is possible to
do \Index{perturbation theory} in the bag model. Hansson and Jaffe 
\cite{Hansson83a} have formulated Feynman rules for quantum
chromodynamics in the bag model using the multiple reflection expansion
for the bag propagator of quarks and gluons. It should therefore be
possible to include both the effect of finite size and the strong
interaction, but only for the volume term are the $\alpha_s$-corrections
known
\cite{Baym76a,Freedman77a,Freedman78a,Baluni78a,Baluni78b,Farhi84a,Goyal91a}.
To first order $\alpha_s$-corrections are
repulsive, resulting in a higher energy per baryon for strange matter.
However, it was shown by Farhi and Jaffe \cite{Farhi84a} that a non-zero
value of the strong \Index{fine-structure constant}
$\alpha_s$ can be absorbed in
a decrease in the bag constant $B$. This rescaling of the parameters
does probably not hold exactly, especially when finite size effects are
included, but for the moment there seems to be no alternative, so I will
assume $\alpha_s=0$ throughout this thesis.

Non-zero $\alpha_s$ corrections were taken into account in a shell model
approach by Ishii and Tamagaki \cite{Ishii93a} using the MIT bag model,
and the effect of $\alpha_S$-corrections have also been studies by Ishhi
\textit{et al.}\ \cite{Ishii92a,Ishii94a} in a constituent quark model
of strange quark matter. In these studies the finite size effects
\emph{were} taken into account, but not in the liquid drop approach.

The \Index{Coulomb energy} of a strangelet can be estimated as being that of a
sphere with a uniform charge density $Z/V$, where $Z/e=\frac{2}{3}N_u
-\frac{1}{3}(N_d+N_s)$, and $V=4/3\pi R^3$ is the volume of the sphere,
giving $E_{\text{Coulomb}}=\alpha Z^2/(10R)$, where $\alpha$ is the
fine-structure constant. It was shown by Madsen \cite{Madsen93b} that the
Coulomb energy is negligible for strangelets compared to the surface and
curvature energies.

\index{stablility!of strange quark matter|(}
The criteria for the stability or metastability of strange quark matter
was discussed in Chapter \ref{chap:introduction}. Here I will briefly
mention what this translates to in terms of the parameters $B^{1/4}$ and
$m_s$ of the MIT bag model, assuming $\alpha_s=0$. First of all there is
an important lower limit on $B^{1/4}$ set by the fact that two flavor
quark matter must be unstable, \viz, $B^{1/4}\gtrsim 145$ MeV. This limit
assumes $m_u=m_d=0$ and \emph{does} include Coulomb effects which are
important when only two flavors are present. 

Absolute stability relative to nuclei such as ${}^{56}$Fe means that the
energy per baryon for strange quark matter must not exceed 930~MeV, and
whether stranglets are absolutely stable can be learned from
\Fig{fig:ld_and_shell}. But even if strange quark mattter has a higher
energy per baryon, it could be very long lived. The next decay mode
\index{decay modes}
which comes into play is neutron (and to a lesser extent proton)
emission. As soon as the energy per baryon for a bulk system becomes
larger than the neutron mass $m_n\approx 940$~MeV it is favorable to emit
neutrons. For $m_s=0$ this corresponds to $B^{1/4}\gtrsim 165$~MeV,
and for $m_s=150$~MeV it corresponds to $B^{1/4}\gtrsim 157$~MeV.
If the energy per baryon of strange quark matter is larger than the
$\Lambda$ mass $m_\Lambda \approx 1116$~MeV it would rapidly emit
$\Lambda$ particles, quickly to be totally dissolved. This happens for
$B^{1/4}\gtrsim 195$~MeV at $m_s=0$ and for $B^{1/4}\gtrsim 188$~MeV
at $m_s=150$~MeV.
For small systems the change in energy per baryon due to particle emission
becomes important, and also \Index{shell effects} cannot be neglected. Berger
and Jaffe \cite{Berger87a} discussed the various decay modes of strange
quark matter in detail.
\index{stablility!of strange quark matter|)}

\clearemptydoublepage
\chapter{Strangelets at Finite Temperature\label{chap:temperature}}
\index{liquid drop model!at finite temperature|(}
In this chapter I briefly discuss how to treat strange quark matter at
non-zero temperature in the liquid drop model. The results are an
application of the formalism set up in section \ref{sec:drop}, and serve
as a link between Chapter \ref{chap:mit} and Chapter \ref{chap:equilibrium}
in which a hadronic phase is also present. The important effect of color
singletness is not taken into account here---it is the subject of
Chapter \ref{chap:projection} and Chapter \ref{chap:color}.

Strangelets at finite temperature (without the color singlet
constraint) have been studied in the liquid drop
model by Mustafa and Ansari \cite{Mustafa93b}, Jensen and Madsen
\cite{Jensen95a}, and by He \textit{et al.} \cite{He96a} who present
results for a large range of parameters. Shell model
\index{shell model!at finite temperature} calculations at
finite temperature have
been performed by Mustafa and Ansari \cite{Mustafa96a}, but
unfortunately---even though the approach is sound---the results
displayed in Ref.~\cite{Mustafa96a} contain an error.

\section{Equilibrium Conditions}
At finite temperature, it is the free energy $F=\Omega+\mu_qN_q$,
rather than the energy that must attain a minimum, for the system to be
in equilibrium.%

Quarks and their antiquarks are assumed to be in equilibrium through
the processes $q\bar{q} \leftrightarrow gg$ and $q\bar{q}
\leftrightarrow \gamma\gamma$, so that the antiquark chemical potential
is given by
\be
	\mu_{\bar{q}} = -\mu_q .
\ee

The optimal composition is still governed by the weak equilibrium
processes \Eq{eq:weakprocesses}, but depending on the physical
conditions, these processes may not be fast enough to keep (or attain)
equilibrium between flavors. Hence if weak equilibrium is reached, the
chemical potentials for the different quark flavors will be equal just
as at zero temperature. In the \Index{quark-gluon plasma}, thought to be
produced in high-energy \Index{heavy-ion collision}s, flavor equilibrium
is not
likely to be reached, and so each flavor is described by a separate
chemical potential. This situation is dealt with in Chapter
\ref{chap:equilibrium}, while in this chapter flavor equilibrium will be
assumed.

Gluons are not subject to number
conservation and are ascribed zero chemical potential. Photons will be
present at $T>0$, but since, unlike gluons, they are not confined to the
bag volume, photons will be present not only in the quark matter phase,
but also in the surrounding phase. Accordingly the photon contribution to
the free energy is left out, since it will be countered by a like term
in the surrounding phase.

In flavor equilibrium the thermodynamic potential for a gas of
$\mcal{N}_q$ massless quark
flavors, their antiquarks, and gluons in an MIT bag is given by
(see Eqs.~(\ref{eq:gluonvolume})--(\ref{eq:quarkantiquarkcurvature})
\begin{eqnarray} \label{eq:omegamasslessquarksandgluons}
        \Omega(T,\mu,V,C) &=& - \left[ \left( \frac{7{\cal N}_q}{60}
        + \frac{8}{45} \right) \pi^2T^4 + \frac{{\cal N}_q}{2} \left(
        \mu^2T^2 + \frac{\mu^4}{2\pi^2} \right) -B \right] V
        \nonumber \\
        &\phantom{=}&
        + \left[ \left( \frac{{\cal N}_q}{24} + \frac{4}{9} \right)
        T^2  + \frac{{\cal N}_q}{8\pi^2}\mu^2 \right] C ,
\end{eqnarray}
where $\mu$ is the common quark chemical potential. If massive quarks
are present they contribute not only volume and curvature terms, but also
a surface term, and at $T>0$ these must be evaluated numerically.

The net quark number derived from the above thermodynamic potential is
\be
	N(T,\mu,V,C) = \mcal{N}_q\left[
		\left( \mu T^2 + \frac{1}{\pi^2}\mu^3 \right) V
		- \frac{1}{4\pi^2} \mu C \right] 
\ee

At fixed temperature and baryon number, the condition for a minimum in
the free energy is
\comment{$-\mcal{S}dT$ and $\mu dN$ terms disappear}
\be
	-pdV + \gamma dC + \sigma dS = 0 ,
\ee
which for a \emph{spherical} MIT bag leads to
\be \label{eq:mecheq}
	\Omega_V(T,\mu) + \frac{2}{R}\Omega_S(T,\mu)
	+ \frac{2}{R^2}\Omega_C(T,\mu) = 0 .
\ee
This equation expresses mechanical equilibrium,
between quarks and gluons on the one hand, and the bag pressure $B$ on
the other hand. At fixed $T$ this has to be solved for $R$ and $\mu$
along with the equation expressing fixed baryon number $A$, \viz ,
\be \label{eq:numequality}
	3A = N(T,\mu, R) .
\ee
This is a cubic equation in $R$, which can be solved by use of Cardano's
formula, to give an expression in terms of $\mu$ (and $T$ and $A$) which
can then be substituted into the mechanical equilibrium condition, which
may then be solved numerically for $\mu$ at fixed $T$ and $A$. In the
more general case where massive quarks are present, a wholly numerical
approach is necessitated.

\section{Mass Formula}
\index{mass formula!finite temperature|(}
\begin{figure}
	\centering
	\includegraphics[width=.9\textwidth]{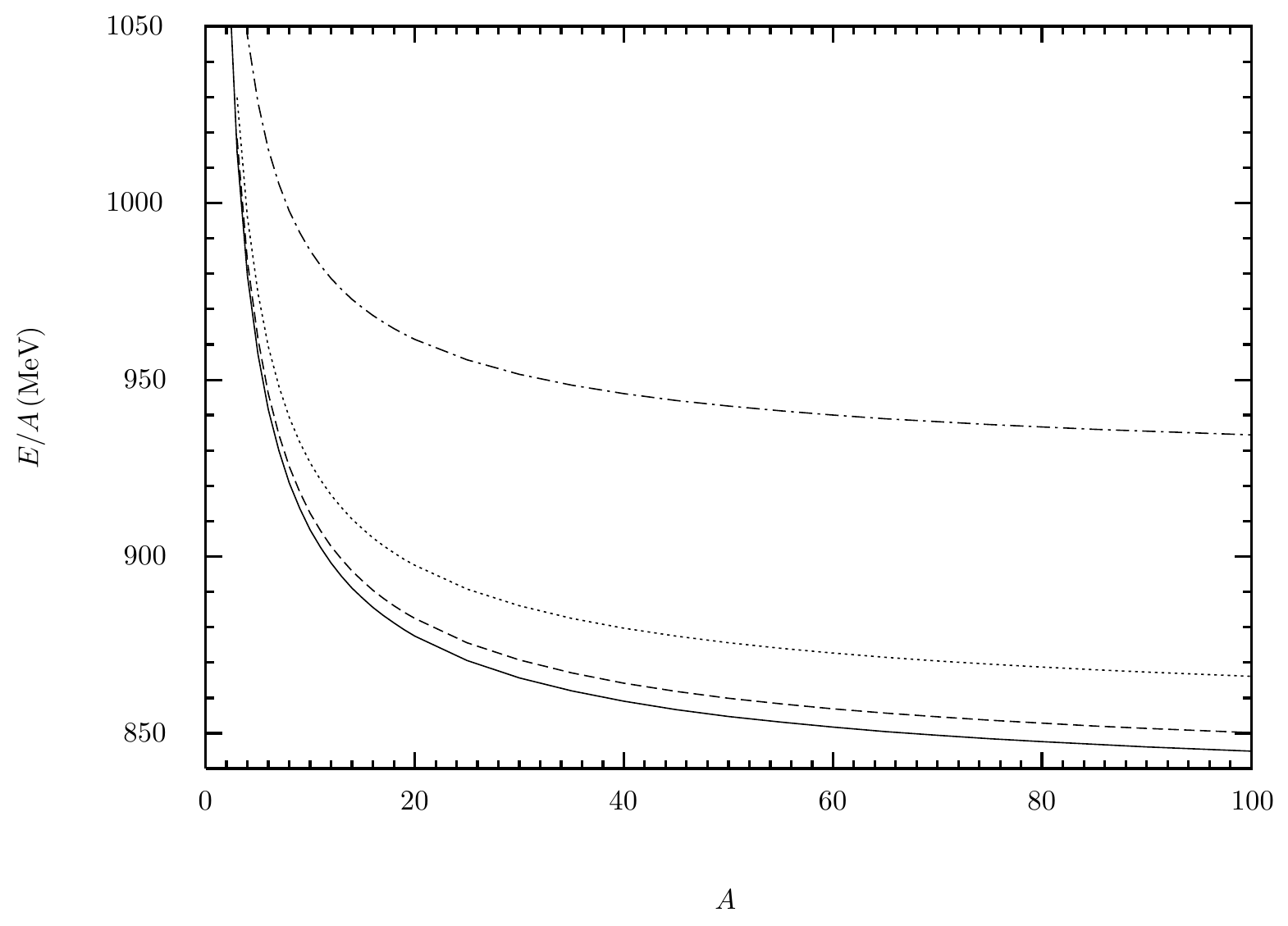}
	\caption{ \label{fig:masslessTplot}
		Energy per baryon for three massless quark flavors
		($\mcal{N}_q=3$). From bottom to top: $T=0$ MeV, 10 MeV, 20 MeV,
		and 40 MeV. $B^{1/4} =  145$ MeV, but
		results for other values can easily be obtained, since $E/A$ scales
		proportionally to $B^{1/4}$. }
\end{figure}
Solving the equations \Eq{eq:mecheq} and \Eq{eq:numequality}, enables us
to calculate the mass of a strangelet at finite temperature, by
inserting the found values for $\mu$ and $R$ into the expression for the
energy
\be
	E = \Omega + \mu N + T\mcal{S} ,
\ee
where the entropy is obtained from $\Omega$ as $\mcal{S} =
-\partial\Omega/\partial T$. For $\mcal{N}_q$ \emph{massless} quark
flavors the entropy is
\be
	\mcal{S} = \left[ \left( \frac{7\mcal{N}_q}{15} + \frac{32}{45}
		\right)\pi^2 T^3 + \mcal{N}_q \mu^2 T \right] V
		- \left(\frac{\mcal{N}_q}{12} + \frac{8}{9} \right) T C ,
\ee
giving an expression for the energy which is
\bea
	E &=& \left[ \left( \frac{7\mcal{N}_q}{20} + \frac{8}{15} \right)
		\pi^2T^4 -\frac{3\mcal{N}_q}{2}\left( \mu^2 T^2 +
		\frac{1}{2\pi^2}\mu^4 \right) +B \right] V \nonumber \\
		&-&
		 \left[ \left( \frac{\mcal{N}_q}{24} + \frac{4}{9} \right) T^2 
		+ \frac{\mcal{N}_q}{8\pi^2}\mu^2 \right] C
\eea

Notice that with the expressions given for massless quarks and gluons
$E_V-B = -3(\Omega_V-B)$, which is the usual equation of state
($\varepsilon = 3p$) valid for massless relativistic particles. In
addition $E_C = -\Omega_C$, and the condition of mechanical equilibrium
is $\Omega_CC = -3\Omega_VV$. Using these three equations gives
\cite{Jensen96a}
\be
	E = E_VV + E_CC = \left(4B-3\Omega_V\right)V + 3\Omega_VV = 4BV ,
\ee
for the energy \emph{in equilibrium}. It is stressed that this result is
valid \emph{only for massless quarks}, since for massive quarks the
equation of state $\varepsilon = 3p$ no longer holds.

\subsection{Fixed Temperature}
A plot of the energy per baryon at  fixed temperature and baryon number,
as a function of baryon number is seen in \Fig{fig:masslessTplot}
for massless quarks. The energy per baryon is seen to increase rapidly with
temperature, reflecting the $T^4$ dependence of the energy. For massless
quarks the only energy scale in the equations is $B^{1/4}$, so all
quantities scale with $B^{1/4}$ according to their energy dimension. For
instance the energy per baryon scales proportionally to $B^{1/4}$, while
the radius scales inversely proportional to $B^{1/4}$.

\begin{figure}
	\centering
	\includegraphics[width=.9\textwidth]{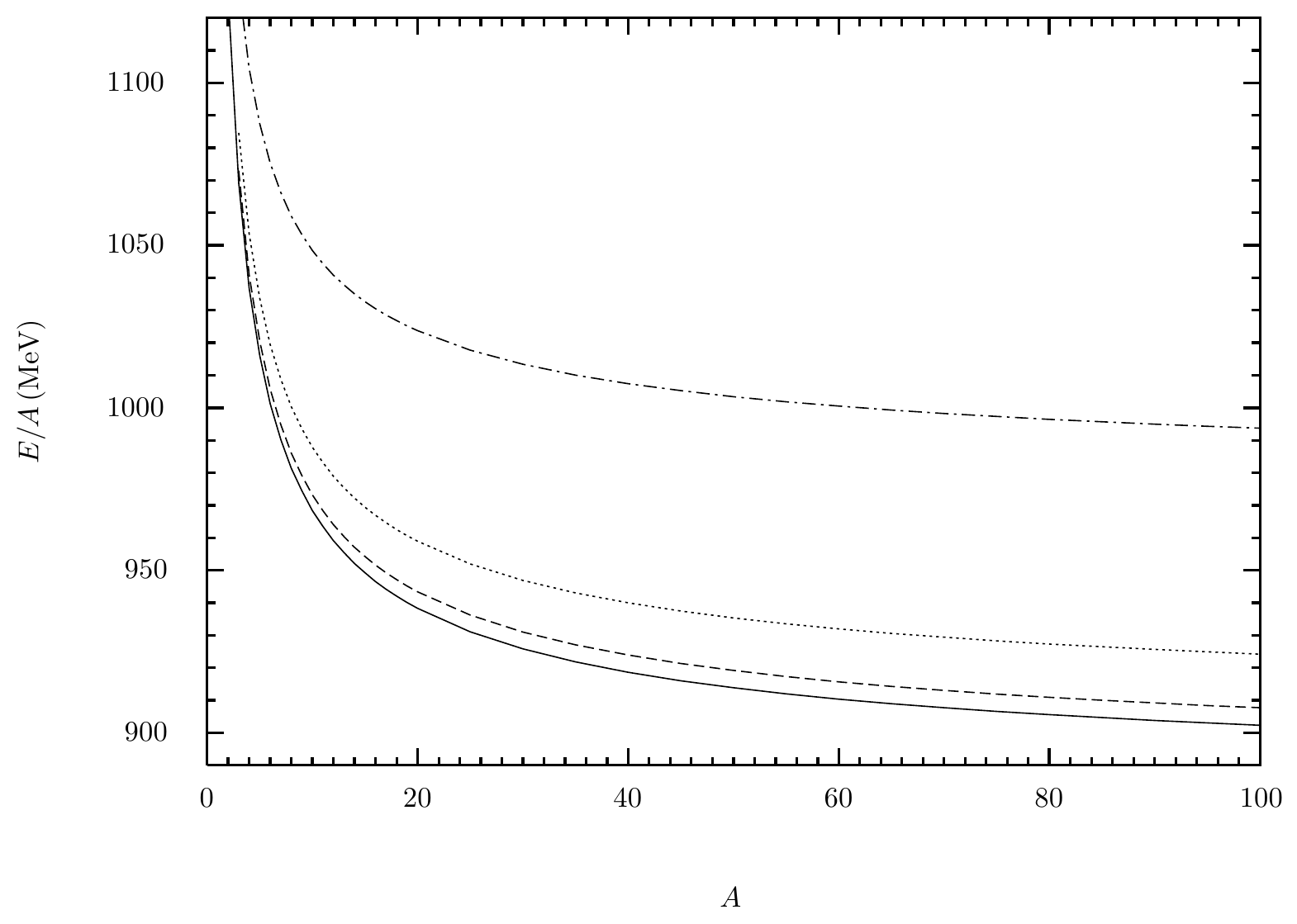}
	\caption{ \label{fig:massTplot}
		Energy per baryon for strangelets, with a strange-quark mass of
		150 MeV.  From bottom to top: $T=0$ MeV, 10 MeV, 20 MeV,
		and 40 MeV. $B^{1/4} =  145$ MeV.
		}
\end{figure}

An equivalent plot for a strange-quark mass $m_s=150$ MeV is shown in
\Fig{fig:massTplot}, and the curves resemble those in the $m_s=0$
case---the major difference being the shift in energy scale.

\subsection{Fixed Entropy}
Instead of fixing the temperature, one may consider the case where the
entropy and the baryon number (and thus also the entropy per baryon) are
fixed. The equilibrium state of the system is then found by solving
$dE=0$, since the energy has natural variable $\mcal{S}$ rather than
$T$. The condition of mechanical equilibrium \Eq{eq:mecheq}, and
\Eq{eq:numequality} expressing fixed baryon number are unchanged, but
must be accompanied by the supplementary equation
\be \label{eq:fixentro}
		\frac{1}{A} \left( \frac{\partial\Omega}{\partial T}
		\right)_{V,C,\mu} = \frac{\mcal{S}}{A} ,
\ee
which is to be solved in conjunction with \Eq{eq:mecheq} and
\Eq{eq:numequality} for $T$, $\mu$ and $R$.

\begin{figure}
	\centering
	\includegraphics[width=.9\textwidth]{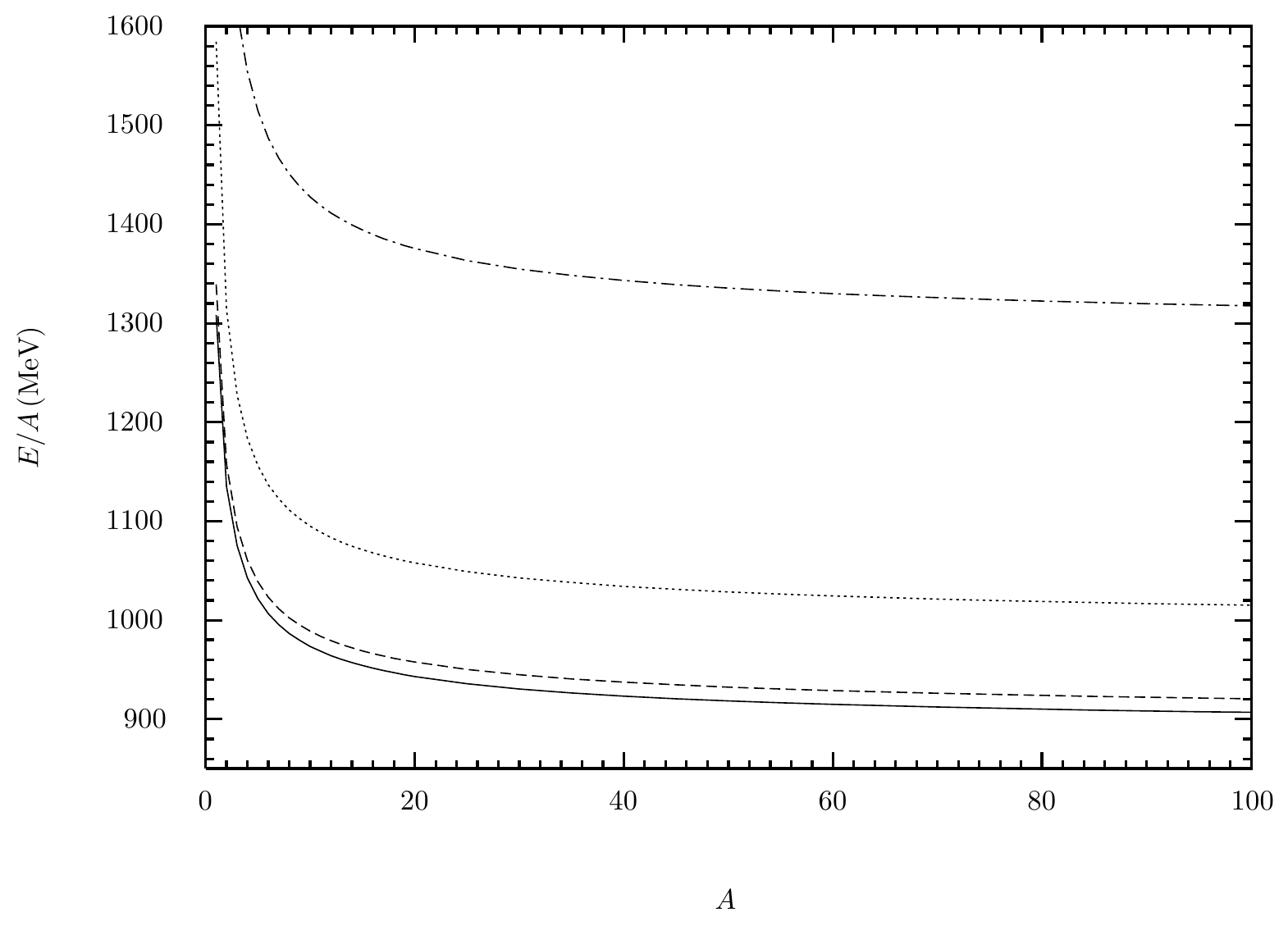}
	\caption{ \label{fig:SAplot}
		Energy per baryon for a strangelet with fixed entropy per
		baryon. From bottom to top: $\mcal{S}/A=1$, 2, 5, and 10.
		$B^{1/4} = 145$ MeV and $m_s = 150$ MeV.
		}
\end{figure}

Numerical solutions to these three equations are displayed in
\Fig{fig:SAplot}, for the same set of parameters ($m_s=150$ MeV,
$B^{1/4}=145$ MeV) as
in \Fig{fig:massTplot}.

\index{mass formula!finite temperature|)}

\section{Phase Diagram}
\begin{figure}
	\centering
	\includegraphics[width=.8\textwidth]{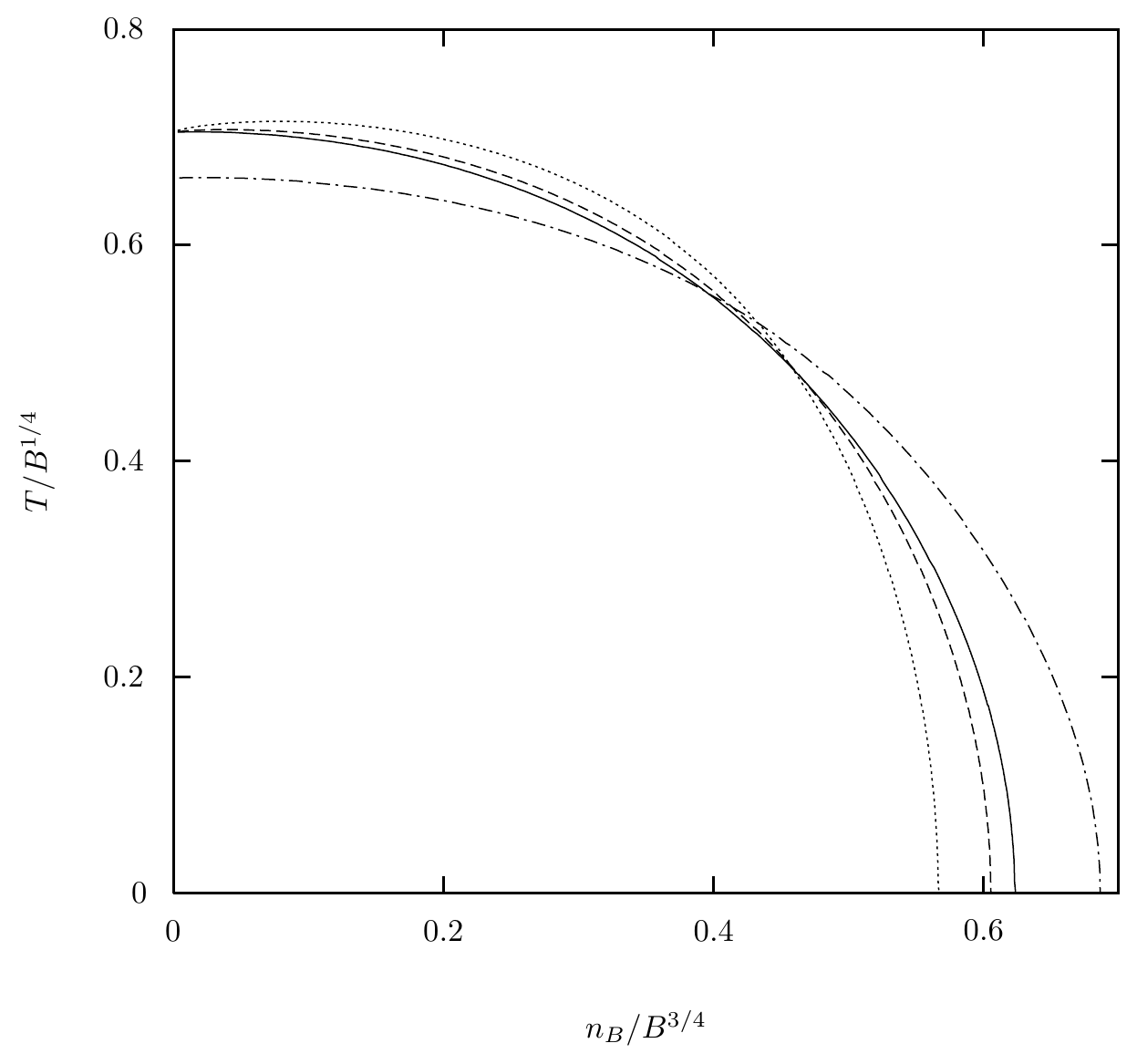}
	\caption{\label{fig:phase_finite}
			 The phase diagram for quark matter in the MIT bag model,
			 with massless quarks. The solid line is for $A=100$, the
			 dashed line for $A=20$, and the dotted line for
			 $A=5$; all for $\mcal{N}_q=2$. The dotted-dashed line is for
			 $\mcal{N}_q=3$ and $A=100$.}
\end{figure}

\index{phase diagram|(}
By solving \Eq{eq:mecheq} and \Eq{eq:numequality} at fixed temperature
the chemical potential $\mu$ and the radius $R$, so found, can be used
to calculate the baryon number density
\be
	n_B \equiv \frac{A}{V} .
\ee
Doing this for all possible temperatures, it is possible to construct a
phase diagram for quark matter, \ie , a plot of $T$ \textit{vs.} $n_B$.
Such a phase diagram is shown in \Fig{fig:phase_finite}, for both 2 and
3 massless flavors, and a range in baryon number.

All curves with the same value of $\mcal{N}_q$ intersect the $T$-axis
at the same temperature, \viz\ the \Index{critical temperature} $T_c$.
This is because at zero baryon density $\mu=0$, and the solution to
\Eq{eq:mecheq} is
\comment{For $\mcal{N}_q = 3$ $T_c\approx 0.66 B^{1/4}$.}
\be \label{eq:critT}
	T_c =  \left( \frac{\pi^2}{90} g_\star
			\right)^{-1/4} B^{1/4} ,
\ee
where $g_\star$ is the effective number of degrees of freedom, defined
as
\be \label{eq:gstar}
	\frac{\pi^2}{90} g_\star=  
		\left(\frac{7\mcal{N}_q}{60} + \frac{8}{45}\right)\pi^2 .
\ee
In the bulk limit the \Index{critical temperature} is also the highest
attainable temperature, but when \Index{finite size effects} are taken into
account, the maximum $T$ occurs for a finite baryon number, as noted by
Mustafa and Ansari \cite{Mustafa93b}.

Comparing the curves for $A=100$ and $\mcal{N}_q=2, 3$ it is seen that
adding an extra flavor decreases the critical temperature. This is
because the critical temperature is determined by the effective number
of degrees of freedom $g_\star$.
Adding an extra flavor increases the effective number of degrees of
freedom, so that pressure equilibrium is obtained at a lower temperature.
On the other hand the baryon number density at $T=0$ increases when an
extra flavor is added. This is because the baryon number can be packed
closer with three flavors than with two flavors, due to the Pauli
exclusion principle.

The effect of finite baryon number is also clearly seen in
\Fig{fig:phase_finite} as a decrease in baryon number density at low
$T$ with decreasing baryon number. This is a consequence of the finite
size effects which cause a decrease in the density of states, as seen in
\Fig{fig:numofstates}, and a therefore a larger volume is required to
hold the same baryon number, resulting in a smaller density.

Quark matter is in equilibrium along the curves shown in
\Fig{fig:phase_finite}, whereas outside the region delimited by such a
curve the combined pressure of the quarks and gluons exceeds the bag
pressure, and matter in this part of the phase diagram will be in 
the form of a deconfined \index{deconfinement} \Index{quark-gluon
plasma}. The region lying within the confines of a particular curve does
not correspond to a physical system unless there is a hadronic phase
present. This case will be treated in Chapter \ref{chap:equilibrium}.
\index{phase diagram|)}

\section{Summary and Discussion}
In this chapter I have shown how to include effects of both the finite
system size and non-zero temperature, or entropy. Both of these effects
have been studied individually before, but here a consistent treatment
of both effects, for both massless and massive quarks, has been
presented. For massless quarks many quantities can be evaluated
analytically, while for massive quarks numerical evaluation is
necessary. For massless quarks all quantities scale with some power of
the bag constant $B$ according to their dimension. This is because in
the absence of quark masses the only dimensional parameter in the bag
model is the bag constant.
The introduction of quark masses break this simple scaling behavior.

The chapter was concluded with a discussion of the phase diagram in the
MIT bag model, which was discussed in a qualitative manner in Chapter
\ref{chap:introduction}. Effects of finite size was clearly seen in the
phase diagram, and this will be further studied in Chapter
\ref{chap:equilibrium}, where a massive strange quark and a hadroninc
phase are included.

\index{liquid drop model!at finite temperature|)}

\clearemptydoublepage
\chapter{\label{chap:equilibrium}Quark Matter -- Hadronic Matter Equilibrium}
In this chapter I present a framework for the inclusion of finite size
effects in phase equlibrium calculations. This is a generalization of
the work by Lee and Heinz \cite{Lee93a}, who presented results for the
equilibrium between two bulk phases. It should be noted that this
work \cite{Jensen95a} was brought to its natural conclusion by He
\textit{et al.}~\cite{He96c}; a point to which I will return
at the end of this chapter.

\section{Scenario}
The collision of two heavy nuclei \index{heavy-ion collision}
is thought to give rise to a
\Index{quark-gluon plasma} phase if the energy density in the collision becomes
sufficiently large. In such a plasma the quarks would be free to roam the
entire phase, as a consequence of the asymptotic freedom\index{asymptotic
freedom} of quarks
predicted by QCD. So far no convincing experimental evidence for this
phase has been produced, but there are indications of a very hot, dense,
and spatially extended system created during the early stages of the
collision. If a quark gluon plasma \emph{is} formed it will expand and
cool. When the temperature reaches the transition temperature 
the system will start
to hadronize. If a quark-gluon plasma droplet has a sufficient
\index{strangelet!creation in heavy-ion coll.}
strangeness content it may cool to form a (meta-) stable strangelet. For
the quark-gluon plasma to develop a net amount of strangeness it is
necessary that strange quarks are produced in the collision. Strange
quarks are only significantly produced in strong interaction processes,
since the weak interaction time scale is much too large to play any role.
This means that strange quarks are created as part of $s\bar s$-pairs.
$s \bar s$ production is favored over $u \bar u$ and $d \bar d$
production because of Pauli blocking of $u$ and $d$ quarks, since the
light quark chemical potential is initially much larger than the strange
quark chemical potential. Also the mass of the strange quark is believed
to  be lower at the high temperatures prevailing due to restoration of
chiral symmetry\index{chiral symmetry!restoration of}.
In order to develop a net strangeness\index{strangeness!distillation
of} the plasma must rid
itself of some of the $\bar s$ quarks during the hadronization process.
Since there is a positive net baryon number, there will be more $u$ and
$d$ quarks around to combine with one of the $\bar s$ to form a
meson with strangeness 1 $(K^0,K^+)$, than there will be $\bar u$ and
$\bar d$ which might combine with an $s$ quark to form a meson with
strangeness $-1$ $(\bar{K^0},K^-)$.
Thus more $\bar s$ would escape the system than $s$, building up a net
negative strangeness. After this process of $s\bar s$ separation the
\index{strangeness separation}
total strangeness of the system would be conserved, again since the weak
interactions are too slow to ensure flavor equilibrium. This is a very
schematic description of how a strangelet might emerge in a heavy ion
collision. This scenario for strangelet production in heavy-ion
collisions was proposed by Greiner
\textit{et al.}~\cite{Greiner87a,greal88,Greiner91a}, and requires a
relatively high baryon number density in the plasma; a condition present
in the AGS, and to a lesser degree in the SPS. The next generation of
cooliders, the RHIC and LHC, will obtain much higher energies, resulting
in a higher degree of \Index{transparency} of the colliding nuclei and
hence a low baryon number density. A different mechanism is thus needed
to form strangelets under such circumstances. It was recently proposed
by Spieles \textit{et al.}~\cite{Spieles96a} that fluctuations might be
a decisive factor in the formation of strangelets at low baryon number
density.

\section{Description of the Model}
\subsection{The Quark Phase}
The quark matter phase is modeled as a collection of non-interacting
$u,d$, and $s$ quarks, their anti-quarks, and gluons in a \emph{spherical}
MIT bag.
I have  assumed massless $u$ and $d$ quarks, and
a strange quark with a mass of 150 MeV.
Since $u$ and $d$ are degenerate and electromagnetic interactions are
ignored there is isospin symmetry, and the two light quark species are
characterized by a single chemical potential, $\mu_q$. The $s$ quark has
its own chemical potential, $\mu_s$.

The quark phase is thus described by the thermodynamic potential
\begin{eqnarray}
  \Omega(T,V,\mu_q,\mu_s)
  &=& -\left[  \frac{37}{90}  \pi^2T^4
  +\mu_q^2T^2+\frac{1}{2\pi^2}\mu_q^4 +\Omega_{s\bar{s},V}(T,\mu_s)
	-B \right] V
  \nonumber \\
  &{\:}&+ \Omega_{s\bar{s},S}(T,\mu_s)S 
  \\
   &{\:}&+\left[ \frac{57}{108}T^2
	+ \frac{1}{
  4\pi^2}\mu_q^2 +\Omega_{s\bar{s},C}(T,\mu_s) \right] C.
	\nonumber
\end{eqnarray}
Here the light quarks contribute to the thermodynamic potential with
expressions given by \Eq{eq:quarkantiquarkvolume} and
\Eq{eq:quarkantiquarkcurvature}, whereas the strange quark contribution
must be evaluated numerically.

Instead of working with $\mu_q$
and $\mu_s$ it is possible to use the chemical potentials $\mu_B$ and
$\mu_{\msf{S}}$ associated
with the conserved charges baryon number, $A$, and strangeness,
\index{strangeness!chemical potential} $\msf{S}$
(not to be confused with $\cal S$ or $S$). The connection between these is
\be 
  \mu_q=\frac{1}{3}\mu_B \qquad\qquad \mu_s=\frac{1}{3}\mu_B-\mu_\msf{S}.
\ee
These chemical potentials have a well defined meaning in the hadron
phase as well.

The baryon number of the strangelet is given by
\begin{eqnarray}
  A&=&-\left( \frac{\partial \Omega}{\partial \mu_B}
  \right)_{R,T,\mu_\msf{S}} \nonumber \\
  &=&\frac{1}{3}\left[ (N_{q,V}+N_{s,V})V+N_{s,S}S
  +(N_{q,C}+N_{s,C})C \right] ,
\end{eqnarray}
where the net number densities of the light quarks are given by
\Eq{eq:nummasslessV} and \Eq{eq:nummasslessC}.

\subsection{The Hadronic Phase}
The hadronic phase consists of
the known hadronic spectrum, as obtained from the summary tables of the
Particle Data Group's \textit{Review of Particle Properties}~\cite{PDG}
(see also \cite{PDG96}). Interactions are included by means of a
hard core repulsion, leading to an excluded volume effect\index{excluded
volume effect}. The longer
range attractive part of the potential has been neglected. There are
several ways to incorporate excluded volume effects (see \eg\
Refs.~\cite{Cleymans93a,Rischke91a}, but
I use the statistical bootstrap model\index{statistical bootstrap model}
\cite{Hagedorn81a} of Hagedorn and
Rafelski, since this allows for a comparison with the results of
Lee and Heinz \cite{Lee93a}.

In this model thermodynamic quantities such as pressure,
energy density, and baryon number density are all obtained from the
corresponding ideal gas quantities by multiplication with the factor
$(1+\varepsilon^{\rm pt}/4B)^{-1}$, where ``pt'' denotes a point
particle quantity, and $\varepsilon$ and
$B$ are the energy density of the hadron gas and the bag constant
respectively. The following physical quantities are thus obtained
\begin{eqnarray}
  p^{(H)}=\frac{1}{1+\varepsilon^{\rm pt}/4B}p^{\rm pt}\\
  n_B^{(H)}=\frac{1}{1+\varepsilon^{\rm pt}/4B}n_B^{\rm pt}.
\end{eqnarray}
The superscript $(H)$ refers to the hadronic phase. The corresponding
point particle expressions are those of an ideal gas, for which the
pressure, energy density, and baryon number density, are given by
\begin{eqnarray}
  p^{\rm pt}&=&\sum_i p_i^{\rm pt} \, , \qquad p_i^{\rm pt}= \frac{g_i}{
    6\pi^2}\int_0^\infty dk\, \frac{k^4}{\epsilon_i(k)}
    \frac{1}{e^{\beta(\epsilon_i(k)-\mu_i)} \pm 1} \\
  \varepsilon^{\rm pt}&=&\sum_i \varepsilon_i^{\rm pt} \, , \qquad
    \varepsilon_i^{\rm pt}= \frac{g_i}{
    2\pi^2}\int_0^\infty dk\, \frac{k^2 \epsilon_i(k)}
    {e^{\beta(\epsilon_i(k)-\mu_i)} \pm 1} \\
  n_B^{\rm pt}&=&\sum_i b_i n_i^{\rm pt} \, , \qquad n_i^{\rm pt}=
    \frac{g_i}{2\pi^2}\int_0^\infty dk\, \frac{k^2 }{
    e^{\beta(\epsilon_i(k)-\mu_i)} \pm 1}, 
\end{eqnarray}
where $i$ runs over the hadronic states. $g_i$ is the degeneracy factor
of the species, and $b_i$ is the baryon number, which is $\pm 1$ for
baryons/anti-baryons and zero for mesons. $\epsilon_i(k)=\sqrt{k^2+m_i^2}$
is the energy of a particle with momentum $k$ and mass $m_i$.
The signs in the denominator
of the integrands correspond to Fermi-Dirac and Bose-Einstein
statistics. 

\subsection{Equilibrium Criteria\index{equilibrium criteria}}
The system consisting of a strangelet and a bulk hadronic phase is
considered to be in chemical and thermal equilibrium, so the temperature
and the strangeness and baryon chemical potentials
\index{strangeness!chemical potential} are the same in the
two phases
\be \label{eq.1}
  T^{(Q)}=T^{(H)}, \qquad \mu_B^{(Q)}=\mu_B^{(H)}, \qquad 
  \mu_\msf{S}^{(Q)}=\mu_\msf{S}^{(H)}.
\ee
Minimizing the free energy leads to the pressure equilibrium condition
\be \label{eq.2}
  \Omega_V+\frac{2}{R}\Omega_S+\frac{2}{R^2}\Omega_C+p^{(H)}=0.
\ee
The combined effect of pressure, surface tension, and curvature forces
from the quarks are countered by the bag pressure, $B$, plus the
hadronic pressure. In equilibrium the chemical potential for a hadronic
species, $i$, is given by its net light quark and strange quark content
$\nu_{q,i}$ and $\nu_{s,i}$ as
\be
  \mu_i=\nu_{q,i}\mu_q+\nu_{s,i}\mu_s.
\ee
The equilibrium conditions \Eq{eq.1} and \Eq{eq.2} can then be solved
numerically, yielding a set of triplets, $(T,\mu_q,\mu_s)$, defining 
the phase diagram.

\section{Results}
I have solved the above system of equations on a computer for a variety
of parameters. Bulk systems as well as finite quark
matter phases both at zero and non-zero temperature have been treated in
order to check against known limits.

The quark phase is parametrized
entirely in terms of the bag constant, $B$, and the quark masses. In
all cases I have used a strange quark mass $m_s=150 \,{\rm
MeV}$,
while $u$ and $d$ have been assumed massless in all calculations. The
bag constant has been varied between $B^{1/4}=145\,{\rm MeV}$, where $u,d$
matter is  marginally stable in the bulk, and $B^{1/4}=195\,{\rm MeV}$,
where $u,d,s$ matter is unstable relative to a gas of $\Lambda$'s, at
zero temperature.

With the chosen model
the hadronic phase is characterized solely in terms of the bag constant
and the particle
spectrum used. In most cases I have only included the ground state
baryons $N,\Lambda,\Sigma,\Xi,\Omega$ and $\Delta(1232)$, while the
mesons included are $\pi,K$ and $\eta$. This gives practically identical
results compared to the case in which a full spectrum of all resonances
below $\sim$2000~MeV was used, except for $B^{1/4}$ well in excess of 200
MeV. All the results depicted here are for the limited spectrum (see
Table~\ref{table:spectrum}).

\begin{table}
\begin{center}
\begin{tabular}{|c||c|c|c|c|r|c|} \hline
Name & $M$ (MeV) & $I$ & $J$ & $g$ &
\multicolumn{1}{c|}{$\msf{S}$}& Composition \\ \hline \hline
\multicolumn{7}{|c|}{Baryons} \\ \hline
$N$ & 939 & 1/2 & 1/2 & 4 & 0 & $p=uud \quad n=udd$ \\ \hline
$\Lambda$ & 1116 & 0 & 1/2 & 2 & -1 & $\Lambda=uds$ \\ \hline
$\Sigma$ & 1194 & 1 & 1/2 & 6 & -1 & $\begin{array}{c}\Sigma^0=uds \\
\begin{array}{cc}\Sigma^+=uus & \Sigma^-=dds\end{array} \end{array}$\\ \hline
$\Xi$ & 1317 & 1/2 & 1/2 & 4 & -2 & $\Xi^0=uss \quad \Xi^-=dss$ \\ \hline
$\Omega$ & 1672 & 0 & 3/2 & 4 & -3 & $\Omega^-=sss$ \\ \hline
$\Delta$ & 1232 & 3/2 & 3/2 & 16 & 0 & $\begin{array}{lr} \Delta^{++}=uuu &
 \Delta^+=uud \\ \Delta^0=udd & \Delta^-=ddd \end{array}$ \\ \hline
\multicolumn{7}{|c|}{Mesons} \\ \hline
 $\pi$ & 138 & 1 & 0 & 3 & 0 & $\begin{array}{c}
\pi^0=\frac{1}{\sqrt{2}}(u\bar{u}-d\bar{d}) \\
\begin{array}{cc} \pi^+=u\bar{d} & \pi^-=d\bar{d} \\
\end{array} \end{array} $ \\ \hline
$\eta$ & 549 & 0 & 0 & 1 & 0 & $\eta=c_1(u\bar{u}+d\bar{d})+c_2s\bar{s}$
\\ \hline
$\bar{K}$ & 496 & 1/2 & 0 & 2 & -1 & $K^-=s\bar{u} \quad \bar{K}^0=s\bar{d}$\\ \hline
\end{tabular}
\end{center}
\caption{\label{table:spectrum} The spectrum of low lying hadrons and
hadronic resonances used in the calculations for the hadron gas. For
each isospin multiplet with average mass $M$ the isospin $I$,
spin $J$, statistical weight $g$, and strangeness $\msf{S}$ are listed.}
\end{table}

The strangeness contents of the system is parametrized by the
\Index{strangeness fraction}, $f_s$, defined as the ratio of minus
the strangeness
density to the baryon density
\be
  f_s\equiv \frac{-n_\msf{S}}{n_B}.
\ee

\subsection{The Phase Diagram at $T=0$}
\index{phase diagram|(}
\begin{figure}
  \begin{center}
    \includegraphics[width=\linewidth]{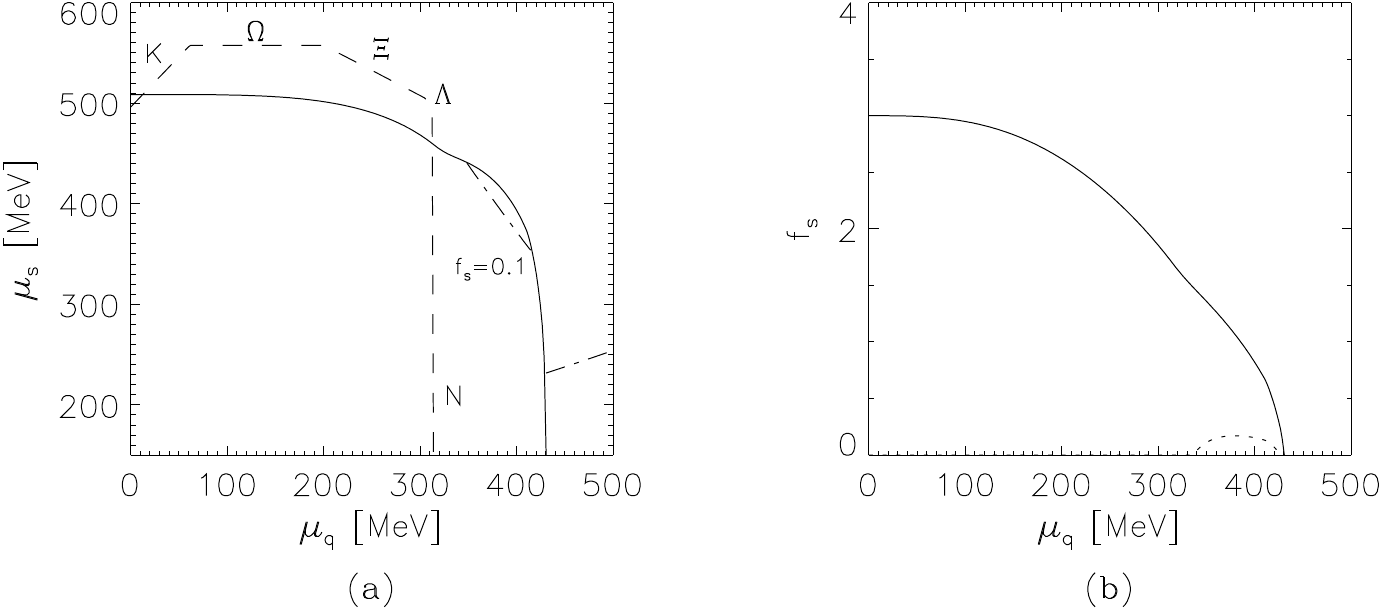}
    \caption{\label{fig:T0bulk}(a) Phase diagram for bulk
    strange quark matter and hadronic matter at $T=0$. (b) The
    strangeness fraction in the quark (full line) and hadronic (dotted
    line) phases as a function of the light quark chemical potential.
    $B=(190\,{\rm MeV})^4,\,\,m_s=150\,{\rm MeV}$.}
  \end{center}
\end{figure}
At zero temperature the strange quark thermodynamic potential is given
by the analytic expressions in Eqs.~(\ref{eq:omegaVT0})--(\ref{eq:omegaST0}).
The integrals occurring in the
expressions for the hadronic properties can also be evaluated
analytically. There are no non-strange mesons, and no anti-baryons.
Solving the equilibrium conditions \Eq{eq.1} and \Eq{eq.2} gives the
phase diagram in the $\mu_q$-$\mu_s$ plane as depicted in
Fig.~\ref{fig:T0bulk} for a \emph{bulk} quark matter phase. The solid line is
the coexistence line, along which quark matter and hadronic matter are
in equilibrium. Hadronic matter is the preferred phase inside the
region bounded by the coexistence line, while the system is in the quark
matter phase outside this region. Baryons are only present at $T=0$ if
the chemical potential $\mu_i$ of the species is greater than the
mass $m_i$. This is the case in the region outside the dashed line.
Each segment of the line is marked with a symbol denoting the particle
species appearing when crossing the line from the inside. When the
chemical potential for $\bar{K^0},K^-$ becomes equal to the kaon mass,
the kaons will undergo a \Index{Bose-Einstein condensation}. Since this cannot
easily be included in the model without a more detailed knowledge of the
potential acting between the kaons, the region of the phase diagram to
the left of the line marked $K$ is not accessible within the present
framework. Kaons will be present to the right of this line only at
finite temperature, so at zero temperature the region inside the dashed
line will be devoid of hadrons. Hadrons may thus be present in the
region outside the dashed curve and inside the full curve.
Coexistence curves for lower
values of $B$ will enclose a smaller region of the phase diagram. At a
value of $B^{1/4}\simeq 165\,{\rm MeV}$ the ``coexistence curve'' will
lie totally within the dashed curve, since quark matter will then be
stable relative to a hadron gas. At the point where the phase coexistence line
crosses the line where nucleons first appear there is a visible ``kink''
due to the fact that nucleons start to contribute to the pressure. The
dashed-dotted line represents systems with a total strangeness fraction
of 0.1. In the hadronic phase such a system would consist entirely of
$N$'s and $\Lambda$'s, for the parameter values chosen.

\begin{figure}
	\includegraphics[width=\linewidth]{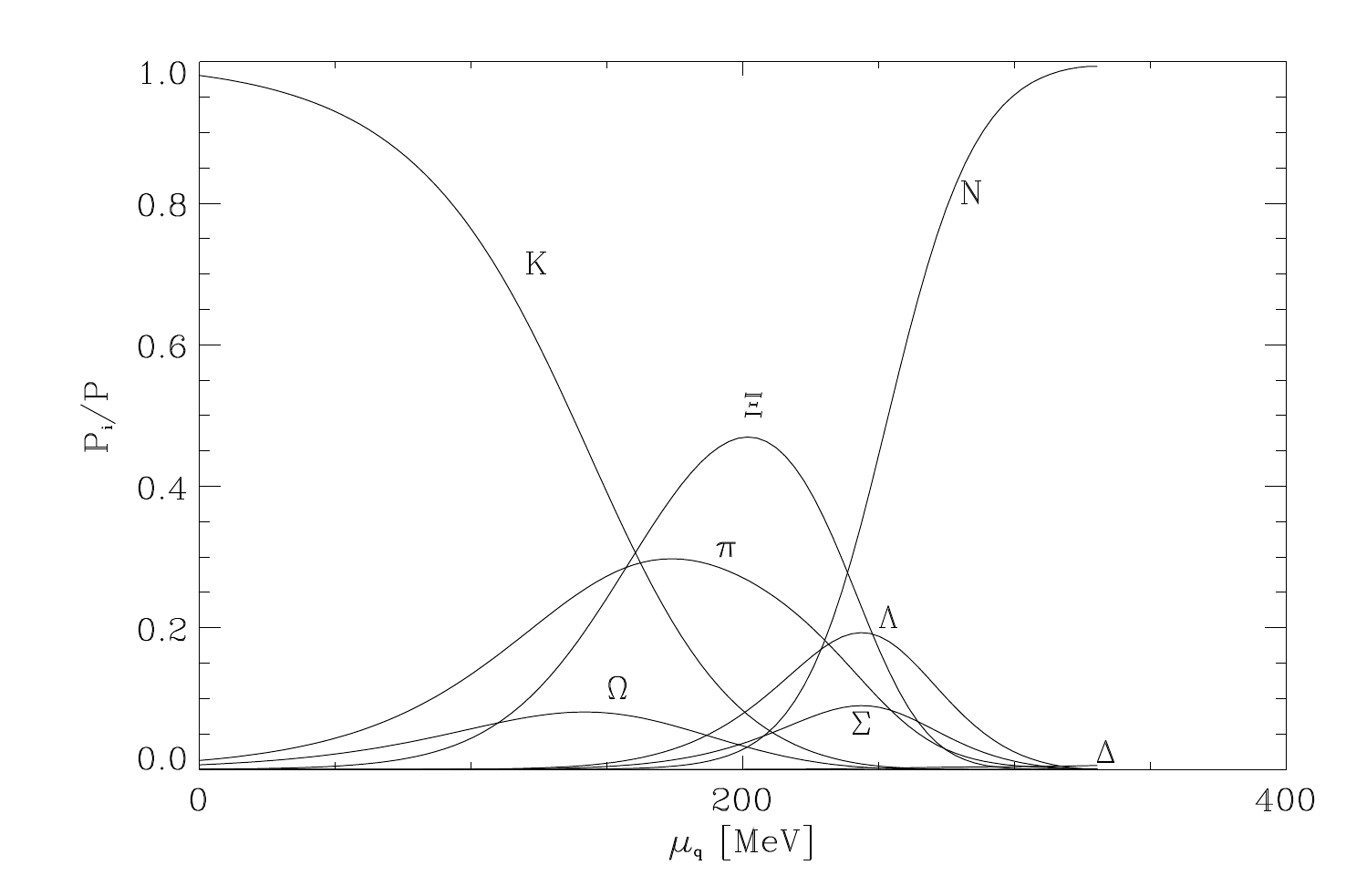}
	\caption{\label{fig:pressure}
		The contribution to the hadronic pressure from
	     different hadronic species. $B^{1/4}=165$ MeV, $T=40$
		 MeV, $A=20$, and $m_s=150$ MeV.}
\end{figure}

\subsection{The Phase Diagram for $T\not= 0$}
At finite temperature the quark matter phase contains thermal gluons and
anti-quarks, while the hadronic phase contains non-strange mesons and
anti-baryons. A phase diagram similar to the one in
Fig.~\ref{fig:T0bulk} can be constructed for a fixed temperature. The effect
of increasing the temperature is that the hadronic phase occupies a
smaller region of the phase diagram. This is because there is a large
number of massless degrees of freedom in the quark matter phase
($g_\star$ defined in \Eq{eq:gstar} has a value of $g_\star
=37$ for $\mcal{N}_q=2$), whereas the lightest constituent in the
hadronic phase is the pion with a mass of $\sim$140~MeV. This means
that the pressure increases faster in the quark phase than in the
hadronic phase, favoring the former over the latter. From this argument
it is also clear that an increase in the bag constant will have the
opposite effect, since $B$ effectively contributes to the hadronic
pressure. 

The reletive importance of the different hadronic species can be seen in
\Fig{fig:pressure}, where the fractional contribution to the total
hadron pressure is plotted for all the species listed in
Table~\ref{table:spectrum}. The plot is for values of $\mu_q$ and $\mu_s$
along the phase coexistence line, which has been parametrized by $\mu_q$
in the plot. For low $\mu_q$ there is a large strangeness fraction
($\mu_s$ is large) and the dominant pressure contribution is from kaons.
As $\mu_q$ increases other species begin to contribute---mainly the
light and/or strangeness rich species, such as the pion and $\Xi$, while
the $\Omega$ is somewhat suppressed due to its larger mass. As the system
becomes less strange the $\msf{S}=-1$ baryons $\Lambda$ and $\Sigma$ are
the main strange contenders. The nucleon dominates at the very highest
values of $\mu_q$ with the $\Delta$ resonance only just visible in this
plot. The $\eta$-meson's contribution to the pressure is too small to be
seen at all.

\begin{figure}
  \begin{center}
    \includegraphics[height=0.4\textheight]{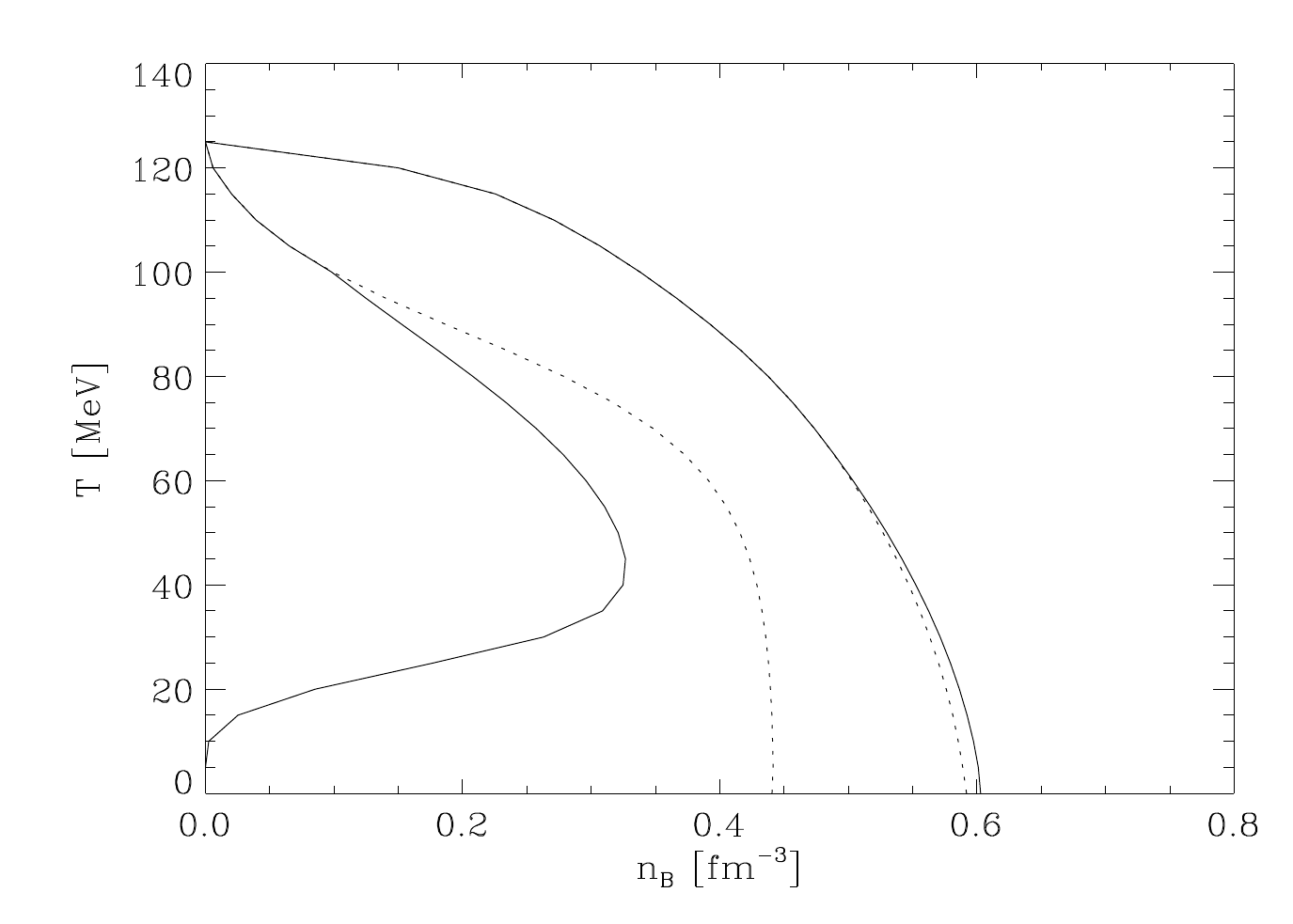}
    \caption{\label{fig:nbta20} Phase diagram for a
    strangelet with baryon number $A=20$, and a bulk hadronic phase. 
    The full curve is for a strangeness fraction, $f_s=0.1$, while the
    dashed curve is for $f_s=0.0$.
    $B^{1/4}=180\,\mathrm{MeV}$, $m_q=0$, $m_s=150\,\mathrm{MeV}$.
    }
  \end{center}
\end{figure}

\begin{figure}
  \begin{center}
    \includegraphics[width=\linewidth]{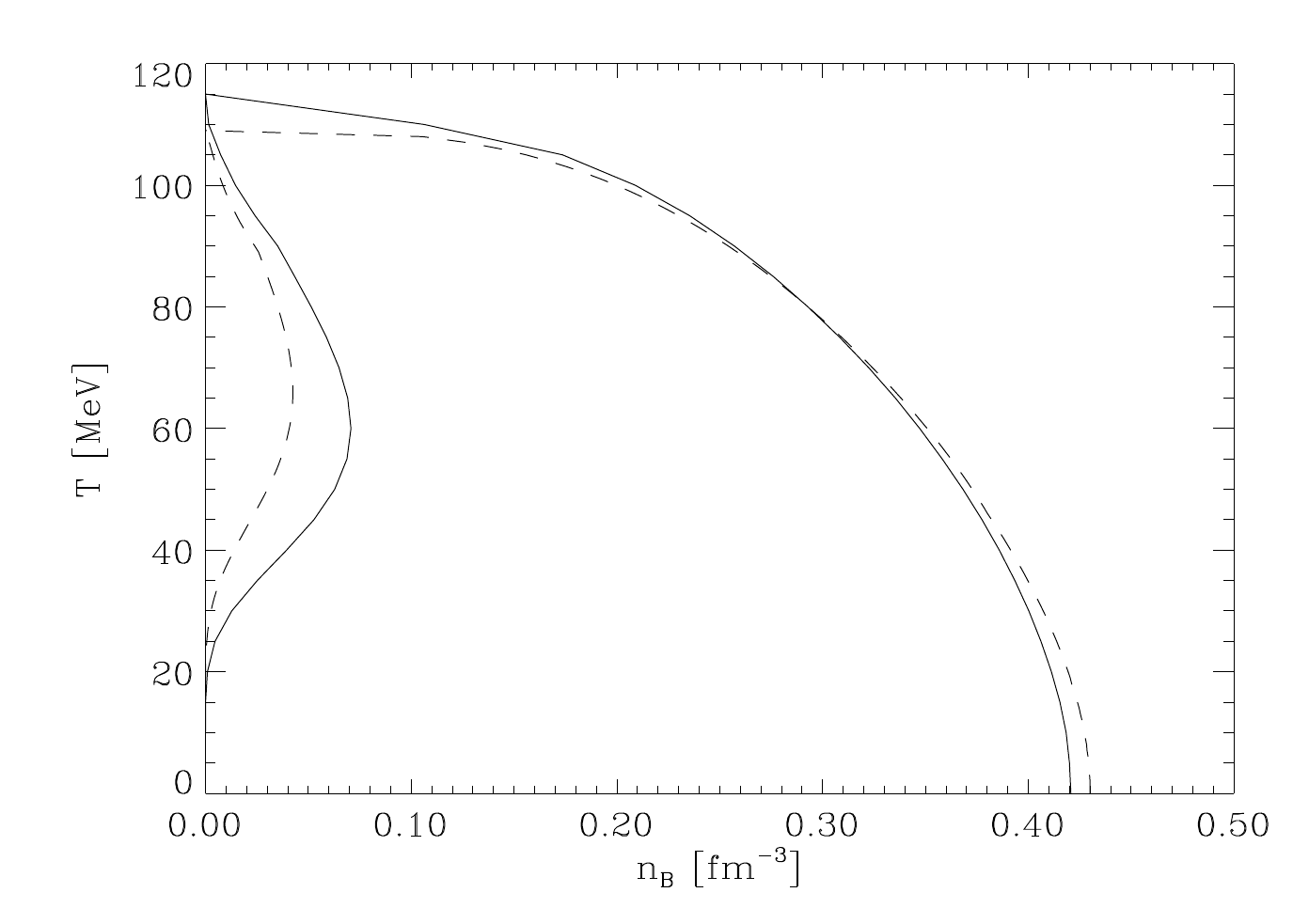}
    \caption{\label{fig:nbt.finite}
    Phase diagram for a bulk hadronic phase in equilibrium with a bulk
    quark phase (dashed line), and a quark phase with baryon number
    $A=20$ (full line). $B^{1/4}=165\,\mathrm{MeV}$, $m_q=0$,
    $m_s=150\,\mathrm{MeV}$, $f_s=0.1$.
    }
  \end{center}
\end{figure}

The quark matter phase benefits from an addition of strange quarks, as
opposed to the hadronic phase where the presence of strange hadrons tends
to increase the free energy, because of their larger masses. Thus
increasing the \Index{strangeness fraction} will favor the quark phase. This
effect is clearly seen in Fig.~\ref{fig:nbta20}, where the phase
diagram for a fixed strangeness fraction has been projected onto the
$n_B$-$T$ plane. The solid curves are for a strangeness fraction of 0.1,
whereas the dashed curves are for $f_s = 0$. The area between two like
curves correspond to a mixed phase. The quark-phase mixed-phase boundary
is the curve at high baryon number density, and the other curve is the
boundary between the mixed phase and the hadronic phase.

The effect of finite size of the quark matter phase is to destabilize
it against the hadronic phase, since the quarks and gluons must now
exert a pressure not only balancing the bag pressure $B$ and the
hadronic pressure, but also the surface tension and curvature forces of
the bag.
This effect is seen in
\Fig{fig:nbt.finite}, where phase diagrams for a bulk quark phase and a
quark phase with baryon number $A=20$ have been superimposed. The most
notable difference is that the hadronic phase grows on account of the
mixed phase when going from a bulk to a finite quark phase.

\section{Conclusion and Discussion}
In this chapter it has been demonstrated how to include the finite size
of the quark phase into phase equilibrium calculations. The effect of
varying the variuos parameters of the model on the equilibrium between
the two phases has been discussed, and most of these effects can be
understood intuitively by simple arguments.

There is a shortcoming of the approach to phase equlibrium described in
this chapter, \viz, that the quark phase is treated as having fixed
baryon number, while the hadron phase is considered as being a bulk
phase. Without lifting this unnatural constraint and instead fixing the
total baryon number of both phases it is not possible to study the
phase transition dynamics.

This shortcoming was adressed by He~\textit{et al.}~\cite{He96c}, who
considered an isentropic expansion of the system, as did Lee and
Heinz~\cite{Lee93a} for bulk systems.
They found that low
$B$, high $f_s$, and low specific entropy ${\cal S}/A$ favored
the production of a strangelet.

Isentropic expansion is only an approximation to the development of
the fireball created in a heavy-ion collision, and a more realistic
approach could be to solve the rate equations governing the abundances
of the hadronic species due to hadronization and reactions among the
different species, as was done by Barz, Friman, Knoll, and
Schulz~\cite{Barz88a,Barz90a,baral91}. These authors conclude that the
distillation of a strangelet is unlikely to happen unless
$B^{1/4}\lesssim 150$~MeV.

An approach intermediate between that of Barz~\textit{et al.}\ and that
of Lee and Heinz, and He~\textit{et al.}\ was suggested by Greiner
and St{\"o}cker~\cite{Greiner91a} who considered a thermodynamic
approach to the hadronization, rather than the more complicated rate
equation solution of Barz~\textit{et al.}  Greiner and St\"ocker obtain
more optimistic estimates of the chance for strangelet distillation than
Barz~\textit{et al.}, especially at higher initial entropy per baryon.

The inclusion of finite size effects in the models of Barz~\textit{et
al.}\ and Greiner and St\"ocker would be interesting, and in addition to
the work by He~\textit{et al.}\ where the finite baryon number of the
system is consistently treated, the inclusion of finite size effects on
the hadron phase as well as the quark phase seems a natural next step.
This could be done by imposing Dirichlet boundary conditions, and using
a multiple reflection expansion density of states for the hadrons.

As mentioned in the beginning of this chapter, Spieles~\textit{et al.}\
\cite{Spieles96a} recently proposed a scenario for the production of
strangelets from an initially almost baryonless plasma, such as is
belived to be produced by the next generation of heavy-ion colliders,
the RHIC and the LHC. Unfortunately the authors of
Ref.~\cite{Spieles96a} disregarded the finite size effects, which have
been demonstrated here to be important. Later findings by Spieles
\textit{et al.}\ \cite{Spieles96b} show that the inclusion of
finite size effects makes it practically impossible to produce
strangelets assuming the scenario proposed in Ref.~\cite{Spieles96a},
except perhaps if \Index{shell effects} are taken into account.

\index{phase diagram|)}

\clearemptydoublepage
\chapter{The Color Singlet Projection\label{chap:projection}}

In statistical physics, one is often interested in a description of a
system where one or more quantities are conserved overall.  The
appropriate partition function for such a system is said to be
canonical with respect to the conserved quantity (referred to as
a \Index{conserved charge}). It is also possible to construct a grand canonical
partition function where the quantity in question is allowed to
fluctuate in such a manner that its average value is well defined.
A well known example is the conservation of particle number.

When dealing with QCD\index{QCD} the situation is somewhat more
complicated. Here we want to constrain the system to be overall color
singlet. The added complication arises from the fact that the symmetry
describing the color interactions is not a simple
symmetry, but a higher-dimensional symmetry, \textit{viz.},
$SU(3)$\index{SU(3)@$SU(3)$}.  Demanding that
the system be color singlet is thus equivalent to saying that we wish
the system to be in a state that transforms according to a particular
(unitary irreducible) representation\index{representation} of the
symmetry group.

As early as 1936 Bethe \cite{Bethe36a} calculated the density of states
for a compound nucleus of a given total angular momentum. His method was
peculiar to the case of $SU(2)$ symmetry, and cannot therefore be
generalized to the problem of color singletness. The method used here
was developed by Redlich and Turko \cite{Redlich80a,Redlich80b,Turko81a},
and it is applicable to any symmetry described by a compact semi-simple
Lie group.

\section{Some Results from Group Theory}
In this section I will briefly outline some results from the theory of
Lie groups\index{Lie group} (see \textit{e.g.}, 
\cite{Jones90a,Huang92a,symmetries,Borchsenius82a})
that will be needed in the following sections.
This also serves to introduce the notation used in this \chapapp.
A word of warning:
Not all of the results quoted here are valid in general, but I have not
always bothered to explicitly list all the necessary assumptions. This
is mainly in order not to overburden the reader with ``irrelevant''
information. As an example of this kind of omission it is implicitly
assumed that there exist finite dimensional unitary representations of
the group. This is not the case if the group is non-compact, but as any
compact group is also semi-simple, and semi-simple groups are the focus
of attention in this \chapapp\ it is of no significance for the further
applications. 

\subsection{Lie Groups}

Let $\mcal{G}$ denote an $n$-parameter Lie group with elements $U$
and generators $L_i$ connected through
\be
	U = e^{i\theta_iL_i}, \qquad i=1,\ldots,n
\ee
via the real parameters $\theta_i$.
The generators satisfy the commutation relations
\be
	[L_i,L_j] = i C_{ijk}L_k \qquad i=1,\ldots,n
\ee
called the Lie algebra\index{Lie algebra}. The constants $C_{ijk}$ are
called the structure constants\index{structure constants} of $\mcal{G}$.

A subgroup $\mcal{A}$ of $\mcal{G}$ is called invariant\index{invariant
subgroup} if
\be
	UA_iU^{-1} = A_j	\qquad U\in\mcal{G}\quad A_i,A_j\in\mcal{A},
\ee
and is called an Abelian invariant subgroup if furthermore
\be
	[A_i,A_j] = 0 \qquad A_i,A_j\in\mcal{A}.
\ee
A group that does \emph{not} contain an invariant subgroup is called
simple\index{simple group}, and a group that does not contain an Abelian
invariant subgroup is called semi-simple. If a group \emph{does} contain
a continuous invariant subgroup it follows that there exists a
subalgebra of generators $\{\Lambda_i\}$ satisfying
\be
	[L_i,\Lambda_j] = i D_{ijk}\Lambda_k.
\ee
Such a subalgebra is called an \emph{ideal}\index{ideal}, and a Lie
algebra is called simple\index{simple Lie algebra} if it does \emph{not}
contain an ideal (other than the null ideal $\{0\}$). If an algebra does
not contain an Abelian ideal it is called semi-simple. A semi-simple
algebra (or group) can be simple, but the reverse is not true.

The maximal Abelian subalgebra of a Lie algebra is called the Cartan
algebra\index{Cartan algebra}, and the number of generators $r$ in the
Cartan algebra is called the \emph{rank}\index{rank} of the group.
According to a theorem by Racah \cite{Racah51a}---valid for any
semi-simple Lie group---there exist exactly $r$ so-called \emph{Casimir
operators}\index{Casimir operator} $C_i$ that commute with all
generators of the group (and therefore also among themselves).
Generally the Casimir operators are non-linear functions of the
generators.

In the case of $SU(N)$\index{SU(N)@$SU(N)$} the number of generators
is $n=N^2-1$ and the rank\index{rank!of SU(N)@of $SU(N)$} is $r=N-1$.

\subsection{Representations}
A representation\index{representation|(} $\Gamma$ of a group $\mcal{G}$
is a homomorphism\index{homomorphism} of
$\mcal{G}$ into $SL(N)$, the group of linear transformations with
determinant 1 of an $N$-dimensional vector space. A unitary
representation\index{representation!unitary} is a homomorphism
of $\mcal{G}$ into $U(N)$, the group of unitary transformations of an
$N$-dimensional vector space. $N$ is called the dimension of the
representation.  By the very definition of a homomorphism we have
\be
	\Gamma(U_1U_2) = \Gamma(U_1)\Gamma(U_2)	\qquad U_i \in \mcal{G}.
\ee
If $\mcal G$ and $\Gamma(\mcal{G})$ are isomorphic\index{isomorphism},
\textit{i.e.} $\Gamma(U_1)\not= \Gamma(U_2)$ if $U_1\not= U_2$,
$\Gamma$ is said to be \emph{faithful}. It can be shown that every
representation is equivalent  to a unitary representation (\textit{i.e.},
$\Gamma'=T \Gamma T^{-1}$,
where $T$ is a linear transformation, and $\Gamma'$ is unitary). Thus it
is only necessary to consider unitary representations, and in the
following all representations will be assumed unitary unless otherwise
noted.

Since the group $U(N)$---to which the elements $\Gamma(U)$ of a unitary
representation of $\mcal G$ belongs---is isomorphic to the group of
unitary $N\times N$ matrices, a representation $\Gamma$ establishes a
connection between an element $U$ of $\mcal G$ and a matrix
$D^{(\Gamma)}_{ij}(U)$.

A (unitary) representation is said to be reducible if there exists a
basis of the $N$-dimensional vector space in which the representative
matrices $D^{(\Gamma)}_{ij}(U)$ all have the form
\be \label{eq:blockdiag}
	D^{(\Gamma)}(U) = \left(
	\begin{array}{cc}
		D^{(1)}(U) & 0  \\
		0 & D^{(2)}(U)
	\end{array} \right).
\ee
If the representative matrices of a representation can not be brought
into this block diagonal form it is said to be
\emph{irreducible}\index{irreducible representation}%
\index{representation!irreducible}. If the submatrices
$D^{(1)}$ and $D^{(2)}$ are themselves reducible, $D^{(\Gamma)}(U)$ can
be further reduced. The full reduction of a reducible representation
into irreducible ones is written as the direct sum\index{direct sum}
\be
	D^{(\Gamma)}(U) = D^{(1)}(U) \oplus D^{(2)}(U) \oplus \cdots,
\ee
and the representation is said to be a direct sum of irreducible
representations
\be \label{eq:irrepdecomp}
	\Gamma = \Gamma^{(1)} \oplus \Gamma^{(2)} \oplus \cdots.
\ee
A given irreducible representation $\Gamma^{(i)}$ may appear more than
once in the decomposition (\ref{eq:irrepdecomp}).

We may interpret Eq.~(\ref{eq:blockdiag}) as to say that there are two
invariant subspaces of the $N$-dimensional vector space, under the
action of the elements in $\mcal G$. If $D^{(1)}$ has dimension $k$ the
first $k$ basis vectors are said to form a basis of the representation
$\Gamma^{(1)}$. The invariant subspace is said to generate a
representation, and a vector in the invariant subspace is loosely
referred to as a vector in the representation. The irreducible invariant
subspaces (those generating the irreducible representations) are called
\textit{multiplets}\index{multiplet}. Because the Casimir
operators commute with all generators and therefore with all group
elements, the eigenvalues of the Casimir operators are the same for
all members of the same multiplet. For a semi-simple Lie group the
multiplets are uniquely classified by the eigenvalues of the $r$
Casimir operators\index{Casimir operator!and multiplets}.

A \textit{character}\index{character} $\chi^{(\Gamma)}(U)$
of a representation
$\Gamma$ is the trace of the representative matrix $D^{(\Gamma)}_{ij}(U)$
\be \label{eq:chardef}
	\chi^{(\Gamma)}(U) = \Tr D^{(\Gamma)}_{ij}(U)
		= \sum_{i=1}^N D^{(\Gamma)}_{ii}(U),
\ee
which has the property that it is invariant under a change in basis of
the $N$-dimensional vector space.
The characters of the irreducible representations satisfy the important
orthogonality relation\index{orthogonality!of
characters}\index{character!orthogonality of}
\be \label{eq:charortho}
	\int d\mu(\{\theta_i\})\; \chi^{(m)}(U(\{\theta_i\}))
	\;	\chi^{(n)}(U(\{\theta_i\}))^\ast = \delta_{m,n},
\ee
where $d\mu(\{\theta_i\})$ is the so-called (normalized) Haar
measure\index{Haar measure|textit}. It depends only on the parameters
$\{\theta_i\}$ corresponding to generators in the Cartan subalgebra.


The representative matrices $D^{(\Gamma)}(U)$ of dimension $N$ are
conveniently obtained by the representations of the generators
$L_i$ by Hermitian $N\times N$ matrices.
Of special importance in gauge theories are the so-called
\textit{fundamental}\index{representation!fundamental} and \textit{adjoint}%
\index{representation!adjoint} representations. The fundamental
representation is the faithful representation of lowest dimensionality. The
adjoint representation is an irreducible  representation whose dimension
equals the number of generators, and whose matrix elements are given
by the structure constants
\index{structure constants!and adjoint representation} as
\be \label{eq:adjointrep}
	\lambda^{(i)}_{jk} = -iC_{ijk}.
\ee
The matrices $\frac{1}{2}\boldsymbol{\sigma}$, where
$\boldsymbol{\sigma}$ are the \emph{Pauli matrices}\index{Pauli matrices!as
representations of SU(2)@as representations of $SU(2)$},
constitute the fundamental representation of the generators of $SU(2)$.
\index{representation|)}

\section{The Group Theoretical Projection Method}
Consider a system, described by a Hamiltonian \Ham, that obeys a
symmetry described by the group $\mcal{G}$. As stated above we assume
$\mcal{G}$ to be a compact semi-simple Lie group. Since $\mcal{G}$ is a
symmetry of the system $\mcal{H}$ commutes with all the elements of
$\mcal{G}$ and therefore also with the generators, $L_i$
\be
	[\Ham,L_i] = 0 \qquad i=1,\ldots ,n.
\ee
A maximally commuting set of operators is constituted by \Ham\ along with
the $r$ generators in the \Cartan\ subalgebra, and  $r$ \Casimir\
operators of the group. 

The grand canonical partition function for the system is
\be
	\Z (T, \{\mu_i\}) = \Tr \exp \left[-\beta (\Ham -
				\sum_{i=1}^r \mu_i Q_i)\right],
\ee
where $Q_i$ are representations of the generators in the \Cartan\
algebra (\Index{charge operators}), and $\mu_i$ are the corresponding chemical
potentials.  The eigenvalues of the operators $Q_i$ are the additive
quantum numbers of the system.

In general we can
construct a partition function which is canonical with respect to a
particular representation $\Gamma$ of $\mcal{G}$, by including only
states in the representation  $\Gamma$. This
partition function will be denoted 
\be
	\mcal{Q}_\Gamma(T) = \Tr\nolimits_\Gamma e^{-\beta \Ham}
\ee
The objective of the group theoretical projection method is to obtain
$\mcal{Q}_\Gamma$ from \Z .

Now consider the function \Ztil , defined as
\be
	\Ztil (T,\{\gamma_i\}) = \Tr \exp \left[ -\beta \Ham
		+ i \sum_{i=1}^r \gamma_i Q_i \right].
\ee
It is related to the grand canonical partition function through the Wick
rotation $\gamma_i \to -i\beta \mu_i$.
Since a representation of a general group element in $\mcal{G}$ is 
\be
	D(U) = \exp( i\sum_{i=1}^n \theta_i L_i ),
\ee
where $L_i$ are representations of the generators, \Ztil\ may be written
\be
	\Ztil (T,\{\gamma_i\}) = \Tr\left[ D(U) e^{-\beta \Ham}
				\right],
\ee
where $D(U)$ is an element in the \Cartan\ subgroup.
The representation $D$ is now decomposed into irreducible
representations
, and the trace is
split up into traces over the invariant subspaces corresponding to the
irreducible representations (multiplets). Denoting a vector (state) in the
multiplet generating the irreducible representation $\Gamma^{(i)}$ by
$\left|\nu_i\xi_i\right\rangle$ the function \Ztil\ is now written
\be
	\Ztil = \sum_i \sum_{\nu_i,\xi_i}
		 \left\langle \nu_i\xi_i \right|
		D^{(i)}(U) e^{-\beta\Ham}
		\left| \nu_i\xi_i \right\rangle .
\ee
The multiplet is characterized by the eigenvalues of the Casimir
operators, for which $i$ may be taken as a shorthand notation. States
within each multiplet are characterized by the eigenvalues of the
Cartan operators denoted by $\nu$, and by any extra quantum numbers
$\xi$ needed to uniquely specify the state. Inserting a complete set
of states within the $i$th representation $1=\sum_{\nu'_i,\xi'_i}
|\nu'_i\xi'_i\rangle\langle\nu'_i\xi'_i|$ gives
\be
	\Ztil = \sum_{i} \sum_{\nu_i,\nu'_i} \sum_{\xi_i,\xi'_i}
                 \left\langle \nu_i\xi_i \right|
                D^{(i)}(U) \left|\nu'_i\xi'_i\right\rangle
		\left\langle \nu'_i\xi'_i\right| e^{-\beta\Ham}
                \left| \nu_i\xi_i \right\rangle .
\ee
The Hamiltonian is diagonal in $\nu$, and $D^{(i)}(U)$ is diagonal in $\xi$ 
since the charge operators act only on the $\nu$ part of the state.
\begin{eqnarray}
	\Ztil &=& \sum_i \sum_{\nu_i,\xi_i}
		\left\langle \nu_i\xi_i \right|
		D^{(i)}(U) \left|\nu_i\xi_i\right\rangle
		\left\langle \nu_i\xi_i\right| e^{-\beta\Ham}
		\left| \nu_i\xi_i \right\rangle \nonumber \\
	&=& \sum_i \sum_{\nu_i}
		D^{(i)}_{\nu_i\nu_i}(U) \sum_{\xi_i}
		\left\langle \nu_i\xi_i\right| e^{-\beta\Ham}
		\left| \nu_i\xi_i \right\rangle \nonumber \\
	&=& \sum_i \chi^{(i)}(U) \sum_{\xi_i}
		\left\langle \nu_i\xi_i\right| e^{-\beta\Ham}
                \left| \nu_i\xi_i \right\rangle.
\end{eqnarray}
Here the definition (\ref{eq:chardef}) of the character has been used.
Since \Ham\ is degenerate on each multiplet, it is possible to insert a
spurious sum over members in the multiplet
\begin{eqnarray}
	\Ztil &=& \sum_i \chi^{(i)}(U)
		\frac{1}{d_i} \sum_{\nu_i} \sum_{\xi_i}
		\left\langle \nu_i\xi_i\right| e^{-\beta\Ham}
		\left| \nu_i\xi_i \right\rangle \nonumber \\
	&=& \sum_i \chi^{(i)}(U) \frac{1}{d_i}\Tr\nolimits_i e^{-\beta\Ham}
		= \sum_i \chi^{(i)}(U)\frac{1}{d_i} \mcal{Q}_i ,
\end{eqnarray}
where $d_i$ is the dimension of the $i$th representation.
Using the orthogonality of characters (\ref{eq:charortho}) one readily
obtains
\be \label{eq:projection}
	\mcal{Q}_j(T) = d_j \int d\mu(\{\theta_i\})
	\Ztil(T,\{\gamma_i\}) \chi^{(j)}(\{\gamma_i\})^\ast
\ee
This is the formula connecting the grand canonical partition function
with the canonical partition function $\mcal{Q}_i$. The effect of the
integral over the group with the complex conjugate of the character of
the $j$th irreducible representation is to project out from the total
trace only that part which is a trace over states in the multiplet
generating the $j$th representation.

\section{A Simple Example: $U(1)$}
As an example of an application of the group theoretical projection
method just derived, I will use it to obtain the formula
\be \label{eq:gc2c}
	\mcal{Q}_N(T,V) = \frac{1}{2\pi} \int_{-\pi}^{+\pi}
		d\lambda \; e^{i\lambda N} \mcal{Z}(T,V,iT\lambda),
\ee
where $\mcal{Q}_N(T,V)$ is the canonical partition function for a
system of $N$ particles, and $\mcal{Z}(T,V,\mu)$ is the grand canonical
partition function with chemical potential $\mu$. This expression is the
inversion of the very definition of the grand canonical partition
function
\be \label{eq:gcdef}
	\mcal{Z}(T,V) = \sum_{N=0}^\infty e^{\beta\mu N} \mcal{Q}_N(T,V).
\ee
In statistical physics texts Eq.~(\ref{eq:gc2c}) is usually obtained by
inverting the series (\ref{eq:gcdef}) by first performing a Wick
rotation and then using Fourier theory (see \textit{e.g},
\cite{GreinerThermo}), or by introducing a delta
function $\delta(N-\bar{N}) = \int\frac{d\lambda}{2\pi}
\exp [i\lambda (N-\bar{N})]$ in the density matrix. Here it will follow
from the general result of Eq.~(\ref{eq:projection}) by using $U(1)$ as
the symmetry group.

In gauge field theory number conservation arises as a consequence of
invariance under phase changes of state vectors of the type
\be
	\psi'(x) = e^{i\theta(x) \mcal{N}} \psi(x) ,
\ee
where $\mcal{N}$ is the number operator satisfying $[\Ham,\mcal{N}]=0$,
where \Ham\ is the Hamiltonian.
This is recognized as a $U(1)$ symmetry, with $\mcal{N}$ being the sole
generator and therefore also both the Casimir operator and the Cartan
generator of the group.
The irreducible representations are labeled by the eigenvalue $N$ of
$\mcal{N}$, and since the dimension is 1, they are identical with the
characters
\be
	\chi^{(N)} = D^{(N)} = e^{i\theta(x) N} ,
\ee
where
\be
	\mcal{N} |N \xi\rangle = N |N\xi\rangle .
\ee
States are labeled by $N$ and $\xi$, the latter denoting all other
quantum numbers of the state.
The grand canonical partition function is the unrestricted trace
\be
	\mcal{Z} = \Tr e^{-\beta (\Ham -\mu\mcal{N}) }
		= \sum_{N\xi} \langle N\xi |
		e^{-\beta (\Ham -\mu\mcal{N}) }
		| N \xi \rangle ,
\ee
and the canonical partition function is a trace over states in the
representation labeled by $N$
\be
	\mcal{Q}_N = \Tr\nolimits_N e^{-\beta\Ham} 
		= \sum_{\xi} \langle N\xi |
		e^{-\beta\Ham}
		| N \xi \rangle .
\ee
The function \Ztil\ is given by
\be
	\Ztil = \Tr e^{-\beta \Ham +i\theta\mcal{N} } .
\ee
Given the  Haar measure\index{Haar measure!of U(1)@of $U(1)$} of
$U(1)$, which  is $d\mu = d\theta/2\pi$ a
direct insertion in Eq.~(\ref{eq:projection}) gives
\begin{eqnarray}
	\mcal{Q}_N &=& \frac{1}{2\pi} \int_{-\pi}^{+\pi} d\theta\;
		\chi^{(N)}(\theta)^\ast \; \Ztil(\theta)
	= \frac{1}{2\pi} \int_{-\pi}^{+\pi} d\theta\;
		e^{-i\theta N} \Ztil(\theta) \nonumber \\
	&=& \frac{1}{2\pi} \int_{-\pi}^{+\pi} d\theta\;
		\Tr e^{-\beta \Ham +i\theta(\mcal{N} -N)} ,
\end{eqnarray}
which with the substitution $\lambda=-\theta$ gives
\be
	\mcal{Q}_N = \frac{1}{2\pi} \int_{-\pi}^{+\pi}
	d\lambda\; e^{i\lambda N} \Z(iT\lambda), 
\ee
which is the promised result.

\section{The Color Singlet Partition Function}
After having seen the group theoretical projection method in action in a
simple case it is now time to proceed to the actual goal,
\textit{viz.}, the calculation of the canonical color singlet partition
function of a quark-gluon plasma (or hot strangelet).

\subsection{Background}
\comment{General references? Itzykson and Zuber, Huang, Pokorski}
Quantum chromodynamics\index{QCD} is a Yang-Mills
\index{Yang-Mills theory} gauge theory based on the  $SU(3)$
\index{SU(3)@$SU(3)$} symmetry group. The Lagrangian density of QCD
is invariant
under the simultaneous local gauge transformations of quark spinors
$\Psi$ and the gluon field tensor $F^{\mu\nu}$ according to
\be
   \begin{array}{rcl}
	\Psi(x) &\to&  D(x) \Psi(x) \\
	& & \\
	F^{\mu\nu}(x) &\to& D(x)F^{\mu\nu}(x)D^{-1}(x)
   \end{array}
\ee
where
\be
	 D(x) = e^{i\theta_i(x)\Lambda_i}
\ee
is the fundamental representation of an element in $SU(3)$, in which the
generators are represented by the Gell-Mann matrices \index{Gell-Mann
matrices} $\Lambda_i =\frac{1}{2}\lambda_i$
\be
	\begin{array}{lll} 
	
	\lambda_1 = \left( \begin{array}{ccc}
				0 & 1 & 0 \\
				1 & 0 & 0 \\
				0 & 0 & 0 
		\end{array} \right) 
	&
	\lambda_2 = \left( \begin{array}{ccc}
				0 & -i & 0 \\
				i & 0 & 0 \\
				0 & 0 & 0 
		\end{array} \right) 
	&
	\lambda_3 = \left( \begin{array}{ccc}
				1 & 0 & 0 \\
				0 & -1 & 0 \\
				0 & 0 & 0 
		\end{array} \right) 

	\\ & & \\ 

	\lambda_4 = \left( \begin{array}{ccc}
				0 & 0 & 1 \\
				0 & 0 & 0 \\
				1 & 0 & 0 
		\end{array} \right) 
	&
	\lambda_5 = \left( \begin{array}{ccc}
				0 & 0 & 1 \\
				0 & 0 & 0 \\
				1 & 0 & 0 
		\end{array} \right) 
	&
	\lambda_6 = \left( \begin{array}{ccc}
				0 & 0 & 0 \\
				0 & 0 & 1 \\
				0 & 1 & 0 
		\end{array} \right) 

	\\ & & \\ 

	\lambda_7 = \left( \begin{array}{ccc}
				0 & 0 & 0 \\
				0 & 0 & -i \\
				0 & i & 0 
		\end{array} \right) 
	&
	\lambda_8 = \frac{1}{\sqrt{3}}
		    \left( \begin{array}{ccc}
				1 & 0 & 0 \\
				0 & 1 & 0 \\
				0 & 0 & -2 
		\end{array} \right) 
	& 

	\end{array}
\ee
It is usually more convenient to consider the infinitesimal
transformations
\be \label{eq:infinitrans}
   \begin{array}{rcl}
	\Psi(x) &\to& \Psi(x) +i\theta_i(x)\Lambda_i\Psi(x)  \\
	& & \\
	F^{\mu\nu}(x) &\to& F^{\mu\nu}(x) + i\left[ \theta_i(x)\Lambda_i,
	F^{\mu\nu}(x) \right]
   \end{array}
\ee
The field tensor is given by the gauge vector potential $A^\mu(x)$ as
\be
	F^{\mu\nu} = \partial^\mu A^\nu(x) - \partial^\nu A^\mu(x)
		-ig\left[ A^\mu(x) , A^\nu(x) \right] ,
\ee
which also introduces the gauge coupling constant $g$. $A^\mu(x)$ has
an index of the adjoint representation \index{adjoint
representation!gauge potential}
\be
	A^\mu(x) = A^\mu_i(x)L_i ,
\ee
where $L_i$ is the adjoint representation of the generators.
Rewriting Eq.~(\ref{eq:infinitrans}) to display the index of the Lie
algebra corresponding to the color degree of freedom gives
\be
    \begin{array}{rcl}
	\Psi_a(x) &\to& \Psi_a(x) + i\theta_b(x) (\Lambda_b)_{ac}\Psi_c(x)
	\\ & &  \\
	F^{\mu\nu}_a(x) &\to& F^{\mu\nu}_a(x) -
	 C_{bca}\theta_b(x)F^{\mu\nu}_c(x) .
    \end{array}
\ee
Since $\Lambda$ is the fundamental representation of dimension 3
(usually denoted by \triplet ) the quark field color index $a$
can take 3 values, say red, green, and blue. The gluon field tensor is
seen to transform according to the adjoint representation given by the
structure constants of $SU(3)$. The adjoint representation has dimension
8 and is denoted \octet . There are thus 8 varieties of gluons.
\comment{Antiquarks are described by
$\tilde{\Psi}=CPT\Psi=\gamma_0i\gamma_2\left[i\gamma_1\gamma_3
\left(\Psi\right)^\ast\right]^\ast$}
Antiquarks are described by the conjugate spinor $\bar{\Psi}$, which
transforms according to the representation \antitriplet\
conjugate to \triplet , which has generators $-\frac{1}{2}\lambda_i^\ast$.
These
two representations (triplet and anti-triplet) are both fundamental,
but they are not equivalent.

A single particle state (in one of the three representations \triplet ,
\antitriplet , or \octet )
is labeled by the eigenvalues of the \Cartan\ operators, which 
can by chosen as $L_3$ and $L_8$
\be
    \begin{array}{rcl}
	L_3 \left| I_3 Y_8 \xi \right\rangle &=&
		I_3 \left| I_3 Y_8 \xi \right\rangle \\
	& & \\
	L_8 \left| I_3 Y_8 \xi \right\rangle &=&
		\frac{\sqrt{3}}{2}Y_8 \left| I_3 Y_8 \xi \right\rangle ,
    \end{array}
\ee
where $I_3$ is the third component of ``strong isospin'' and $Y_8$ is
the ``strong hypercharge'', in analogy with the nomenclature used in
the connection with $SU(3)_\mrm{flavor}$\index{SU(3)@$SU(3)$!of flavor}
symmetry. $\xi$ denotes all other quantum numbers that are needed to
specify the state. The possible values of $I_3$ and $Y_8$ are found by
diagonalizing the matrix representations of $L_3$ and $L_8$.

For quarks
in the \triplet\ representation the pairs of eigenvalues can immediately
be read from the Gell-Mann matrices as
\be
    \begin{array}{rcrrr}
	I_3 &=& \frac{1}{2}, & -\frac{1}{2}, & 0\phantom{,} \\
	& & & & \\
	Y_8 &=&  \frac{1}{3}, & \frac{1}{3}, & -\frac{2}{3},
    \end{array}
\ee
and the corresponding states may by labeled by an index $c=r,g,b$.
The triplet of quark states in the \triplet\ representation is thus
$|I_3 Y_8 \rangle = |\frac{1}{2},\frac{1}{3}\rangle_r$,
$|-\frac{1}{2},\frac{1}{3}\rangle_g$, $|0, -\frac{2}{3}\rangle_b$.

For antiquarks, the eigenvalues are minus the above, due to the opposite
sign on the Gell-Mann matrices in relation to the representative
matrices. Antiquark states will be labeled by $c=\bar{r},\bar{g},\bar{b}$.

The matrices $L_3$ and $L_8$ in the adjoint representation
\index{adjoint representation!and gluons} relevant for
gluons can be found from the prescription in Eq.~(\ref{eq:adjointrep}),
where the structure constants can be found by calculating commutators of
the Gell-Mann matrices. The result of simultaneously diagonalizing the
matrices $L_3$ and $L_8$ in the adjoint representation is the pairs of
eigenvalues
\be \label{eq:octeteigenvalues}
    \begin{array}{rcrrrrrrrr}
	I_3 &=& 0, & 0, & 1, & -1, & \frac{1}{2}, & -\frac{1}{2}, &
		\frac{1}{2}, & -\frac{1}{2}\phantom{,} \\
	& & & & & & & & &\\
	Y_8 &=& 0, & 0, & 0, & 0, & 1, & -1, & -1, & 1,
    \end{array}
\ee
and the corresponding eigenstates may be labeled by $c=1,\ldots,8$.

This shows that
instead of using the eigenvalues of the two Cartan operators to label
states within the irreducible representations it is possible to label
them by a single ``color index'' $c$ which for quarks and antiquarks
has three possible values, and for gluons takes  on eight different
values. This is just an arbitrary enumeration of substates within each
multiplet, with the index taking on as many different values as the
dimension of the representation.

The multiplets (and the corresponding unitary irreducible
representations) are characterized by the eigenvalues of the two Casimir
operators, which like the Hamiltonian are degenerate on a multiplet. For
$SU(3)$ it is usual practice \cite{symmetries} to characterize the irreducible
representations $D^{(p,q)}$ by two non-negative integers $p$ and $q$
which are given by the maximum values of the eigenvalues of the two
Cartan generators within the multiplet as
\be
	I_3^{(\mrm{max})} = \frac{p+q}{2} \qquad \qquad
	Y_8^{(\mrm{max})} = \frac{p-q}{3} .
\ee
Thus the triplet representation \triplet\ is $ D^{(1,0)}$, the
anti-triplet \antitriplet\ is $D^{(0,1)}$, and the octet \octet\ is
$D^{(1,1)}$.

\subsection{The Partition Function}
The grand canonical partition function for a system of non-interacting
quarks, antiquarks, and gluons taking into account the internal $SU(3)$
symmetry is
\be
	\mcal{Z} = \Tr e^{-\beta (\Ham -\mu_i\mcal{N}_i - \mu_3 L_3
		- \mu_8 L_8 )} ,
\ee
where $\mu_i$ ($i=u,d,\ldots$)  are chemical potentials related to the
$U(1)$ symmetries giving rise to conservation of flavor, and $\mu_3$ and
$\mu_8$ are the chemical potentials related to the third component of
color isospin and color hypercharge respectively. Since only
non-interacting systems are considered the Hamiltonian is a sum of
single particle Hamiltonians. The partition function therefore
factorizes into independent contributions from each species
\be
	\mcal{Z} = \prod_i \mcal{Z}_i .
\ee

A multiparticle state that transforms according to an irreducible
representation of $SU(3)$ can in principle be constructed, but this is
a complex task. Luckily it is also an unnecessary task, since the trace
is independent of the chosen basis. This fact can be used to evaluate
the trace in a convenient basis of multiparticle states in which a state
of the whole system is a product of states of each species, so that in
evaluating $\mcal{Z}_i$ only substates of the species $i$ are needed.

The grand canonical partition function with Wick \index{Wick rotation}
rotated chemical potentials is
\be \label{eq:Ztilde}
	\Ztil = \Tr e^{-\beta (\Ham -\mu_i\mcal{N}_i) + i\theta_3 L_3
		+i \theta_8 L_8 } .
\ee
\Ztil\ is used to construct the color singlet canonical partition function
by projecting onto the subspace of many-particle states transforming
according to the singlet representation $D^{(0,0)}$. The singlet
representation has dimension 1, and since it corresponds to $I_3=0$ and
$Y_8=0$ every element is mapped onto the $1\times 1$ unit matrix,
\textit{i.e.} 1. The canonical color singlet partition function is thus
given by Eq.~(\ref{eq:projection}) as 
\be \label{eq:Zsinglet}
	\mcal{Z}_{\singlet} = \int d\mu(\theta_3,\theta_8)\,
		\Ztil(\theta_3,\theta_8) .
\ee
It is more convenient in the following to use the eigenvalues of $I_3$
and $Y_8$ rather than those of $L_3$ and $L_8$, so two parameters
$\phi$ and $\psi$ are introduced by
\be
	\phi = \theta_3 \qquad \qquad
	\psi = \frac{\sqrt{3}}{2} \theta_8 ,
\ee
so that $\theta_3L_3 = \phi I_3$ and $\theta_8L_8 = \psi Y_8$. The Haar
measure of $SU(3)$ \index{Haar measure!of SU(3)@of $SU(3)$} is most
conveniently expressed in terms of yet another parametrization,
\viz , the parameters corresponding to $\lambda_3$ and
$\sqrt{3}\lambda_8$, which are 
\be
	\gamma_3 = \frac{1}{2} \phi = \frac{1}{2} \theta_3 \qquad
	\gamma_8 = \frac{1}{3} \psi = \frac{1}{2\sqrt{3}} \theta_8 .
\ee
In terms of this set of parameters the Haar measure is
\index{Haar measure!of SU(3)@of $SU(3)$}
\cite{symmetries}
\comment{Check this again?}
\be \label{eq:HaarSU3}
	d\mu(\gamma_3,\gamma_8) = \frac{8}{3\pi^2}\left[
		\sin \gamma_3 \sin \frac{1}{2}(3 \gamma_8 + \gamma_3)
		\sin \frac{1}{2}(3 \gamma_8 - \gamma_3) \right]^2 
	d\gamma_3 d\gamma_8 ,
\ee
where $\gamma_3, \gamma_8 \in [-\pi,\pi]$.

The equations (\ref{eq:Ztilde}), (\ref{eq:Zsinglet}), and
(\ref{eq:HaarSU3}) together give a
concise prescription of how to construct the color singlet partition
function $\mcal{Z}_{\singlet}$. The actual calculation of \Ztil\ and
$\mcal{Z}_{\singlet}$ is the topic of the following sections. In the
preceding I have called $\mcal{Z}_{\singlet}$ the canonical color
singlet partition function, in order to distinguish it from
$\Ztil(\gamma_3=0,\gamma_8=0)$, which might be called the grand
canonical color singlet partition function, since it represents a system
which although color singlet in mean, allows fluctuations with respect
to color. On the other hand $\mcal{Z}_{\singlet}$ is still grand
canonical with respect to the $U(1)$ symmetries related to the
conservation of each quark flavor, \ie\ it still depends on the chemical
potentials $\mu_i$. In order to stress this grand canonical nature of the
ensemble, the function $\mcal{Z}_{\singlet}$ will later be referred to
as the grand canonical color singlet partition function. Since this is
the only color singlet partition function referred to there should be no
chance of misunderstanding. All this notational trouble stems from the
fact that $\mcal{Z}_{\singlet}$ is really mixed canonical/grand
canonical partition function \cite{Elze86a}.

\subsubsection{Quarks}

The  contribution to \Ztil\ from a single quarks species, say $q$, with
chemical potential $\mu_q$ is
\be
	\tilde{Z}_q = \Tr e^{-\beta (\Ham_q - \mu_q\mcal{N}_q)
		+ i\phi \hat{I}_3 + i\psi \hat{Y}_8} ,
\ee
where there is \emph{no} summation over $q$. A single particle state
$|\mbf{k}, \sigma, c \rangle$ is
characterized by the quark momentum $\mbf{k}$, the helicity
\index{helicity!for quarks} $\sigma=\pm \frac{1}{2}$,
and its color $c=r,g,b$. The creation operator 
$a^\dagger_{\mbf{k}, \sigma, c}$, when acting on the
vacuum ket $|0\rangle$,  creates the state $|\mbf{k}, \sigma, c \rangle$.
Multiparticle states $|n\rangle = |n_{\mbf{k}_1, \sigma_1, c_1},\ldots,
n_{\mbf{k}_l, \sigma_l, c_l},\ldots \rangle$ are constructed by successively
applying the creation operators
\be
	|n\rangle = \prod_i \left( a^\dagger_{\mbf{k}_i, \sigma_i, c_i}
		\right)^{n_i} |0\rangle .
\ee
The number of quarks $n_i$ with quantum numbers $\mbf{k}_i, \sigma_i,
c_i$ is either 0 or 1.
\comment{The momentum spectrum is considered to be discrete.}
The anticommutation relations $\{ a^{\phantom{\dagger}}_{\mbf{k_1},
\sigma_1, c_1} ,
a^\dagger_{\mbf{k_2},\sigma_2, c_2} \} = \delta_{\mbf{k}_1 \mbf{k}_2}
\delta_{\sigma_1 \sigma_2} \delta_{c_1 c_2}$ ensures that the resulting
state has the correct symmetry. 
The Hamiltonian, the number operator, and the color isospin and
hypercharge operators have the representations
\begin{eqnarray}
	\Ham_q &=& \sum_{\mbf{k},\sigma,c} \epsilon_q(\mbf{k})
		a^\dagger_{\mbf{k},\sigma,c}
		a^{\phantom{\dagger}}_{\mbf{k},\sigma,c} \\
	\mcal{N}_q &=& \sum_{\mbf{k},\sigma,c}
		a^\dagger_{\mbf{k},\sigma,c}
		a^{\phantom{\dagger}}_{\mbf{k},\sigma,c} \\
	\hat{I}_3 &=& \sum_{\mbf{k},\sigma,c} I_3(c)
		a^\dagger_{\mbf{k},\sigma,c}
		a^{\phantom{\dagger}}_{\mbf{k},\sigma,c} \\
	\hat{Y}_8 &=& \sum_{\mbf{k},\sigma,c} Y_8(c)
		a^\dagger_{\mbf{k},\sigma,c}
		a^{\phantom{\dagger}}_{\mbf{k},\sigma,c} ,
\end{eqnarray}
where $\epsilon_q(\mbf{k}) = \sqrt{\mbf{k}^2 + m_q^2}$ is the energy of
a quark.

The function $\Ztil_q$ can now be written
\be
	\Ztil_q = \Tr \exp  \left( \sum_{\mbf{k},\sigma,c}\left\{
		-\beta \left[ \epsilon_q(\mbf{k}) - \mu_q \right]
		+i \left[ \phi I_3(c) + \psi Y_8(c) \right] \right\} 
		 \right) . 
\ee
The trace is now written as a sum over states in the Fock space of
quarks, and the exponential of the sum is written as a product of
exponentials, to give
\be
	\Ztil_q = \sum_{|n\rangle} \prod_{\mbf{k},\sigma,c} \langle n|
		e^{ \left\{ -\beta \left[ \epsilon_q(\mbf{k}) - \mu_q \right]
		+i \left[ \phi I_3(c) + \psi Y_8(c) \right] \right\}
		a^\dagger_{\mbf{k},\sigma,c}
		a^{\phantom{\dagger}}_{\mbf{k},\sigma,c}}
		| n \rangle .
\ee
Instead of summing over all possible states $|n\rangle$, it is possible
to sum over all possible occupation numbers (\ie\ 0 and 1) in effect
interchanging the sum and product
\begin{eqnarray}
	\Ztil_q &=& \prod_{\mbf{k},\sigma,c} \; 
		 \sum_{n_{\mbf{k},\sigma,c}=0}^1
		 e^{ \left\{ -\beta \left[ \epsilon_q(\mbf{k})
		- \mu_q \right]
		+i \left[ \phi I_3(c) + \psi Y_8(c) \right] \right\}
		n_{\mbf{k},\sigma,c} } \nonumber \\
	&=& \prod_{\mbf{k},\sigma,c} \left( 1 +
		e^{ -\beta \left[ \epsilon_q(\mbf{k}) - \mu_q \right]
		+i \left[ \phi I_3(c) + \psi Y_8(c) \right] } \right) .
\end{eqnarray}
It is more convenient to work with the logarithm of this expression
\be \label{eq:lnZtilq}
	\ln \Ztil_q = 2 \sum_{\mbf{k},c} \left( 1 +
		e^{ -\beta \left[ \epsilon_q(\mbf{k}) - \mu_q \right]
		+i \left[ \phi I_3(c) + \psi Y_8(c) \right] } \right) ,
\ee
where the factor 2 represents the sum over the degenerate spin states
$\sigma = \pm \frac{1}{2}$.

\subsubsection{Antiquarks}

The treatment of antiquarks is very similar to that of quarks. The only
difference is that antiquark states transform according to the
representation \antitriplet\ rather than \triplet\ which gives different
eigenvalues of $\hat{I}_3$ and $\hat{Y}_8$.  The result is therefore
almost identical with Eq.~(\ref{eq:lnZtilq}), \viz ,
\be \label{eq:lnZtilqbar}
	\ln \Ztil_{\bar{q}} = 2 \sum_{\mbf{k},c} \left( 1 +
		e^{ -\beta \left[ \epsilon_{\bar{q}}(\mbf{k})
		- \mu_{\bar{q}} \right]
		+i \left[ \phi I_3(c) + \psi Y_8(c) \right] } \right) ,
\ee
where $c=\bar{r}, \bar{g}, \bar{b}$, instead of $r,g,b$.

\subsubsection{Gluons}

Gluons differ from quarks in the respect that they are bosons 
(massless spin-1 particles) rather
than fermions, and again the different transformation properties under
the $SU(3)$ transformations also makes a difference relative to quarks.

The Fock space 
may be built by acting on
the vacuum state $|0\rangle$ with creation operators
$b^\dagger_{\mbf{k},\lambda,c}$ creating a single particle state
\be
	| \mbf{k}, \lambda, c \rangle =
		b^\dagger_{\mbf{k},\lambda,c} |0\rangle ,
\ee
describing a gluon with momentum $\mbf{k}$, helicity $\lambda = \pm 1$
\index{helicity!for gluons}, and color index $c=1,\ldots,8$.
The commutation relation 
\be
	\left[ b^{\phantom{\dagger}}_{\mbf{k},\lambda,c} ,
	b^\dagger_{\mbf{k}',\lambda',c'} \right] =
	\delta_{\mbf{k}\mbf{k}'}\delta_{\lambda\lambda'}\delta_{cc'}
\ee
 ensures
the right symmetry of multiparticle states 
\be
	|n\rangle=|n_{\mbf{k}_1,
	\lambda_1, c_1},\ldots, n_{\mbf{k}_l, \lambda_l, c_l},\ldots \rangle,
\ee
in which each occupation number can take on the values
$n_{\mbf{k}_i, \lambda_i, c_i}=0,\ldots,\infty$.

The number of gluons is not conserved, so gluons have vanishing
chemical potential, and since they are massless the dispersion relation
is simply $\epsilon(\mbf{k}) = |\mbf{k}|$. Taking into account these
differences we can immediately write the gluon contribution to \Ztil\ in
the form
\be
	\Ztil_g = \prod_{\mbf{k},\lambda,c}
		\sum_{n_{\mbf{k},\lambda,c}=0}^\infty
		e^{\left\{ -\beta |\mbf{k}| + i\left[ \phi I_3(c)
		+ \psi Y_8(c) \right] \right\} n_{\mbf{k},\lambda,c} } .
\ee
Performing the sum, and taking the logarithm gives
\comment{Use: $\sum_{n=0}^\infty x^n = \frac{1}{1-x}$}
\be
	\ln \Ztil_g = -2 \sum_{\mbf{k},c} \left( 1 - 
		e^{ -\beta |\mbf{k}| + i\left[ \phi I_3(c)
		+ \psi Y_8(c) \right] } \right) ,
\ee
in which the factor two is due to the degeneracy of helicity states.

\section{Evaluation of the Partition Function}
As a first step toward an expression for the color singlet partition
function $\mcal{Z}_{\singlet}$ the grand canonical partition function with
Wick rotated $SU(3)$-chemical potentials, \Ztil\ needs to be evaluated.
In the \Index{liquid drop model} approach the sums over momentum
eigenvalues are approximated by an integral
\be \label{eq:sumtointegral}
	\sum_{\mbf{k}} \longrightarrow \int_0^\infty dk\, \rho(k) ,
\ee
where $\rho(k)$ is the density of states in Eq.~(\ref{eq:DOS}), calculated
by means of the \Index{multiple reflection expansion}.

As was the case in section \ref{sec:drop} there are two limits in which
the partition function can be evaluated analytically. The zero
temperature limit is not interesting here, since the effect of imposing
color singletness disappears in the zero temperature limit.
\comment{Explain the $T\to 0$ limit intuitively}

The other limit in which \Ztil\ can be evaluated analytically is the
limit of zero quark mass. This means that for a proper inclusion of
strange quarks, part of \Ztil\ will have to be evaluated numerically.
Even for massless quarks the integral over the $SU(3)$ group, as given
by the \Index{Haar measure} Eq.~(\ref{eq:HaarSU3}), has not been solved
analytically, but a \Index{saddle-point approximation} is applicable at
high temperature and/or baryon chemical potential. The saddle-point
approximation of the color singlet partition function was first calculated
by Gorenstein, Lipskikh, Petrov, and Zinovjev \cite{Gorenstein83a}, 
including only the volume term in the density of states
(it was pointed out by Auberson \textit{et al.} that there was a mistake
in the group integration in \cite{Gorenstein83a}). Later on Elze
and Greiner \cite{Elze86b} did a similar calculation including the
curvature dependent finite size correction  to the density of states.
Mustafa \cite{Mustafa92a,Mustafa93a} has done an equivalent calculation
based on
the work of Auberson \textit{et al.} \cite{Auberson86a} and Savatier
\cite{Savatier91a}. Recently Shrauner \cite{Shrauner96a} investigated
the effect of color singletness and the use of a continuous density of
states instead of a discrete sum, on the thermodynamic treatment of
fireballs produced in heavy-ion collisions, but the author does not
include terms beyond the volume term in the density of states.

Unfortunately the different derivations mentioned above are not all in
agreement. Since the authors do not all treat the exact same situation
and do not always give intermediate results, it has not been possible
to cross check all of the results, to see where the discrepancies
originate. Because of this state of affairs I have done an independent
calculation of \Ztil\ for massless quarks, antiquarks, and gluons and
obtained results in agreement with those of Elze and Greiner
\cite{Elze86b}. An outline of this calculation will be given in the
following sections.

\subsection{Massless Quarks and Gluons}
Performing the substitution in Eq.~(\ref{eq:sumtointegral}), gives
integrals of the form
\comment{The factor 2 is spin/helicity}
\be \label{eq:logztildei}
	\ln \Ztil_i = \pm 2 \sum_c \int_0^\infty dk\, \rho_i(k) \ln \left(
		1 \pm e^{ -\beta (k -\mu_i ) + i[
		\phi I_3^{(i)}(c) + \psi Y_8^{(i)}(c) ] } \right) ,
\ee
where the upper sign is for fermions (quarks, antiquarks) and the lower
is for bosons (gluons).
Inserting the expressions for the multiple
reflection expansion density of states given in section \ref{sec:DOS},
which for massless quarks and antiquarks are
\be
	\rho(k) = \frac{Vk^2}{2\pi^2} -\frac{C}{24\pi^2} ,
\ee
and for gluons
\be
	\rho(k) = \frac{Vk^2}{2\pi^2} -\frac{C}{6\pi^2} ,
\ee
gives integrals of the form
\be
	\int_0^\infty k^n dk\, \ln\left( 1 \pm e^{-\beta k + z} \right) ,
\ee
with $n=0,2$, where $z=\beta\mu_i+ i(\phi I_3+ \psi Y_8)$ is a complex
number.

For real $z$ it was seen in section \ref{sec:partitionfunction}
that analytic results could be obtained for the combination
$\ln\Z_q+\ln\Z_{\bar{q}}$, provided $\mu_{\bar{q}}=\mu_q$ is assumed.
This is also the case here, which rests on the fact that $I_3^{(q)}(c) =
-I_3^{(\bar{q})}(\bar{c})$ and $Y_8^{(q)}(c) =-Y_8^{(\bar{q})}(\bar{c})$
so that $z_q(c) = -z_{\bar{q}}(\bar{c})$.
For gluons a similar trick is possible, since---as is seen from
Eq.~(\ref{eq:octeteigenvalues})---the eigenvalues of $\hat{I}_3$
and $\hat{Y}_8$ in the \octet\ representation come in pairs of opposite sign.

Hence for both massless quark-antiquark pairs and gluons the integrals
can, after a partial integration, be grouped to form integrals of the type
\be
	\int_0^\infty dk\, k^n \left( \frac{1}{e^{\beta k -z}\pm 1}
		+(-1)^{n+1} \frac{1}{e^{\beta k +z}\pm 1} \right)
\ee
with $n=1$ for the curvature term and $n=3$ for the volume term.

Using the results in Appendix \ref{app:integrals} for the above
integrals, and performing the color sum in Eq.~(\ref{eq:logztildei}), I
obtain the following results
\begin{eqnarray}
	\ln\Ztil_{q\bar{q},V} &\equiv& \ln \Ztil_{q,V} + \ln\Ztil_{\bar{q},V} 
	\nonumber \\
	&=&
	T^3 \left\{ \frac{7\pi^2}{60} + \frac{1}{4\pi^2} (\beta\mu_q)^4
	+ \frac{1}{2}(\beta\mu_q)^2 + \frac{1}{\pi^2}\gamma_3^2\gamma_8^2
	+\frac{1}{6\pi^2}(\gamma_3^4 + 9\gamma_8^4) \right. \nonumber \\
	&+& \left. \frac{2i}{\pi^2}
	\beta\mu_q \gamma_8(\gamma_8^2-\gamma_3^2)
	-\left[ \frac{1}{3} + \frac{1}{\pi^2}(\beta\mu_q)^2\right]
	(\gamma_3^2+3\gamma_8^2) \right\}
	\\&&\nonumber\\
	\ln\Ztil_{q\bar{q},C} &\equiv& \ln\Ztil_{q,C} + \ln\Ztil_{\bar{q},C}
	\nonumber\\
	&=& -T\left[ \frac{1}{24} + \frac{1}{8\pi^2}(\beta\mu_q)^2
	-\frac{1}{12\pi^2}(\gamma_3^2 + 3\gamma_8^2) \right]
	\\&&\nonumber\\
	\ln\Ztil_{g,V} &=&T^3\bigg[ \frac{8\pi^2}{45} -
	2(\gamma_3^2+3\gamma_8^2) + \frac{8}{3\pi}\gamma_3^3
	+\frac{18}{\pi}\gamma_8^3 \nonumber\\
	&\phantom{=}&\phantom{T^3\bigg[}
	-\frac{9}{\pi^2}\gamma_3^2\gamma_8^2
	-\frac{3}{2\pi^2}(\gamma_3^4+9\gamma_8^4) \bigg]
	\\&&\nonumber\\
	\ln\Ztil_{g,C} &=& -T \left[ \frac{4}{9} - \frac{2}{3\pi}
	(\gamma_3+3\gamma_8) + \frac{1}{\pi^2}(\gamma_3^2+3\gamma_8^2)
	\right]
\end{eqnarray}

To get the color singlet partition function for massless quarks,
antiquarks, and gluons in an MIT bag the function \Ztil\
\be
	\Ztil = e^{-\beta BV} \Ztil_g \prod_q \Ztil_{q\bar{q}} ,
\ee
has to be integrated over the $SU(3)$ group, using the \Index{Haar
measure} given in Eq.~(\ref{eq:HaarSU3}). The resulting integral is
obviously very complicated---even a numerical evaluation is not 
straightforward.

Fortunately, as has already been mentioned, there is a limit in which
an approximate evaluation of $\mcal{Z}_{\singlet}$ is applicable, \viz
, in the limit of high temperature and/or chemical potential.

\subsection{Saddle-Point Approximation
\label{sec:saddle}\index{saddle-point approximation|(}}
The idea of the saddle-point method \cite{MathewsWalker} (or the
Gaussian approximation), is that an integral of the type
\be
	\int f(x) e^{-\lambda g(x)} \, dx
\ee
can be evaluated approximately in the limit of large, positive $\lambda$ by
expanding $g(x)$ to second order around its maximum $x_0$, and replacing $f(x)$
by its lowest order expansion (typically $f(x_0)$).

The  maximum of $\Ztil(\gamma_3,\gamma_8)$ is at
$\gamma_3=\gamma_8=0$, when curvature terms are not included. The
linear terms in the gluon curvature contribution will displace the
maximum slightly away from this position, but neglecting this small
perturbation, and including only terms of second order in $\gamma_3$ and
$\gamma_8$, the pre-exponential factor is expanded to leading order, and
the integration is extended to infinity, giving
\be \label{eq:Zsingletsaddleint}
	\Z_{\singlet} = \frac{1}{6\pi^2} \Z \int_{-\infty}^\infty
		d\gamma_3 \int_{-\infty}^\infty d\gamma_8\;
		\gamma_3^2 \left( \gamma_3^2-9\gamma_8^2\right)^2
		e^{-\Lambda (\gamma_3^2+3\gamma_8^2)} ,
\ee
where \Z\ is the grand canonical partition function given by the
expressions in section \ref{sec:partitionfunction}, and
\be \label{eq:Lambda}
	\Lambda(T,V,C,\mu_q) = VT^3 \left[ 2 + \frac{\mcal{N}_q}{3}
		+\frac{1}{\pi^2}\sum_q\left(\frac{\mu_q}{T}\right)^2\right]
		+ CT \frac{12-\mcal{N}_q}{12\pi^2} ,
\ee
where $\mcal{N}_q$ is the number of massless quarks flavors.
The Gaussian integral in \Eq{eq:Zsingletsaddleint}, is easily reduced to a sum
of products of integrals that are given by the \Index{gamma function}
\cite{Abramowitz}
\be
	\Gamma(x) = \int_0^\infty t^{x-1}e^{-t} dt ,
\ee
yielding
\be \label{eq:Zsingletsaddle}
	\Z_{\singlet}(T,V,C,\mu_q) = \Z(T,V,C,\mu_q) \frac{1}{2\sqrt{3}\pi}
		\Lambda^{-4}(T,V,C,\mu_q) .
\ee
This is the color singlet partition function in the saddle-point
approximation. Each of the functions \Z\ and $\Lambda$
is a sum of a volume term and a curvature term, but the resulting
$\Z_{\singlet}$ does not have this simple form. The result
\Eq{eq:Zsingletsaddle} is in 
agreement with the result obtained by Elze and Greiner \cite{Elze86b},
but it differs from the results of Gorenstein \textit{et al.} by a
factor 9, and from the result of Mustafa \cite{Mustafa93a} by a factor
18, in the former case taking into account only the volume term in the
density of states.

It should be noted that this saddle-point approximation can be improved
by using the exact location of the maximum, instead of the approximate
$\gamma_3=\gamma_8=0$.

\begin{figure}[t]
	\includegraphics[width=\textwidth]{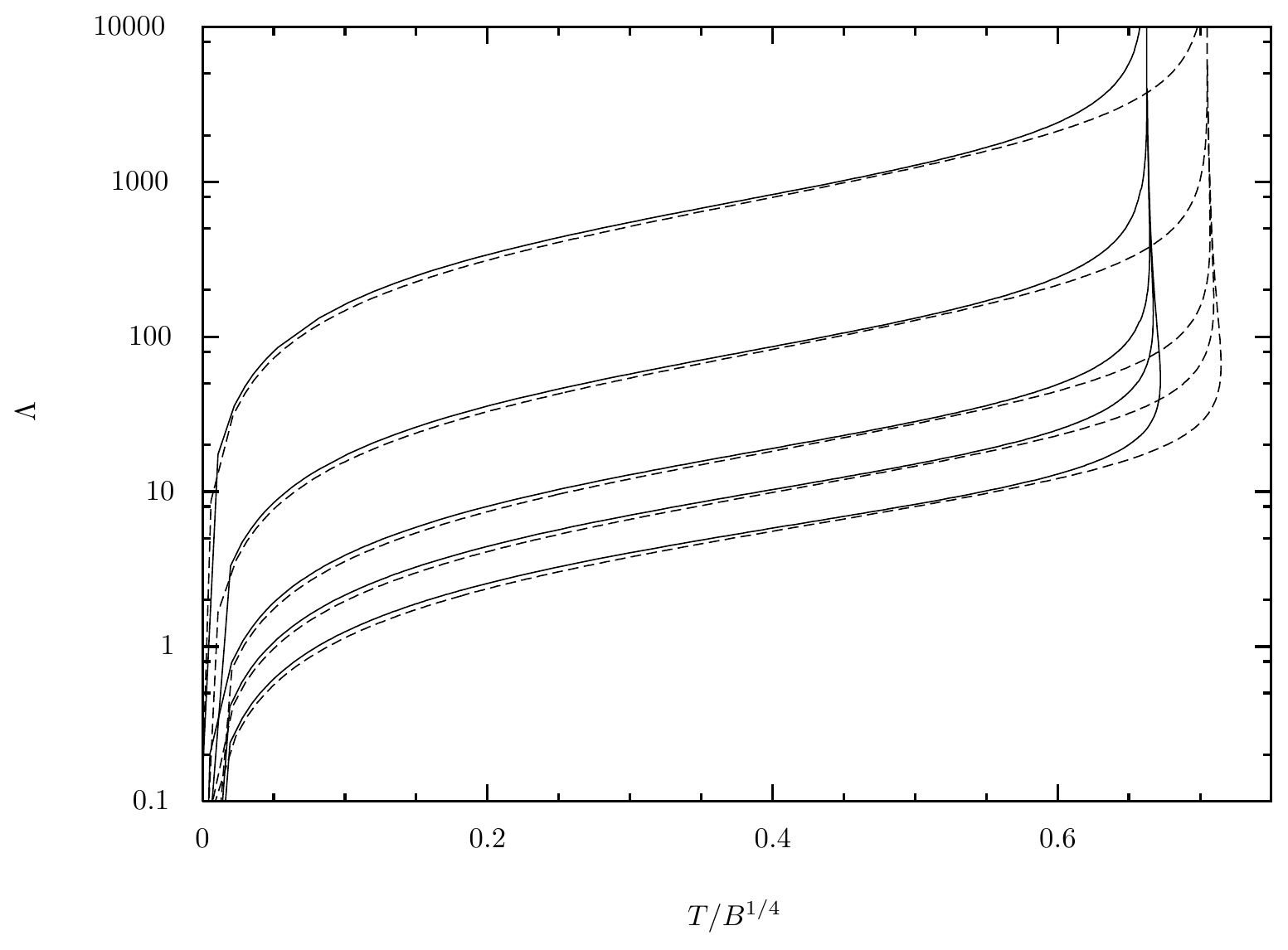}
	\caption{\label{fig:lambda}
		Plot of $\Lambda$ \textit{vs.}\ $T$ for different values of
		the baryon number. From bottom to top: $A=5$, 10, 20, 100,
		1000. Solid lines are for $\mcal{N}_q=3$, while dashed lines
		are for $\mcal{N}_q=2$.
		}
\end{figure}

When $\Lambda$ is big ($\Lambda \gg 1$) the saddle-point approximation
\index{saddle-point approximation!validity of}
is good (for quantitative results, see Chapter \ref{chap:color}). This
corresponds to either high temperature or high chemical potential and
moderate to high temperature. The system size also plays a role---entering
through $V$ and $C$ in \Eq{eq:Lambda}---so the relevant parameters
controlling the validity of the saddle-point approximation are $RT$ and
$RT(\mu/T)^{2/3}$ for a spherical system with radius $R$ and a typical
chemical potential $\mu$. For a baryonless plasma
($\mu=0$), $RT$ is the only parameter, and for a dense system (cool
strangelet) the second parameter is the dominating determinant.

A plot of $\Lambda$, calculated along the phase curve, seen in
\Fig{fig:phase_finite}, is shown in \Fig{fig:lambda} as a function of
temperature, for different values of the baryon number, and for both
2 and 3 massless flavors. Depending on the value of the bag constant and
the interpretation of the condition $\Lambda\gg 1$, the validity of the
saddle point approximation for given $A$ and $T$ can be learned from
this plot.

\index{saddle-point approximation|)}

\subsection{Numerical Evaluation\label{sec:Znum}}

\begin{figure}
	\includegraphics[width=\textwidth]{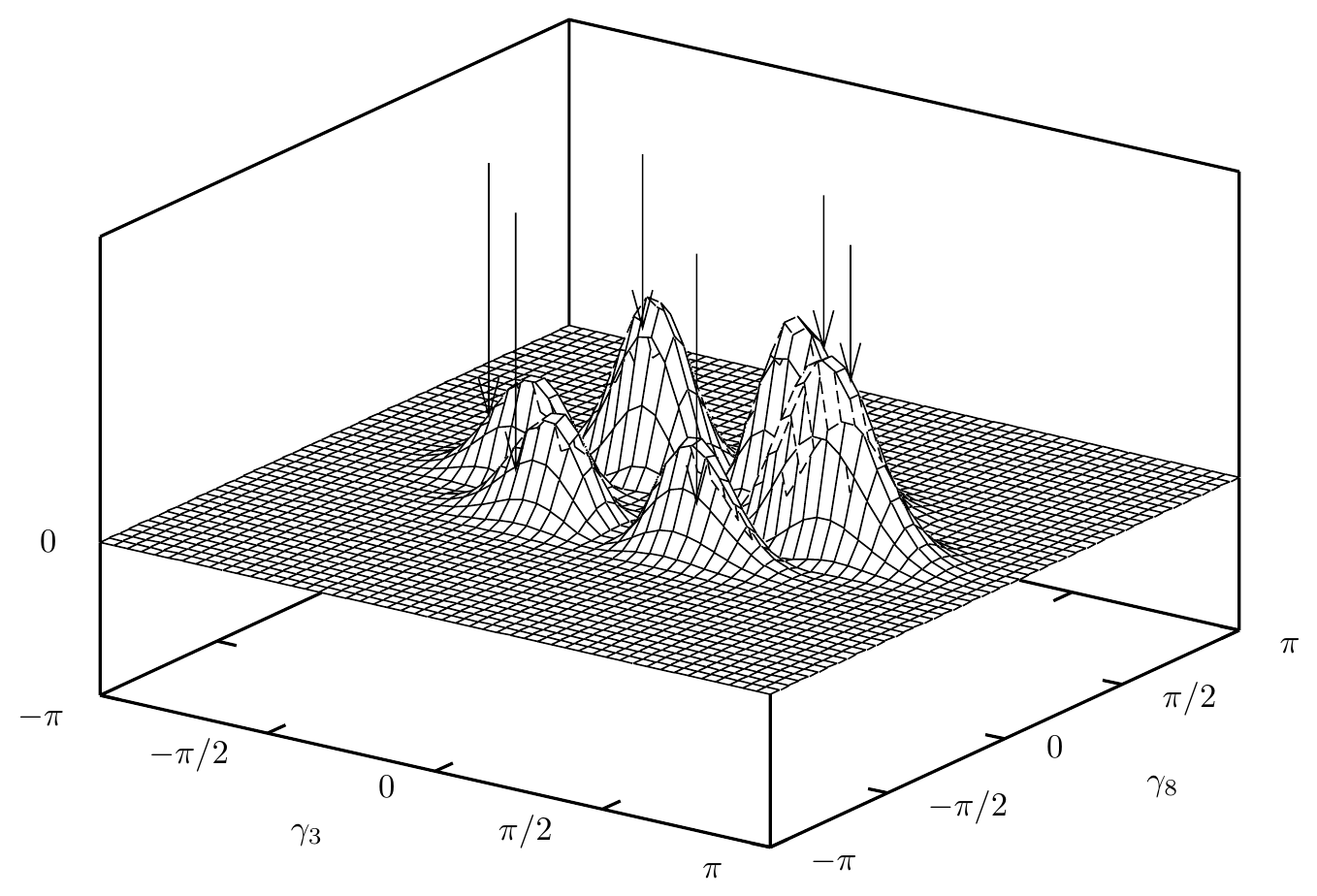}
	\caption{\label{fig:integrand}
		The integrand in the $SU(3)$ integration, for massless quarks,
		in arbitrary units.
		The parameters chosen are: $T=45$ MeV, $\mcal{N}_q=3$, $\mu=0$,
		and $R=5^{1/3}$ fm, corresponding to $\Lambda\approx 1.49$.
		Arrows indicate the
		positions of the maxima as calculated in the saddle-point
		approximation.}
\end{figure}

For low temperatures and non-zero quark masses it is necessary to
evaluate the color singlet partition function numerically. For massive
quarks this entails not only a numerical calculation of the integral
over the group parameters, but also a numerical calculation of the quark
contributions to \Ztil. Hence in the case of massive quarks there
are three nested integrations, making for very long computations.
Needless to say a lot of time has therefore been spent on optimizing the
integration routines.

One of the important things to take into account, not only in order to
speed up execution, but also to improve accuracy is how to evaluate the
group integral. It is generally best to start the integrations at, or
very near a maximum of the function to be integrated. The function
\Ztil\ has its maximum at $\gamma_3=\gamma_8=0$ (again neglecting the
small displacement), where the Haar measure
vanishes, thus generating a set of secondary maxima (see
\Fig{fig:integrand}). The position of these maxima is thus an important
prerequisite for the integration. In order to avoid an expensive search
for the location of these maxima, the values found in the saddle-point
approximation are used. There are six maxima, whose positions in the
saddle-point approximation are given by
\be
	\sqrt{\frac{\Lambda}{3}} \left(\gamma_3,\gamma_8\right) =
	\left(\pm\frac{1}{2}, \pm\frac{1}{2}\right), \left(\pm1,0\right) .
\ee
Figure~\ref{fig:integrand} shows the integrand for massless quarks with
a particular choice of parameters, and the maxima are seen to lie close
to the prediction based on the saddle-point approximation.
It is not obvious that this procedure should work as well when including
massive quarks, but in fact it does. The integrand itself changes by
many orders of magnitude, but the positions of the maxima are almost
unchanged.

\section{Summary and Discussion}

The group theoretical projection method has been explained 
and applied to the construction of a color singlet partition
function for a system of non-interacting quarks and gluons. The idea
of the projection method is to start with a partition function that takes
into account the internal $SU(3)$ symmetry only approximately via chemical
potentials corresponding to the conserved charges. Of all the
possible color states, those that are color singlets are picked out by a
projection onto the color singlet sector.

A $q\bar{q}$ state consisting of a quark in the \triplet\ representation
and an antiquark in the \antitriplet\ representation gives the possible
states $\triplet \otimes \antitriplet = \singlet \oplus \octet$, where
only the singlet state (\singlet ) corresponds to a physical meson.
Likewise a three quark state gives rise to the multiplets $\triplet
\otimes \triplet \otimes \triplet = \singlet \oplus \octet \oplus \octet
\oplus \boldsymbol{10}$, where the physical baryon emerges as the
color singlet combination, and all the other states remain unobserved.

In the case of flavor $SU(3)$ symmetry all of these combinations are
seen, giving rise to the meson octets
(\eg, $\pi^+$,$\pi^0$,$\pi^-$,$K^+$,$K^-$,$K^0$,$\bar{K^0}$,$\eta$) and
singlets (\eg, $\eta'$), and the baryon singlets (\eg, $\Lambda(1520)$),
octets (\eg, $p$, $n$, $\Sigma^+$, $\Sigma^0$, $\Sigma^-$, $\Lambda(1690)$,
$\Xi^0$,$\Xi^-$), and decuplets (\eg, $\Delta^{++}$, $\Delta^+$,
$\Delta^0$, $\Delta^-$, $\Sigma^{\star +}$, $\Sigma^{\star 0}$,
$\Sigma^{\star -}$,$\Xi^{\star 0}$, $\Xi^{\star -}$, $\Omega^-$). But
the flavor $SU(3)$ symmetry is not exact and members in the same
\index{SU(3)@$SU(3)$!of flavor}
multiplet are not degenerate and mixing between multiplets 
also take place (\eg, $\eta$ and $\eta'$ both
contain singlet and octet components).

The non-existence of non-singlets has an analogue in nuclear physics where
the deuteron, which is the isospin singlet combination of the
nucleon-nucleon system exists, but the three states in the isospin
triplet representation containing the di-neutron and the di-proton do
not exist as bound states. The nucleons are members of the fundamental $SU(2)$
duplet generating the representaion $\boldsymbol{2}$. Since
$\boldsymbol{2}\otimes\boldsymbol{2}=\singlet \oplus \triplet$, there
are in principle both an isospin singlet and an isospin triplet. The
nucleon-nucleon interaction has an isospin dependence which gives the
isospin triplet a higher energy than the isospin singlet. The deuteron
is only bound by 2.23~MeV, and the difference in energy between the two
multiplets is enough to make the isospin 1 triplet of states unbound.

In the case of quarks the difference is even more dramatic---the energy
of a quark antiquark pair increases linearly with their separation,
making it impossible to break up a meson into a quark and an antiquark.
The energy of an isolated quark is literally infinite. Once the energy
of the chromoelectric field between two quarks becomes large
enough, as the separation grows, new particles will be created. So it is
believed that all non-singlet states of QCD have infinite energy, and
that therefore they are not observed. Whether the energy is indeed
infinite or merely very large does not really matter here. As long as
these states are not part of the observable spectrum, they should not be
included. The effect of the color singlet projection is exactly to weed
out these states in the partition function.

The group theoretical projection method was applied to the simple case
of a $U(1)$ symmetry, and the well-known connection between the grand
canonical and the canonical partition functions was obtained. The
projection method can also be applied to other symmetries, for example
the calculation of the partition function (and from that the density of
states) for a compound nucleus with fixed isospin.

The color singlet partition function (re-)derived in this chapter is a
complicated integral over the $SU(3)$ group manifold, but a saddle-point
approximation can be evaluated analytically in the case of massless
quarks. For massive quarks there are no other alternatives than a
numerical calculation.

\clearemptydoublepage
\chapter{Color Singlet Strangelets\label{chap:color}}
In this chapter I use the results of Chapter \ref{chap:projection} to
study the effect of the color singlet constraint on the mass of
strangelets with non-zero entropy.

Experiments searching for strangelets are presently taking place
both at the
\comment{Flere referencer?}
Brookhaven AGS \cite{Beavis95a} and at the CERN SPS \cite{Appelquist96a},
but so far only upper limits on strangelet production have resulted from
these experiments. As a step toward realistic calculations of strangelet
yields in heavy-ion collisions it is 
\index{strangelet!creation in heavy-ion coll.}
important to investigate the effect of a hot environment on strangelet
properties, most notably strangelet masses.
In this chapter I assume that the quark flavors are in equilibrium, so
that they are described by a common chemical potential. For some
applications it would be more natural to fix the strangeness fraction,
as was done in Chapter \ref{chap:equilibrium}, and use this to relate the
chemical potential of the light quarks and that of the strange quark.

\section{Massless Quarks}
Massless quarks are so much easier to deal with than their massive
counterparts, that they are a natural first topic of investigation, if
for no other reason then to test the results of calculations done for
massive quarks. But there are plenty of reasons to study massless quarks
in their own right, first because $u$ and $d$ quarks have very low
masses, second because massive quarks are not always present, \eg , in
the nucleation of quark gluon plasma droplets in 
ultrarelativistic heavy-ion collisions, or nucleation of quark matter
droplets in the centers of neutron stars. 

\subsection{Saddle-Point Approximation}
\index{saddle-point approximation|(}
\index{fixed momentum constraint|(}
In addition to the constraint of color singletness, there is another,
though less important constraint, \viz , that the system as a whole
should have a fixed momentum \cite{Elze86b,Auberson86a}, which can be
included in a way similar to the \Index{color singlet constraint}. In the
saddle-point approximation the two constraints `decouple' \cite{Elze86b},
each giving
rise to a (temperature and chemical potential dependent) correction factor 
in the resulting partition function
\be \label{eq:Zsingletp0}
	\Z_{\singlet,p=0} = \Pi_{\singlet} \Pi_{p=0} \Z
\ee
where
\be
	\Pi_{\singlet} = \frac{1}{2\sqrt{3}\pi} \Lambda^{-4}(T,V,C,\mu_q) ,
\ee
is the correction due to the color singlet constraint, already
calculated, with $\Lambda$ given by \Eq{eq:Lambda}, and
\begin{eqnarray} \label{eq:Pip0}
	\Pi_{p=0}^{-2/3}  &=& \pi VT^3\left\{\left[
         \frac{7\mcal{N}_q}{30} + \sum_q\left(\frac{\mu_q}{\pi T} \right)^2
         + \frac{1}{2} \sum_q\left( \frac{\mu_q}{\pi T} \right)^4
         \right] + \frac{16}{45} \right\} \nonumber \\
         &-& {CT\over\pi}\left\{ \frac{1}{72}\left[ \mcal{N}_q+3
         \sum_q\left(
         \frac{\mu_q}{ \pi T}\right)^2 \right] +\frac{4}{27} \right\},
\end{eqnarray}
is the factor arising from the fixed momentum constraint, as given by
Elze and Greiner \cite{Elze86b}, here taken at zero total momentum.
$\mcal{N}_q$ is again the number of massless quark flavors, and terms
proportional to $\mcal{N}_q$ or containing a sum over quark flavors $q$
are due to quark-antiquark pairs, while the remaining terms are gluon
contributions. In the following I will assume a common quark chemical
potential $\mu$ for $\mcal{N}_q=3$ massless quark flavors.
As is the case for $\Lambda$, so is the right hand side
of \Eq{eq:Pip0} seen to be a sum of volume and curvature terms.

The partition function \Z\ in  \Eq{eq:Zsingletp0} is the unprojected
grand canonical partition function, given by the expressions in section
\ref{sec:partitionfunction}.
\index{fixed momentum constraint|)}

To find the the mass of a spherical strangelet at fixed baryon
number $A$ and fixed entropy per baryon $\mcal{S}/A$, the equilibrium
condition
\be \label{eq:coneq}
		\left( \frac{\partial\Omega}{\partial V} \right)_{T,\mu} =0 ,
\ee
has to be solved along with the condition of fixed baryon number
\be \label{eq:connum}
	-\frac{1}{3} \left( \frac{\partial\Omega}{\partial\mu}
		\right)_{T,V} = A ,
\ee
and the condition of fixed entropy
\be \label{eq:conentro}
	-\frac{1}{A} \left( \frac{\partial\Omega}{\partial T}
		\right)_{V,\mu} = \frac{\mcal{S}}{A} .
\ee
In these expressions the thermodynamic potential $\Omega$ can be either
the unconstrained potential, as in Chapter \ref{chap:temperature}, or it
can include either, or both, of the color singlet and fixed (zero) momentum
constraints. In the case where both constraints are included $\Omega$ is
given by
\be
	\Omega_{\singlet,p=0} = -T\ln\Z_{\singlet,p=0} ,
\ee
and leaving out one of the constraints corresponds to setting one of
$\Pi_{\singlet}$ or $\Pi_{p=0}$ equal to 1. For the spherical system
assumed here, the curvature is no longer an independent variable, but is
given by $C\equiv 8\pi(3/4\pi)^{1/3}V^{1/3}$.

\begin{figure}
	\centering
	\includegraphics[width=\textwidth]{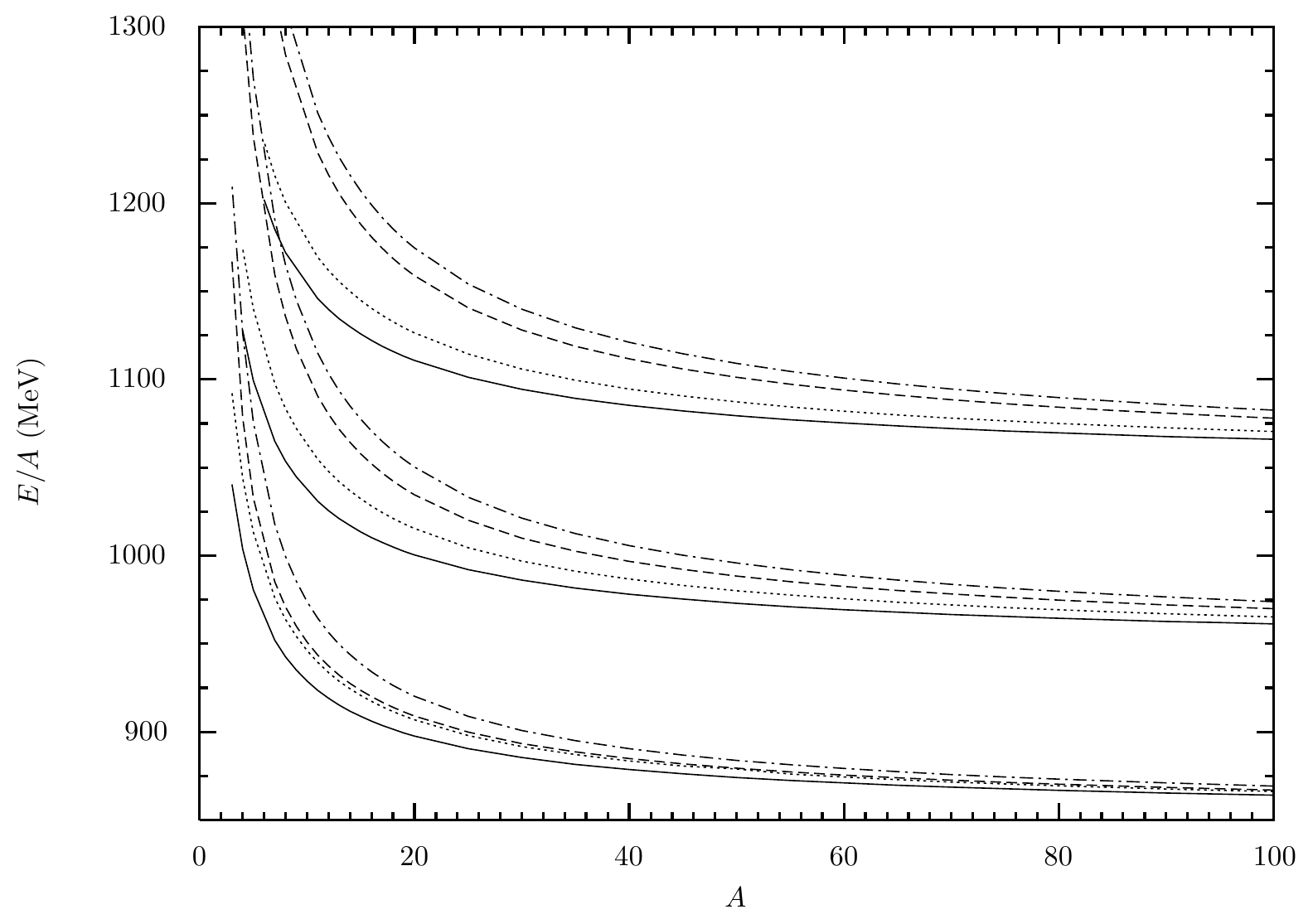}
	\caption{\label{fig:SAm0plot}
		The energy per baryon at constant entropy per baryon
 		for $\mcal{N}_q=3$ and $B^{1/4}=145$. The 3 sets of curves are
		for $\mcal{S}/A=2$, 5, and 7.
		The solid lines are unconstrained calculations, 
		dotted lines are with the fixed momentum constraint only, 
		dashed lines with the color singlet constraint only, and
		dotted-dashed lines are with both constraints.
		}
\end{figure}

The result of solving these equations, and calculating the energy per
baryon for a range in baryon number, is shown in \Fig{fig:SAm0plot},
where dashed curves includes the color singlet constraint, but not the
fixed momentum constraint, dotted curves include the fixed momentum
constraint, but not the color singlet constraint, and dotted-dashed
curves include both constraints. The solid curves are results of
unconstrained calculations, like the results shown in
\Fig{fig:SAplot}, but for massless quarks.

The particular choice of bag constant, $B^{1/4} = 145$ MeV, corresponds
to the most optimistic choice, but results for higher values of $B$ can
easily be deduced from \Fig{fig:SAm0plot}, since, as is always the case
for massless quarks, the energy scales in proportion to $B^{1/4}$.

The effect of both the color singlet constraint and the zero momentum
constraint is to increase the energy. The color singlet constraint is
seen to be the most significant of the two constraints. The effect of
the color singlet constraint grows with increasing entropy, and
decreasing baryon number.  In the bulk limit the effect of both
constraints disappear, which in the case of the color singlet constraint
is due to the fact that the grand canonical ensemble and the
canonical ensemble (with respect to color) become equivalent. 
\comment{Better explanation.}
At zero temperature the effect of the color singlet constraint also
disappears, since it is always possible to choose a combination of
colors using the lowest lying energy levels which gives a color singlet.

At very low baryon number ($A\le 0$--7, depending on $\mcal{S}/A$)
there are no solutions to Eqs.~(\ref{eq:coneq})--(\ref{eq:conentro}).
This is due to the breakdown of the multiple reflection expansion at
\index{multiple reflection expansion!breakdown of}
very low values of the product of particle momentum and bag radius $kR$.
At $kR$ below the ground state value ($kR \approx 2.04$ for massless
quarks; see Table \ref{table:shell}) the density of states becomes
spuriously negative. For low baryon number strangelets this part of the
spectrum is weighted relatively more, than at larger baryon number. This
is a signal of the breakdown of the multiple reflection expansion, and
hence the liquid drop model. This is to be expected since the multiple
reflection expansion is an expansion in powers of $(kR)^{-1}$ (see
Appendix~\ref{chap:mre}).

At low temperature, as already discussed in section \ref{sec:saddle},
the saddle-point approximation breaks down, so it is necessary to use
numerical methods.
\index{saddle-point approximation|)}

%

\subsection{Numerical Calculation\label{sec:numcalc}}
In section \ref{sec:Znum}, I briefly described the numerical
evaluation of the color singlet partition function. Apart from the
partition function (or equivalently the thermodynamic potential
$\Omega=-T\ln \Z$) the derivatives with respect to $T$, $\mu$, and $V$ are
also needed in order to calculate the entropy, baryon number, and pressure,
required for solving Eqs.~(\ref{eq:coneq})--(\ref{eq:conentro}).

For this the partial derivatives of \Ztil\ are needed, so that with
\be
	\Z(T,V,\mu) = \int_{SU(3)} d\mu (\gamma_3,\gamma_8)
		 \Ztil (T,V,\mu,\gamma_3,\gamma_8) ,
\ee
the entropy is given by
\bea
	\mcal{S} &=& -\left( \frac{\partial\Omega}{\partial T} \right)_{V,\mu}
	\nonumber \\
	&=& \ln \Z + T\left( \frac{\partial\ln\Z}{\partial T} \right)_{V,\mu}
	\\
	&=& \ln \Z + \frac{T}{\Z} \int_{SU(3)} d\mu
	\Ztil \left(\frac{\partial\ln\Ztil}{\partial T} \right)_{V,\mu} .
	\nonumber
\eea
and all other derivatives are evaluated in the same fashion.


\begin{figure}
	\includegraphics[width=\textwidth]{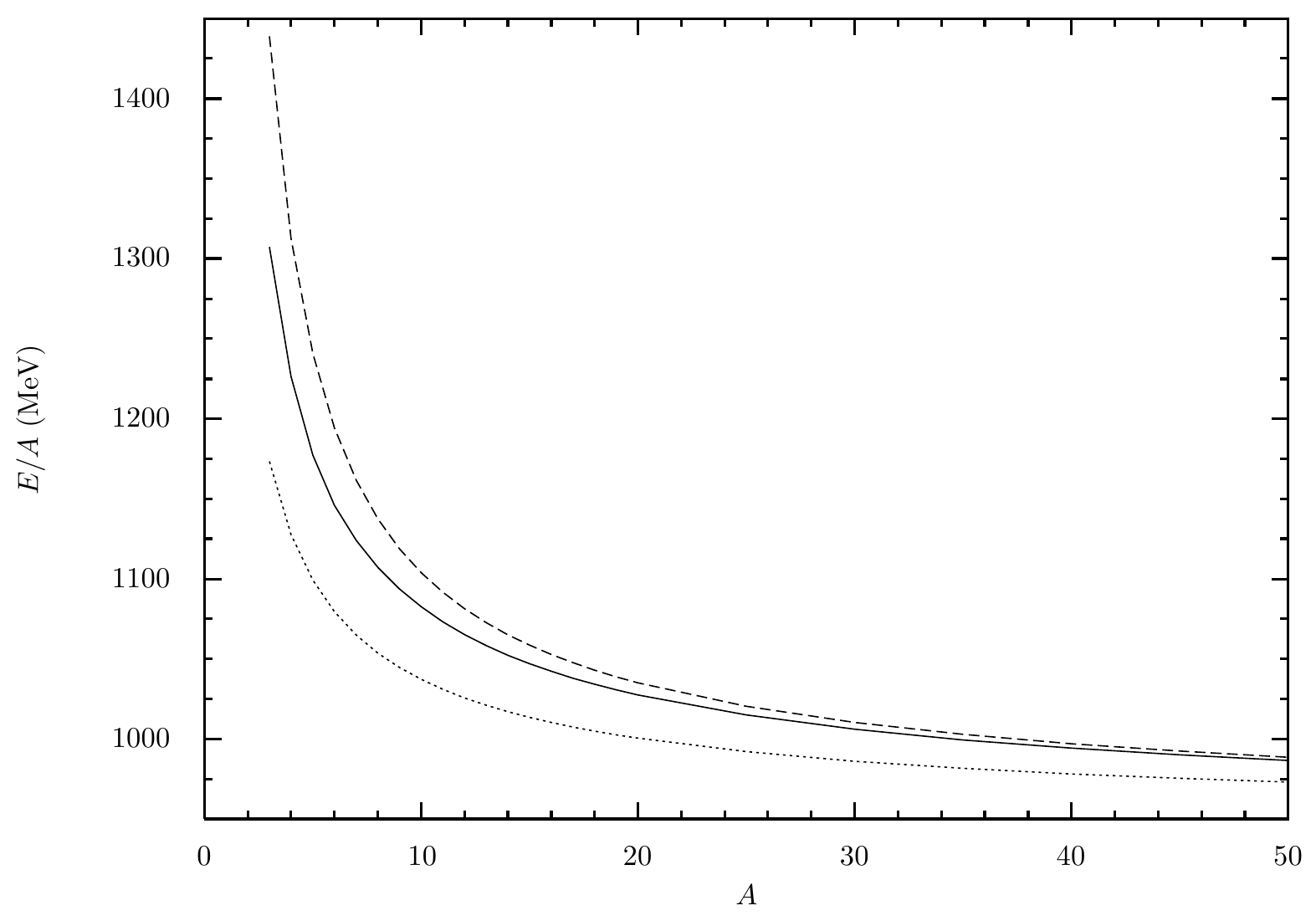}
	\caption{\label{fig:saddle_valid}
		The saddle-point approximation (dashed line) compared to an exact
		numerical calculation (solid line), including only the color
		singlet constraint. An unconstrained calculation (dotted line)
		is shown for comparison. The plot is for constant entropy per
		baryon $\mcal{S}/A = 5$, and $B^{1/4} = 145$ MeV.}
\end{figure}

\index{saddle-point approximation!validity of|(}
\label{sub:saddle_validity}
The result of such numerical integrations is compared to the saddle-point
approximation in \Fig{fig:saddle_valid}, where only the color singlet
constraint has been taken into account.
The agreement is seen to be
rather good, although not as good as reported in Ref.~\cite{Jensen96a},
where there was a mistake in the numerical evaluation\footnote{
\label{foot:error}The odd
terms in $\gamma_3$ and $\gamma_8$ in $\ln\Ztil$, which appear in the
functions $S^{-}_1$ and $S^-_3$ (see Appendix~\ref{app:integrals}) when
the chemical potential has a non-zero imaginary part, were not included
in the numerical calculation in Ref.~\cite{Jensen96a}, which resulted in a 
better agreement with the saddle-point approximation.}.

\begin{figure}
	\includegraphics[width=\textwidth]{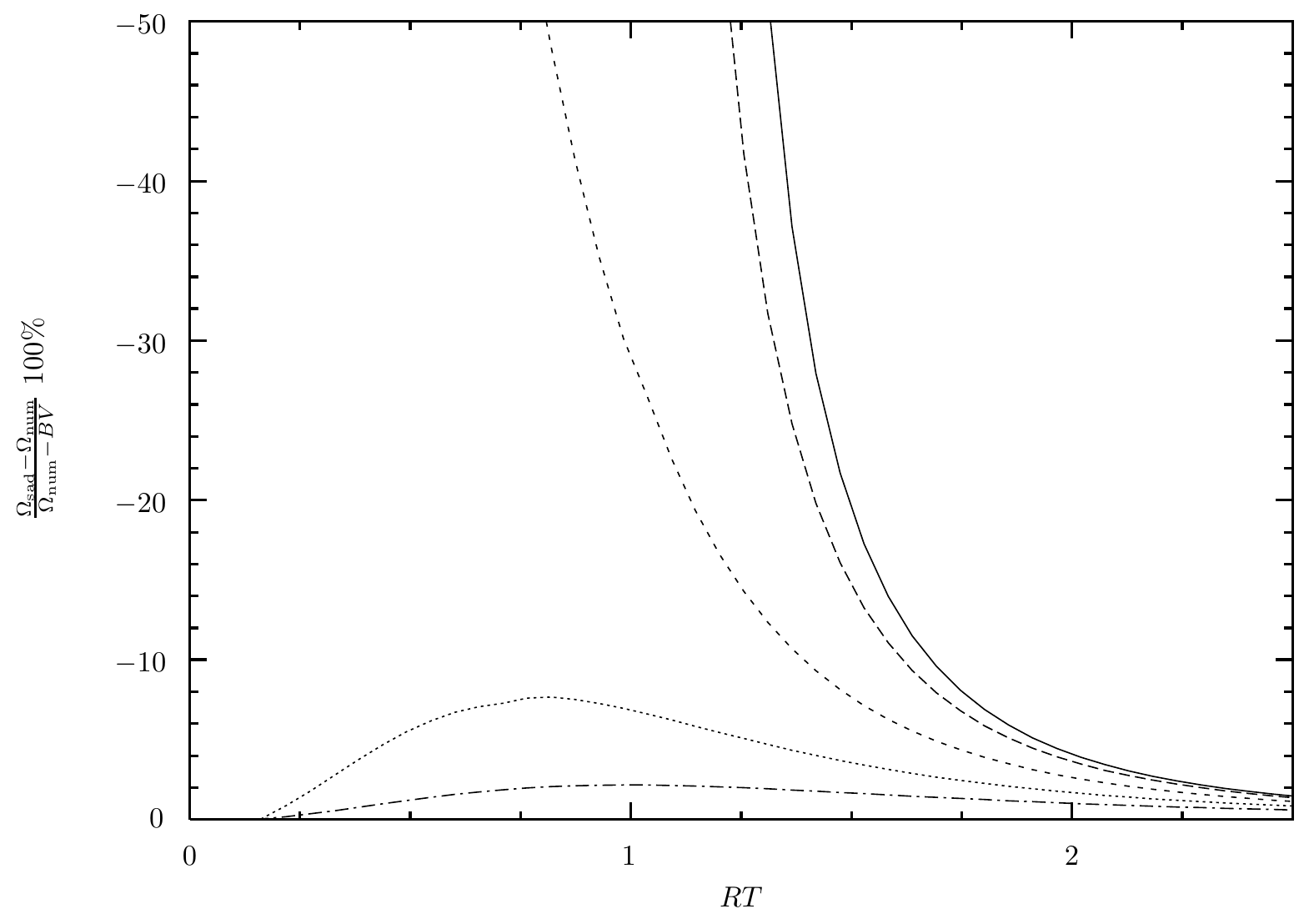}
	\caption{\label{fig:saddle} The deviation (in \%) of the thermodynamic
	potential, as calculated in the saddle-point approximation, from the
	exact numerical result, as a function of $RT$.  The bag contribution
	$BV$ has been subtracted in both potentials. The solid curve is for
	$\mu=0$, the long-dashed curve for $\mu=100$ MeV, the short-dashed
	curve for $\mu=200$ MeV, the dotted curve for $\mu=300$ MeV, and the
	dotted-dashed curve for $\mu=400$ MeV.}
\end{figure}

The validity of the saddle-point approximation can also be seen in
\Fig{fig:saddle}, where the relative error of the saddle-point
approximation is plotted as a function of $RT$, where $R$ is the radius
of the system. Again only the color singlet constraint has been
included. At fixed $R$ and $\mu$ the thermodynamic potential has
been calculated for a range in $T$ to produce this plot. The bag
contribution $BV$ to $\Omega$ has been subtracted, since only the
particle contributions to $\Omega$ are affected by the color singlet
constraint.

It was stated in section \ref{sec:saddle} that the saddle-point
approximation is believed to be good when $RT\gg 1$. Here this statement
can now be made more quantitative. At zero chemical potential the
saddle-point approximation is good to within 4\% for $RT>2$  and a
higher chemical
potential improves the approximation. A typical chemical potential for
strangelet calculations is around 250~MeV, so as long as $RT$ is
sufficiently large, say greater than 1.5, the saddle-point approximation
is expected to be reasonably good.

This should be compared with the findings of Elze, Greiner, and Rafelski
\cite{Elze83a,Elze84a,Elze86b} who state that the saddle
point-approximation is good to within 30\% at $RT=1$ and $\mu=0$,
decreasing toward a few percent at $RT=2$. Their calculation was done
without the inclusion of curvature terms, so a better agreement is to be
expected.

If the add-on in energy due to the color singlet constraint is 10\%, a
5\% error in the  total energy is more like a 50\% error in the energy
difference, so one should be careful not to accept too large deviations
when looking at differences rather than absolute quantities.

\index{saddle-point approximation!validity of|)}

\section{Massive Quarks}

\begin{figure}
	\includegraphics[width=\textwidth]{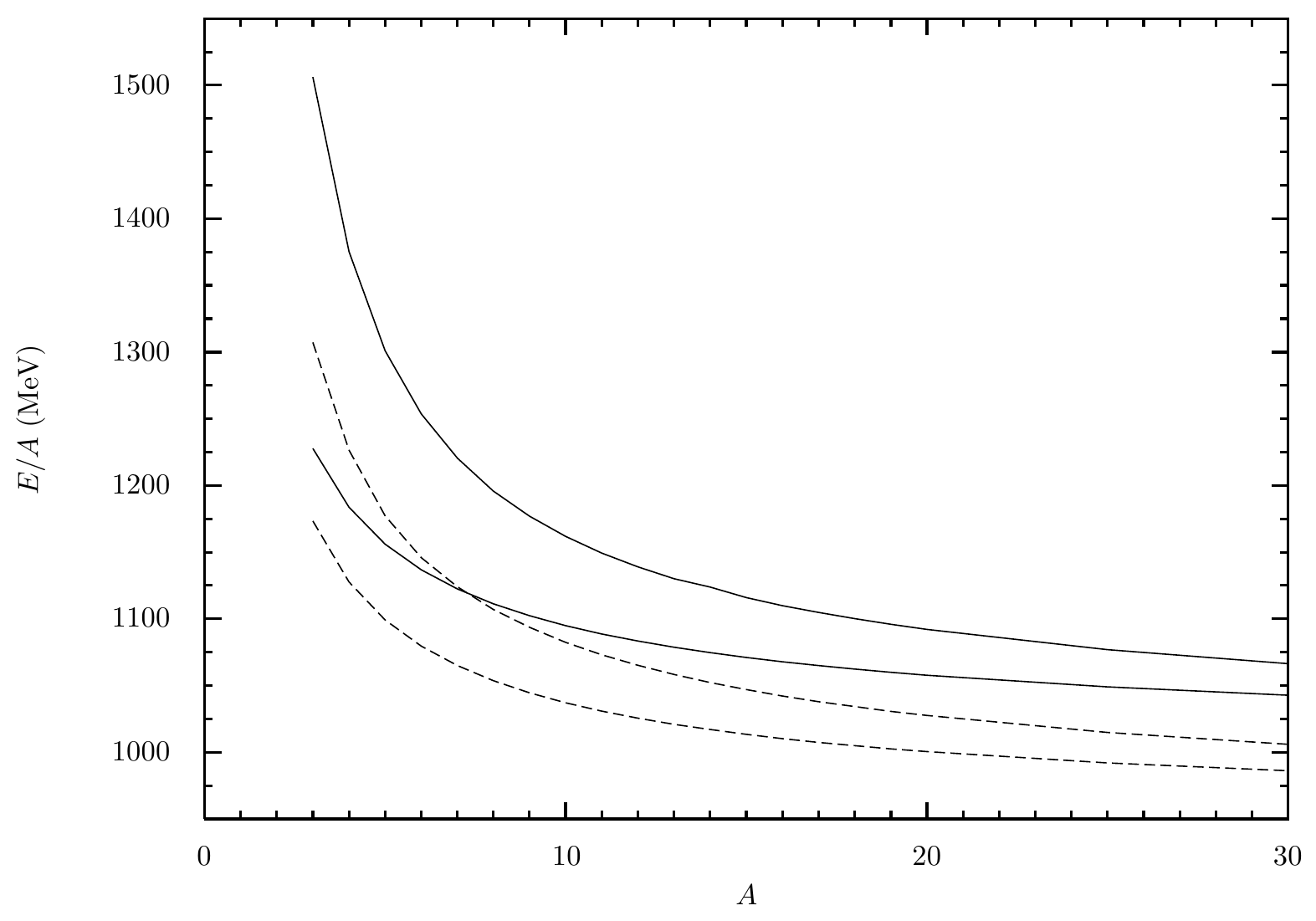}
	\caption{\label{fig:massive}
		The energy per baryon as a function of baryon number, for 
		$\mcal{S}/A = 5$, and $B^{1/4} = 145$ MeV. The solid curves are
		for a strange-quark mass $m_s=150$ MeV, while the dashed curves
		are for a massless strange quark. Upper curves in each set are
		with the color singlet constraint, while lower curves do not
		include it.}
\end{figure}

The inclusion of non-zero quark masses in the calculation of strangelet
masses at non-zero entropy entails a numerical evaluation of the
Fermi-Dirac integrals. The $u$- and $d$-quark masses are sufficiently
small that they may be taken to be zero. The $s$-quark mass, however, is
expected to lie in the range 100--300 MeV, and so a proper treatment
should take a non-zero $s$-quark mass into account. Plots of the energy
per baryon as a function of baryon number for different values of the
entropy per baryon were shown in \Fig{fig:SAplot}, for a strange quark
mass of 150 MeV.

The numerical evaluation of the energy when the color singlet constraint
is taken into account, is in itself a very time consuming. The addition
of a non-zero strange quark mass makes it even more so. Therefore I have
only performed very few such calculations, for a limited set of
different parameters. A plot of the energy per baryon for a strange
quark mass $m_s=150$ MeV, including the color singlet constraint is
shown in \Fig{fig:massive} along with results for the unconstrained
case, and a calculation for $m_s=0$.

The effect of the non-zero strange-quark mass is of course to increase
the energy per baryon. This effect is in itself smaller than the mass of
the strange quark, which is also seen in \Fig{fig:ld_and_shell}.

Except at very low $A$ the effect of the color singlet constraint seems
to be relatively independent of the strange quark mass. The calculation
for non-zero strange quark mass may not be totally reliable at very low
$A$, though, so a refined calculation could give a more uniform difference.

\section{Conclusion and Discussion}
In this chapter it has been shown that the color singlet constraint, and
to a lesser extent the fixed momentum constraint has the effect of
increasing strangelet masses, as calculated at fixed entropy per baryon.

The fixed momentum constraint and the color singlet constraint can both
be derived in a saddle-point approximation as shown by Elze and Greiner
\cite{Elze86b}.  The saddle-point approximation is generally good at
high temperature and chemical potential, but is very sensitive to the
size of the system.

The saddle-point approximation is a useful guide for numerical
calculations, which have also been performed, although not with the fixed
momentum constraint included. 

Finally a calculation for a finite strange-quark mass has been
presented. At present such calculations are rather time consuming, and
further improvement is needed before a systematic study can be
performed.

The effect of color singletness on strangelet masses is important in the
context of strangelet formation in heavy-ion colisions,
\index{strangelet!creation in heavy-ion coll.}
and the results of this chapter could be incorporated in a production
model, \eg\ the thermal model by Braun-Munzinger \textit{et al.}
\cite{Braun-Munzinger95a,Braun-Munzinger96a}.

Another obvious application of the results in this chapter is to study
the effect of the color singlet constraint on the phase equillibrium
between hadroninc matter and quark matter, as was done in
Chapter~\ref{chap:equilibrium} without any constraints.

\clearemptydoublepage
\chapter{Color Singlet Suppressed Nucleation\label{chap:suppression}}

In this chapter the effect of the color singlet constraint on the
nucleation rate of quark matter droplets is investigated. The result is
a lowering of the nucleation rate relative to the unconstrained case.
This has consequences for \Index{quark-gluon plasma} formation in heavy-ion
collisions, and quark matter droplet formation in neutron stars. For the
reverse transition, \ie , the quark-hadron phase transition a similar
effect may play a role, but it is unclear at present how to properly
treat this case.

\section{Nucleation Rate}
\index{nucleation rate|(}
The nucleation is assumed to be thermally induced, and the transition to
be of first order (it is by construction in the bag model), which means
that homogenous nucleation theory \cite{LandauStat1} is applicable.
A common estimate for the nucleation rate $\mcal{R}$, within this
framework, is given by
\be \label{eq:rate}
	\mcal{R} = T^4 e^{-\Delta F/T},
\ee
where $\Delta F$ is the free energy barrier, \ie, the amount of work it
takes to produce a droplet of the critical radius.  The prefactor $T^4$
is merely a dimensional estimate. A more refined treatment would include
a more realistic choice of prefactor, but in the present case of
massless quarks there is no surface term in the free energy, and a
derivation of the prefactor when curvature effects rather than surface
effects dominate has not been made. In any case the exponential is the
dominating factor in the expression for the nucleation rate, so a
modification of the prefactor is only expected to have marginal effects.
\index{nucleation rate|)}

\subsection{Free Energy Barrier}
\index{free energy!barrier|(}
For a droplet in chemical equilibrium with the surrounding hadron phase
(\ie, with a composition determined by the chemical potential of the
hadron phase) the free energy barrier is
\be
	\Delta F = \Omega + P_{\text{hadron}}V + \sigma_{\text{hadron}}S,
\ee
where $\Omega$ is the thermodynamic potential of the quark phase, 
$P_{\text{hadron}}$ is the hadron pressure, $ \sigma_{\text{hadron}}$ is
the surface tension due to hadrons, and $V$ and $S$ are the volume and
surface area of the formed droplet. We will make the seemingly drastic
approximation that $P_{\text{hadron}}=\sigma_{\text{hadron}}=0$. Note,
however, that a non-zero pressure or surface tension of the hadronic
phase will increase $\Delta F$ and thereby lower the nucleation rate.
Since it will be demonstrated that the effect of the color singlet
constraint is to lower the nucleation rate considerably, the inclusion
of the hadronic terms would only strengthen this effect.

An estimate of the effect of the hadron pressure is the pressure of
a massless pion gas, with three degrees of freedom,  at $\mu=0$ which is 
$P_{\pi} = \pi^2T^4/30$. The quark-gluon plasma has a number of degrees
of freedom given by \Eq{eq:gstar}, which for $\mcal{N}_q=2$ amounts to
$g_\star = 37$ giving a pressure $g_\star T^4\pi^2/90$, which is more
than 10 times the pion pressure at the same temperature. The pion
pressure would add to the bag constant to give an effective bag constant
$B[1+\pi^2/30(T/B^{1/4})^4]$ instead of $B$. At the critical temperature
$T_c\approx 0.7B^{1/4}$ given by \Eq{eq:critT} this results in a change
in $B^{1/4}$ which is less than 2\%.

\begin{figure}
	\centering
	\includegraphics[width=\textwidth]{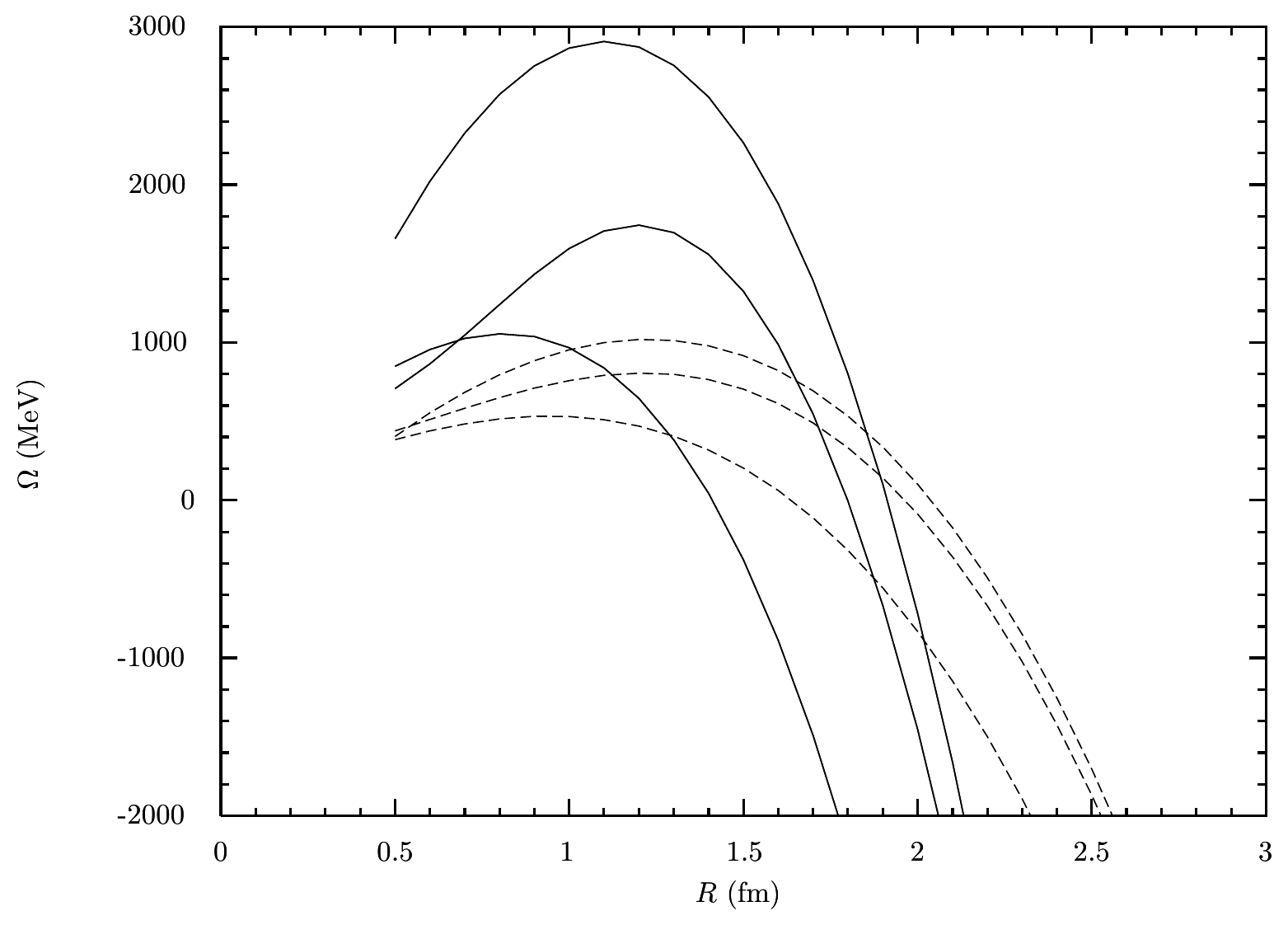}
	\caption{\label{fig:barrier}
		The thermodynamic potential $\Omega$ as a function of droplet
		radius $R$, for $B^{1/4}=200$ MeV. The solid curves are for
		$T=170$ MeV and $\mu=0$, whereas the dashed curves are for
		$T=70$ MeV and $\mu=400$ MeV. The lower curve in each set is the
		unconstrained grand potential, the middle curve is the exact
		numerical calculation including the color singlet constraint,
		and the upper curve is the saddle-point approximation.}
\end{figure}

\begin{figure}
	\centering
	\includegraphics[width=\textwidth]{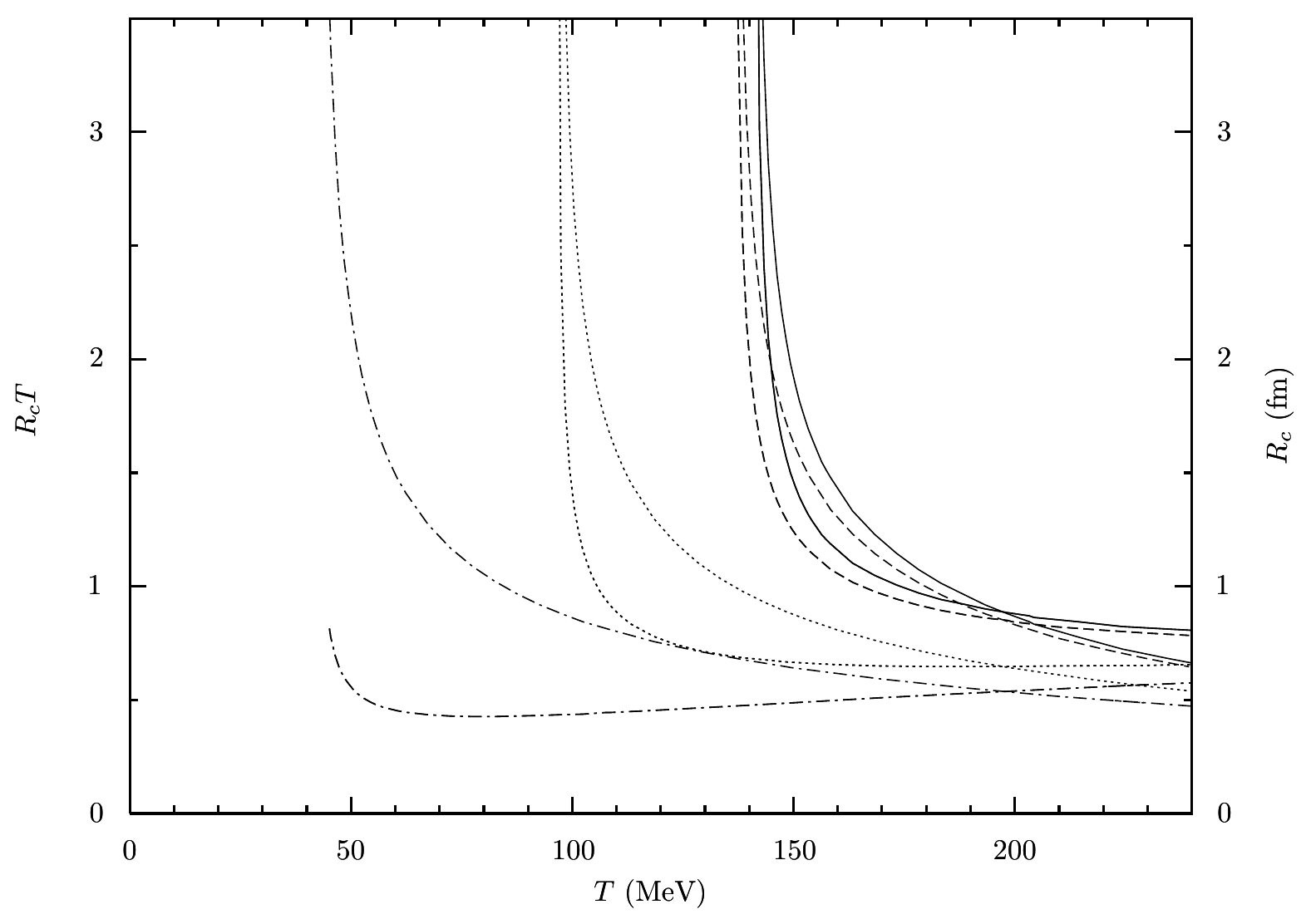}
	\caption{\label{fig:RTplot}
		The critical radius $R_c$ and the product $R_cT$ as a function
		of temperature, including the color singlet constraint.
		Upper (mostly) curves are $R_c$ while (mostly)
		lower curves are $R_cT$. The solid curves are for $\mu=0$, the
		dashed curves for $\mu=100$ MeV, the dotted curves for $\mu=300$
		MeV, and the dotted-dashed curves for $\mu = 400$ MeV, all for
		$B^{1/4}=200$ MeV.  }
\end{figure}

\index{saddle-point approximation!validity of|(}
The free energy barrier $\Delta F = \Omega$, for  a spherical droplet
with two massless quark flavors ($\mcal{N}_q=2$), having a common
chemical potential $\mu$ is shown as a function of
radius $R$ in \Fig{fig:barrier} for $T=170$ MeV, $\mu=0$ corresponding
to about 30 MeV superheating, and for $T=70$ MeV, $\mu=400$ MeV
corresponding to a superheating of $\sim$25 MeV.
It is seen that the inclusion of the color singlet
constraint results in an increase of the free energy barrier. The
critical radii for $\mu=0$ correspond to values of $RT\approx 1$, which
means that
the saddle-point approximation cannot be expected to hold (\textit{cf.}\
the discussion in section~\ref{sec:numcalc}).
It is clear from \Fig{fig:barrier} that
the saddle-point approximation is indeed better for $\mu=400$ MeV than
for $\mu=0$ MeV even though $RT$ is smaller in the latter case. 
As mentioned in Chapter~\ref{chap:projection} this is because another
parameter, \viz, $RT(\mu/T)^{2/3}$, is also important as a determinant
of the validity of the saddle-point approximation. But at
higher temperature  the critical radius gets
smaller (see \Fig{fig:RTplot}) causing the saddle-point approximation
to become less accurate.
So it seems that even for large chemical potential it is not safe to use
the saddle-point approximation.

This is in contrast to what was claimed in Ref.~\cite{Madsen96a}, where
the saddle point approximation was used exclusively. A few comments on
this seem to be in order. When the calculations in Ref.~\cite{Madsen96a}
were done the preliminary numerical calculations indicated that the
saddle point approximation was valid for the high temperatures and
chemical potentials that were used. As mentioned previously (in the
footnote on page~\pageref{foot:error}) there was an error in these
calculations, which caused a much better agreement with the saddle-point
approximation. The saddle-point approximation \emph{is} valid at high
temperature and chemical potential \emph{provided} that the system size
is not too small. It is exactly the very small critical radii that bring
about the fall of the saddle-point approximation in the present context,
despite the favorable values of $T$ and $\mu$. This is clearly seen in
\Fig{fig:RTplot}, where $RT$ is below 1 except very near the critical
temperature. In \Fig{fig:RTplot} and in the rest of this chapter only
results of the numerical calculations are presented.
\index{saddle-point approximation!validity of|)}

In addition to the color singlet constraint the fixed momentum
constraint was also taken into account in Ref.~\cite{Madsen96a}, but at
present this has not been implemented numerically, and so it is left out
here. But were it to be included it would result in a further  increase
in the free energy barrier. As pointed out in Chapter~\ref{chap:color}
the color singlet constraint is the more decisive of the two, as can
also be seen in Ref.~\cite{Madsen96a}, which means that the major effect
is accounted for.

\begin{figure}
	\centering
	\includegraphics[width=\textwidth]{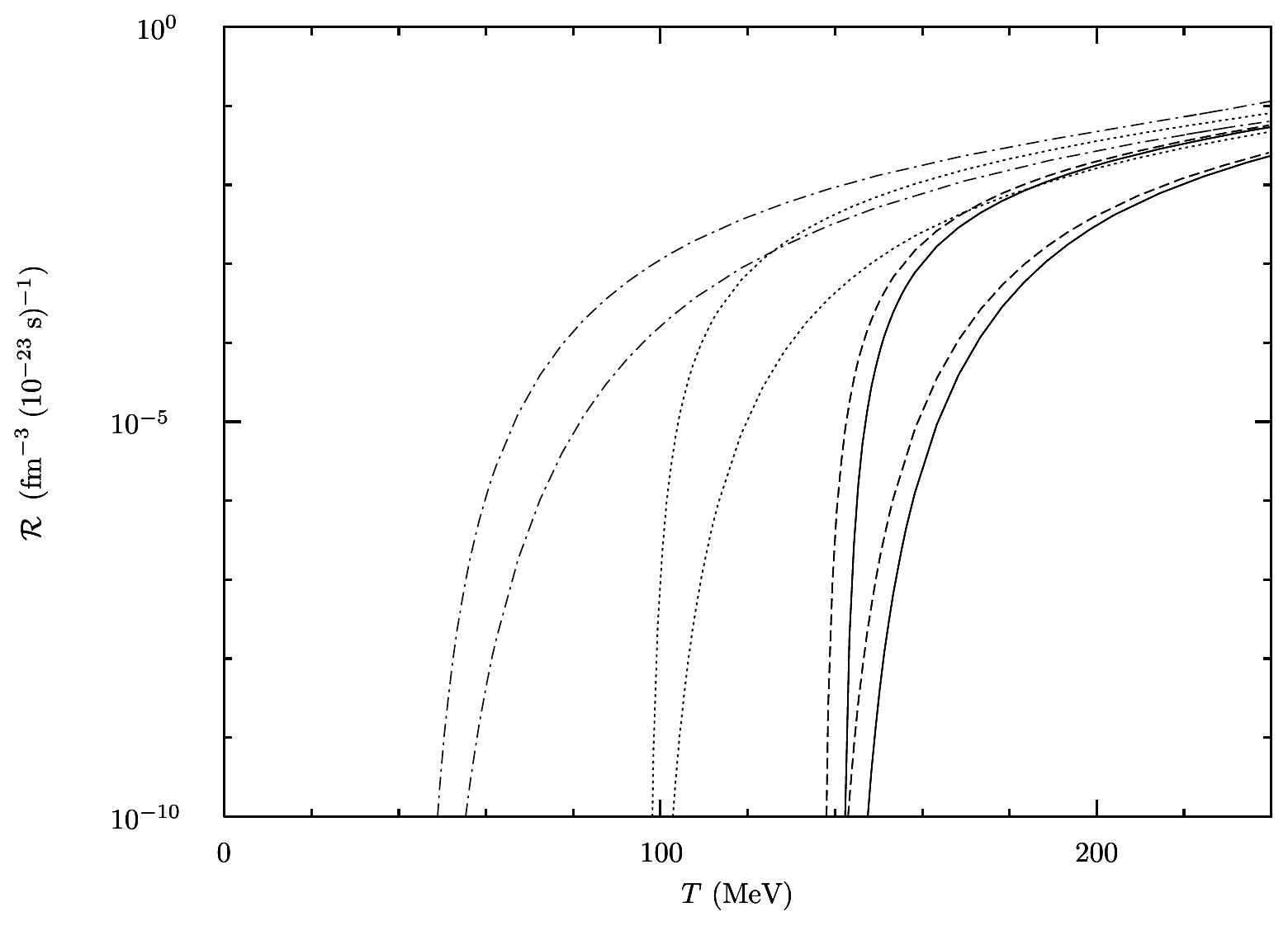}
	\caption{\label{fig:rate}
			The nucleation rate, as a function of temperature,
			with 2 massless quark flavors, for
			different values of the chemical potential, and
			$B^{1/4}=145$ MeV. Upper curves are
			without the color singlet constraint, while lower curves
			include the constraint. Solid curves are for $\mu=0$, dashed
			curves for $\mu=100$ MeV, dotted curves for $\mu=300$ MeV,
			and dotted-dashed curves for $\mu=400$ MeV.}
\end{figure}

\index{free energy!barrier|)}

\section{Rate Suppression}
\index{nucleation rate!suppression of|(}
By finding the maximum in the free energy shown in \Fig{fig:barrier} the
critical radius $R_c$ and the barrier height $\Delta F(R_c)$ are found.
Inserting the value of $\Delta F(R_c)$ in the expression \Eq{eq:rate}
gives the nucleation rate, which can be rewritten
\be
	\mcal{R} \approx 0.2\;\text{fm}^{-3}\;(10^{-23}\;\text{s})^{-1}
			T_{100}^4 e^{-\Delta F/T} .
\ee
Here $T_{100}$ is the temperature in units of 100 MeV. The nucleation
rate is thus given in units of the number of nucleations per fm${}^{3}$
per $10^{-23}$ s ($=3$ fm$/c$), which is a natural scale in the context
of heavy ion collisions.

\begin{figure}
	\centering
	\includegraphics[width=\textwidth]{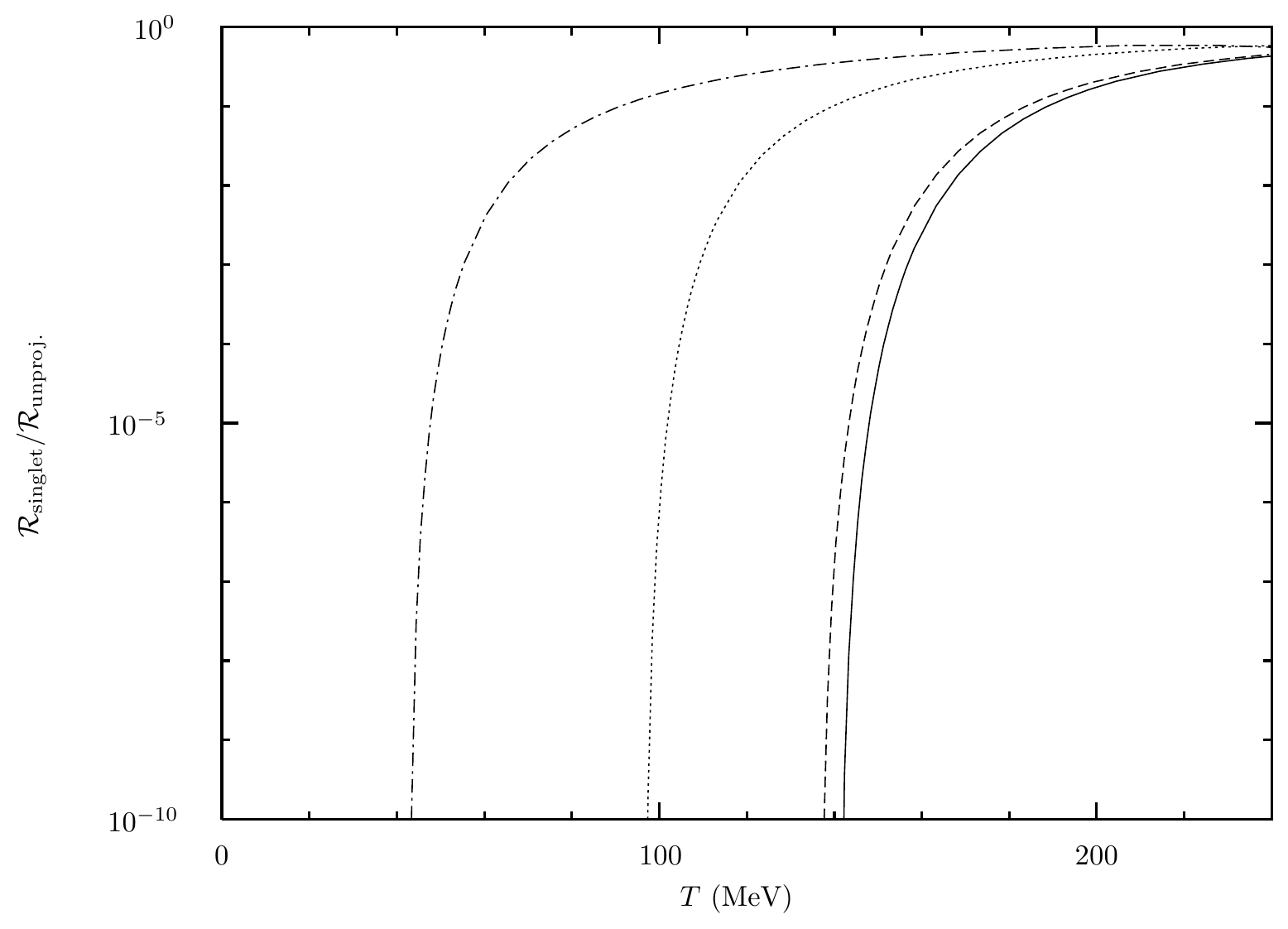}
	\caption{\label{fig:suppression}
			The rate suppression factor, as a function of temperature
			for the same choice of parameters as in \Fig{fig:rate}.
			Again the curves correspond to (from right to left) $\mu=0$, 
			$\mu=100$ MeV, $\mu=300$ MeV, and $\mu=400$ MeV.}
\end{figure}

The nucleation rate is shown in these units in \Fig{fig:rate}, where a
considerable suppression of the rate in the constrained case is seen
relative to the unconstrained case. At moderate superheating the rate is
suppressed by many orders of magnitude, shrinking to 1 order of magnitude
only when the  superheating exceeds $\sim$40 MeV. This clearly seen from
\Fig{fig:suppression}, where the relative rate, \ie , the rate calculated
with the color singlet constraint divided by the rate calculated without
the constraint, is plotted as a function of temperature.

Results for other values of the bag constant $B$ can be obtained by
scaling the temperature in proportion to $B^{1/4}$ and the nucleation
rate in proportion to $B$. The relative rate, or the suppression factor,
shown in \Fig{fig:suppression} is of course unchanged.

The volume and time available in a heavy-ion collision amounts to
perhaps 100 fm${}^3$ fm$/c$, so
in the case where no constraints are included one thus predicts a fair 
probability for bubble nucleation at moderate superheating, say 10--20 MeV.
Conversely, with the color singlet constraint included a superheating of
at least double that amount is needed to achieve a similar probability.

\index{nucleation rate!suppression of|)}

\section{Conclusion and Discussion}
It has been demonstrated that in the framework of homogenous nucleation
theory, there is a significant suppression of the rate for nucleation of
quark-gluon plasma droplets if the formed droplets are subjected to the
restriction that they must be color singlets. This conclusion rests on
several assumptions that will be discussed below.

First of all the phase transition is assumed to be of first order. This
is the case in the bag model, but it could conceivably be second order,
as indicated by lattice calculations (at $\mu=0$). This assumption is
not testable until reliable lattice calculations exist, also for $\mu
\not= 0$. 

The result is of course model dependent. Choosing a different model than
the bag model, would certainly give different results, but the trend
should be the same, as the effect of imposing the color singlet
constraint will always be to reduce the effective number of degrees of
freedom, and thereby to increase the free energy.

At very high superheating, the critical radius becomes very small
which could result in the breakdown of the multiple reflection expansion
on which the liquid drop model is based. This can only be tested by
a direct comparison of the shell model and the liquid drop model. Shell
model calculations at finite temperature have only recently been
attempted \cite{Mustafa96a} (see also Ref.~\cite{Mardor91a} for a
calculation at $\mu=0$), and so this has not been checked. But at more
moderate superheating, which includes the more realistic temperatures,
there should be no problems in using the liquid drop model.

The most seriuos shortcoming of the present approach is probably that
the timescales relevant for a heavy-ion collision are too short for the
applicability of a thermodynamic equilibrium approach. Nevertheless this
is often used as a first approximation, and even if a thermal nucleation
picture is wrong there may still be an effect of the color singlet
constraint, which has to be fulfilled regardless of the nucleation
mechanism.

In \Index{neutron star} cores on the other hand, there is ample time for
thermodynamic equilibrium to be reached, and so the present approach
should not be problematic. A simple model, using a pure neutron gas
\cite{Olesen94a}, could be used to estimate the stability of neutron
stars with respect to quark matter formation.

So the conclusion remains that a significant suppression of the
nucleation rate of quark-gluon plasma droplets in a superheated hadron
gas is a direct consequence of the color singlet constraint. Most of the
effects left out in this simple treatment, like the pressure of the
hadron phase and the fixed momentum constraint will further reduce
the nucleation rate.

\clearemptydoublepage
\chapter{Finale}

\section{Thesis Conclusion and Summary}
This thesis has been centered around a few key themes which have been
staged differently in individual chapters. These themes are the finite
system size, non-zero temperatures, and colorlessness.

The finite size effects were introduced in Chapter \ref{chap:mit} by
use of the multiple reflection expansion method, explained in
Appendix~\ref{chap:mre}, and the liquid drop
model was compared to the shell model, showing that the multiple
reflection expansion approach is valid for strangelets, unless one
discusses subjects such as strangelet decay modes which depend on the
detailed effects of the shells. The mean effects of the shell structure
are well reproduced by the liquid drop model, whose major advantage 
over the shell model is that many quantities can be
evaluated analytically, allowing for more insight into parameter
dependences and general trends, as well as making solutions easier. Even
in those cases where analytic results are not available, the simpler
cases serve as important known limits against which the numerical
results can be checked.

The effect of non-zero temperature, was introduced in Chapter
\ref{chap:temperature}, where the results were used to calculate
strangelet masses and to study the phase diagram of quantum
chromodynamics in the bag model at high temperature and baryon density.
At finite temperature it is only possible to obtain analytic results
for massless quarks, in contrast to the zero temperature case were
massive quarks \emph{can} be treated analytically.

In Chapter \ref{chap:equilibrium} the phase equilibrium between strange
quark matter and a hadron gas was studied with emphasis on the
effects of finite size of the quark phase, and a non-zero strangeness
fraction. The abundance of strangeness in ultrarelativistic heavy-ion
collisions and the mechanism of \Index{strangeness separation}
\cite{Greiner87a,greal88,Greiner91a} makes it necessary to study the
effect of non-zero strangeness fraction on the phase equilibrium. This
was done by Lee and Heinz~\cite{Lee93a} for bulk phases and generalized
by Jensen and Madsen~\cite{Jensen95a} to include the finite size of the
quark phase, which is what was described in
Chapter~\ref{chap:equilibrium}.  This work was
extended by He~\textit{et al.}~\cite{He96c} to allow the calculation of
isentropic expansion trajectories of the system as a whole.

In Chapter \ref{chap:projection} it has been explained how to implement
the effect of exact color singletness, using the group theoretical
projection method. The color singlet partition function was constructed
for a system of non-interacting quarks, antiquarks, and gluons, and the
partition function was evaluated in the saddle-point approximation, in
the case of massless quarks. The
result obtained here was identical with that obtained by Elze and
Greiner \cite{Elze86b}, but other authors have reported different
results. The saddle-point approximation is valid at high
temperature and chemical potential, but these conditions are not always
present, and so a numerical evaluation is necessary. The inclusion
of massive quarks cannot be handled in the saddle-point approximation,
so this case must always be treated numerically.


In Chapter \ref{chap:color} the effect of the color singlet constraint
on strangelet masses was investigated. At fixed entropy per baryon the
color singlet constraint has the effect of increasing strangelet masses.
The effect is most significant for large entropy and small baryon
number, disappearing in the bulk limit as well as at zero temperature.
In the saddle-point approximation another effect---demanding fixed, zero
momentum---was also included, and though not negligible it was shown to
be less important than the color singlet constraint. A thorough
discussion of the saddle-point approximation, and a comparison with
exact numerical calculations was then performed. It was shown that the
saddle-point approximation is generally useful at high temperature and
high chemical potential, provided the system size is not excessively
small. Finally a calculation including massive quarks was presented, and
the effect of the color singlet constraint was seen to be roughly
independent of the quark mass. But further refinements and improvements
to these $m\not=0$ calculations need to be done before a large scale
investigation of color singlet strangelets can be undertaken.

In Chapter \ref{chap:suppression} it was demonstrated that the color
singlet constraint has the effect of diminishing the rate of quark
matter nucleation from a hadronic gas by many orders of magnitude at
moderate superheating, only gradually approaching the result obtained
without the color singlet constraint included. This suppression of the
nucleation rate could have important consequences for the formation of quark
gluon plasma droplets in ultrarelativistic heavy-ion collisions, and the
nucleation of quark matter droplets in neutron stars.

The approach used relies on some simplifications, 
like the neglect of the hadron pressure, but this was argued to strengthen
the conclusion, and it is therefore not a serious omission.  Other
assumptions such as the phase transition being of first order, are not
readily testable, but must await the outcome of more elaborate
\Index{lattice gauge calculations}.

From the studies presented in these chapters it is learned that all
three effects---the effect of finite size, the effect of non-zero
temperature, and the effect of exact color singletness---have a
destabilizing effect on strange quark matter. This conclusion is
\textit{a priory} obvious, and does therefore not come as any surprise.
What is important is the magnitude of the effects, and their impact on
our chances of ever observing strange quark matter.

The
baryon number available in a heavy-ion collision is limited to that of
the colliding nuclei ($2\times 207$ for Pb on Pb), but a strangelet can only
have a small fraction of this number, since most of the baryon number ends up
in hadrons. The lower the baryon number the more likely will a  strangelet
be unstable, unlike nuclei where Coulomb effects give rise to a maximum
binding energy near ${}^{56}$Fe. Shell effects, however, play an important
\index{shell effects}
stabilizing role for strangelets with a magic number of quarks, such as
\index{magic numbers}
the $A=6$ system composed of 6 $u$ quarks, 6 $d$ quarks, and 6 $s$
quarks. 

When increasing the temperature, strangelets acquire thermal energy,
making them prone to surface evaporation of hadrons. This effect is
important in heavy-ion collisions as well as in the early universe.
Shell effects become less important at higher temperature, and so this
possible stabilizing mechanism is to some degree lost.

In addition to the increase in mass due to thermal energy, the demand
of colorlessness also has the effect of increasing the  mass of a
strangelet with non-zero entropy.

In spite of all these  destabilizing effects there is still the chance
that cold strange quark matter is stable or metastable, and the
production of strangelets in heavy-ion collisions cannot be ruled out.
At this point it is relevant again to stress the fact that the MIT bag
model can only give a hint at what to expect. The question of the
stability of strange quark matter must ultimately be settled
experimentally, but model predictions are still important as a guide for
experiments.

It seems that the most favorable conditions for strange quark matter is
in compact objects such as neutron stars. Here truly bulk systems can
exist, and the temperature is low. Even in the case of strange quark
matter being unstable, the pressure in the core of a neutron star may be
great enough to cause strange quark matter to be the preferred phase of
matter.

\section{Outlook}
The results, just summarized, that have been presented in this thesis
point toward some possible directions of future research.
Some of these prospects have been mentioned in the text, others have
not, and so I would like to finish by mentioning some of the topics I
find interesting and worthy of further study.

The color singlet constraint is an important addition to the bag model,
which otherwise implements the well known \Index{confinement} feature of
QCD. Being able to describe both of these two essential features---confinement
and colorlessness---should improve bag model predictions. The color
singlet constraint was implemented in the early to mid eighties, but many
studies would still benefit from taking this constraint into
consideration. With today's more powerful computers, this has become
much easier.

One of the places where the effect of the color singlet constraint has not
yet been investigated is in phase equilibrium calculations such as those
presented in Chapter~\ref{chap:equilibrium}. It will be interesting to
see how this changes the phase diagram and the prospects for forming
strangelets in ultrarelativistic heavy-ion collision experiments.

A related area of research is the cosmological quark-hadron phase
\index{quark-hadron phase transition!cosmological}
transition where the effect of color singletness is also expected to
play a role. A bubble of hadronic matter nucleating in the supercooled
quark matter phase is composed of hadrons that are color singlets. These
hadrons materialize from a combination of quarks antiquarks and gluons
that must come together to form a color singlet. It is unclear at
present how to formulate this, but it is an interesting problem for
future research.

The suppression of quark matter formation described in
Chapter~\ref{chap:suppression} can be applied to nucleation of quark
matter droplets in \Index{neutron star}s, by a modification of the
treatment of Olesen and Madsen~\cite{Olesen94a}. With the present state
of the calculations it should be relatively easy to obtain this goal.

It was mentioned in Chapter~\ref{chap:projection} that the inclusion of
the curvature term shifts the location of the maximum in the
saddle-point approximation. This means that an improved saddle-point
approximation, taking this into account is possible. The feasibility of
such an improved approximation has not been undertaken yet, but if it
proves not to be beyond analytical evaluation it might greatly improve
the saddle-point approximation presented here, and would perhaps obviate
the need to do numerical calculations for massless quarks.

Calculations including massive quarks must be done numerically, and
results of such calculations were presented in Chapter~\ref{chap:color}.
There is still room for improvement of these calculations, and this work
is currently in progress.

\clearemptydoublepage

\appendix
\chapter{The Multiple Reflection Expansion\label{chap:mre}}
\index{multiple reflection expansion|(}
In the previous chapters I have used an asymptotic expansion of the
density of states in powers of $\frac{1}{kR}$, where $k$ is the wavenumber
and $R$ is the radius of the system. This \chapapp\ describes how such an
expansion can be obtained using Green's function methods.
The following is a brief exposition of the general
formalism used, but a few detailed calculations are presented.
The \chapapp\ is organized as follows. In section
\ref{sec:trace} I relate the density of states to the Green's function.
In section \ref{sec:vol}  this relation is used to calculate the
density of states for a non-interacting free field. Next in section
\ref{sec:mre} the multiple reflection expansion for the Green's function
is introduced and discussed. Finally in section \ref{sec:dos} a short
outline of how to obtain the multiple reflection expansion for the
density of states is given.
In this \chapapp\ a generic 4-vector is denoted $x^\mu=(t,{\bf r})$.

\section{\label{sec:trace}Trace Formula for the Density of States}
\index{density of states!trace formula|(}
In this section I derive a formula relating the density of states to
the propagator. The propagator in question is that of a non--interacting 
Dirac field
confined to a finite region of space $(\Omega)$, obeying appropriate boundary
conditions at the surface $(\partial \Omega)$.
The \emph{form} of the result does not depend on the details of the
Hamiltonian or the nature of the boundary conditions, only on the spinor
nature of the field, so I will adopt a general
notation in which the Dirac equation\index{Dirac equation} is written:
\be
  D \Psi =0,
\ee
or
\be
  i \frac{\partial \Psi}{\partial t}=H\Psi,
\ee
where the free particle case corresponds to
\begin{eqnarray}
  D &=& i\gamma ^\mu \partial _\mu -m \\
  H &=&-i \bsa  \cdot {\bf \nabla} + \beta m.
\end{eqnarray}
The particular problem to be solved is that of solving the Dirac
equation for a non--interacting quark confined to an MIT--bag.
\index{Dirac equation!for quark in MIT bag}The Dirac equation with
boundary condition in this case is:
\be
    \begin{array}{r@{\;=\;}lr@{\;\in\;}l}
        \left( i\gamma_\mu\partial_\mu - m\right)\Psi(x) & 0 & \mbf{r}
        & \Omega \\[1ex]
        in_\mu\gamma^\mu\Psi(x)& \Psi(x) \qquad\qquad&\mbf{r}& \dom \\
    \end{array}
\ee
where $\bf n$ is the outward normal to $\dom$ at $\bf r$. The second boundary
condition in Eq.~(\ref{eq:quarkinbag}) has been omitted since it corresponds
to pressure equilibrium, which is taken into account when minimizing the
free energy.
In the rest of this section an accompanying boundary condition is
always assumed when I refer to the field equation.
The propagator or Green's function, $S(x,x')$, for the Dirac equation is
defined through\index{Green's function!for Dirac equation|(}
\be
  DS(x,x')=\delta (x-x'),
\ee
which in terms of the Hamiltonian is
\be
  \left(i\frac{\partial}{\partial t} - H\right) S(x,x')=\gamma ^0 \delta(x-x').
\ee
The time independent Green's function is obtained as the Fourier
transform, with respect to time, of the time dependent Green's
function defined above.
\begin{eqnarray}
  S(x,x')&=&\int_{-\infty}^\infty \frac{d\omega}{2\pi}\, S({\bf r,r'},\omega) e^{-i
\omega t} \\
  S({\bf r,r'},\omega)&=&\int_{-\infty}^\infty dt\, S(x,x')e^{i\omega t},
\end{eqnarray}
where $t=x_0-x'_0$.
Inserting this in the previous equation gives
\be \label{eq:timeindepgreens}
  (\omega -H)S({\bf r,r'},\omega)=\gamma ^0 \delta({\bf r-r'}).
\ee

Denote the eigenfunctions of $H$ by $\Psi _n({\bf r})$ and the
corresponding eigenvalues $E_n$, here assumed to be discrete.
(The generalization to a continuous spectrum is straightforward).
Then the eigenvalue equation is
\be \label{eq:eigenvalue}
  H\Psi _n({\bf r})=E_n\Psi_n({\bf r}).
\ee
The eigenfunctions are assumed to constitute a complete set:
\be \label{eq:completeness}
 \sum_n \Psi_n({\bf r})\Psi_n^\dagger({\bf r'})=\delta({\bf r-r'}).
\ee
Without loss of generality we can assume them to be orthonormal
\index{orthonormality!of spinors}as well, thus satisfying
\be
  \int_\Omega d^3{\bf r}\, \Psi_n^\dagger ({\bf r})\Psi_m({\bf r})=\delta _{n,m}.
\ee
The orthonormality relation can also be written as 
\be \label{eq:orthonormality}
  \int_\Omega d^3{\bf r}\, \Tr \left( \Psi_n({\bf r})\Psi_m^\dagger ({\bf r})
\right)=\delta_{n,m}.
\ee
Using the eigenvalue equation (\ref{eq:eigenvalue}) and the completeness
relation (\ref{eq:completeness})  in the equation (\ref{eq:timeindepgreens})
for the time independent Green's function gives
\be
  S({\bf r,r'},\omega)=\sum_n \frac{\Psi_n({\bf r})\Psi_n^\dagger ({\bf r'})}{\omega
-E_n}\gamma^0.
\ee
\index{Green's function!for Dirac equation|)}
Since $H$ is Hermitian, the eigenvalues are real, and
$S({\bf r,r'},\omega)$ is observed to have its poles distributed along
the real $\omega$--axis.
Taking the trace and integrating over $\Omega$ with ${\rm r'=r}$
(using Eq.~(\ref{eq:orthonormality}) yields
\be
  \int_\Omega d^3{\bf r}\, \Tr \left( S({\bf r,r},\omega)\gamma^0
\right)=\sum_n \frac{1}{\omega - E_n}.
\ee
Now the identity $$\lim_{\varepsilon \to 0_+}\frac{1}{x\pm i\varepsilon}
={\cal P} \frac{1}{x}\mp i\pi \delta (x)$$ is applied, giving the
result
\be
  \int_\Omega d^3{\bf r}\, \Tr \left( S({\bf r,r},\omega \pm i\varepsilon)\gamma^0
\right)
=\sum_n \left( {\cal P}\frac{1}{\omega -E_n} \mp i\pi \delta(\omega -E_n)
\right).
\ee
${\cal P}$ denotes the Cauchy principal value.
Here and in the following the limit $\varepsilon \to 0_+$ is
implicitly assumed. The density of states $\rho(\omega) = \sum_n
\delta(\omega -E_n)$ is thus given by
\be
  \rho(\omega)=\mp \frac{1}{\pi} {\rm Im} \int_\Omega d^3{\bf r}\, \Tr
\left[ S({\bf r,r},\omega \pm i\varepsilon) \gamma^0 \right].
\ee
Since the integration is actually a trace operation performed on the spatial
``indices'' of $S$ we can compactify the notation a little, by
introducing the trace operation
\be
  \tr {\bf A} = \int d^3{\bf r}\, \Tr {\bf A}({\bf r,r})=\int
d^3{\bf r}\, \sum_\mu A_{\mu \mu}({\bf r,r}).
\ee
The final result is then
\be \label{trace}
  \rho(\omega)=\mp \frac{1}{\pi}{\rm Im}\, \tr S(\omega \pm i\varepsilon)
\gamma^0.
\ee
\index{density of states!trace formula|)}

\section{\label{sec:vol}Density of States for a Free Field}
\index{density of states!for a free field|(}
The simplest possible case that we can apply the method of the previous
section to is that of a free field. The density of states could be
obtained from simple arguments (waves in a box) in this case, but it
is instructive and reassuring that the more formal approach yields
the same result. In the case of a free field the Green's function
can be obtained directly from the field equation, since there is no
boundary condition. The basic definition of the Green's function is
\index{Green's function!for Dirac equation|(}
\be
  \left( i\gamma^\mu \partial_\mu -m\right) S^{(0)}(x-x')=\delta(x-x').
\ee
Introducing the Fourier transform as
\be
  \begin{array}{rcl}
  S^{(0)}(x)&=& \displaystyle\int \frac{d^4p}{(2\pi)^4}\,
  S^{(0)}(p)\, e^{-ip\cdot x}
 \\ & & \\
  \displaystyle S^{(0)}(p)&=& \displaystyle\int d^4x\,  S^{(0)}(x)\, e^{ip\cdot x},
  \end{array}
\ee
we get the field equation in momentum space
\be
  \left( \gamma^\mu p_\mu -m\right)S^{(0)}(p)=1
\ee
Thus $S^{(0)}(p)$ is the inverse matrix of the ``Dirac operator''
in momentum space. From the algebra satisfied by the gamma matrices
it is easily shown that
\begin{displaymath}
  (\not p -m)(\not p +m)=p^2-m^2,
\end{displaymath}
which gives the following form of the free space propagator
\be
  S^{(0)}(p)=\frac{\not p +m}{p^2-m^2}.
\ee
The Green's function in configuration space is thus
\be
  S^{(0)}(x-x')=\int \frac{d^4p}{(2\pi)^4}\, e^{-ip\cdot (x-x')}
  \frac{\not p+m}{p^2-m^2}.
\ee
It is useful to compare this to the Green's function for a free
scalar field. The field equation, in this case, is the Klein--Gordon
equation and the Green's function, $\Delta^{(0)}(x-x')$, is defined
by\index{Green's function!for Klein-Gordon equation}
\be
  \left( \square +m^2 \right)\Delta^{(0)}(x-x')=-\delta(x-x').
\ee
Note the minus sign on the right hand side, which has been
introduced for convenience. In analogy to the spinor case we take
the Fourier transform, to obtain the momentum space representation
\be
 \begin{array}{rcl}
  \Delta^{(0)}(k)&=&\displaystyle \frac{1}{k^2-m^2} \\
   & & \\
  \Delta^{(0)}(x)&=&\displaystyle \int \frac{d^4k}{(2\pi)^4}\, e^{-k \cdot x}
  \frac{1}{k^2 -m^2}.
 \end{array}
\ee
Comparing this with the spinor result, we see that the two Green's functions
are connected as
\be \label{spinor:scalar}
  \begin{array}{rcl}
  S^{(0)}(p)&=&(\not p + m)\Delta^{(0)}(p) \\
  & & \\
  S^{(0)}(x-x')&=&(i\! \not \! \partial{} + m)\Delta^{(0)}(x-x').
  \end{array}
\ee
This is a convenient way of obtaining the propagator for the free
Dirac field.
In order to find the density of states we need to know the time
independent Green's function. This was previously defined as the
Fourier transform with respect to time of the time dependent Green's
function. In the scalar case this gives
\begin{eqnarray}
  \Delta^{(0)}({\bf r- r'},\omega)&=& \int_{-\infty}^{\infty}dt\,
    \Delta^{(0)}(x-x')\,e^{i\omega t} \nonumber \\
  &=& \int_{-\infty}^{\infty}dt\, \int\frac{d^4k}{(2\pi)^4}\,
    e^{i{\bf k \cdot (r-r')}} e^{it(\omega -k_0)} \frac{1}{k^2-m^2}
    \nonumber \\
  &=& \int \frac{d^3{\bf k}}{(2\pi)^3}\, e^{i{\bf k \cdot (r-r')}}
    \frac{1}{\omega^2- {\bf k}^2 -m^2}
\end{eqnarray}
Since the integrand has poles for ${\bf k}^2=\omega^2-m^2$, which lie
on the path of integration for real $\omega$, the integral is not
well defined. It is however uniquely specified by the imaginary
part added to the energy in the trace formula for the density of
states. Using the upper sign in Eq.~(\ref{trace}) the pole is pushed into the upper
$k$-plane, and the integral can be evaluated by closing the contour
of integration along an infinite semi circle. The result is
\be
  \Delta^{(0)}({\bf r-r'},\omega+i\varepsilon)=\frac{e^{ik| {\bf
r-r'}|}}{4\pi |{\bf r-r'}|},
\ee
where $k=\sqrt{\omega^2 -m^2 +i\varepsilon}$, is the square root with
positive imaginary part. The above result immediately gives the
corresponding result in the spinor case
\begin{eqnarray}
&&S^{(0)}({\bf r-r'},\omega +i\varepsilon)= \left(\gamma^0\omega +i
\bsg \cdot {\bf \nabla} +m\right) \Delta^{(0)}({\bf r-r'},\omega+
i\varepsilon) \\
&&\qquad =\left( \gamma^0 \omega -i \bsg \cdot {\bf (r-r')} \frac{ik|{\bf
r-r'}| -1}{|{\bf r-r'}|^2} + m \right) \Delta^{(0)}({\bf r-r'}, \omega
+i\varepsilon) \nonumber 
\end{eqnarray}
Multiplying by $\gamma^0$ and taking the trace---using $\Tr
\gamma^\mu\gamma^\nu = 4\eta^{\mu\nu}$, where $\eta^{\mu\nu}={\rm
diag}(1,-1,-1,-1)$ is the Minkowski metric---gives the result
\index{Minkowski metric}
\be
  \Tr S^{(0)}({\bf r-r'},\omega+i\varepsilon)\gamma^0= 4\omega
\Delta^{(0)}({\bf r-r'},\omega+i\varepsilon)
\ee
\index{Green's function!for Dirac equation|)}
The density of states is then
\begin{eqnarray}
  \rho(\omega)&=&\frac{1}{\pi}\int_{\Omega}d^3{\bf r}\, \lim_{\rm r' \to r}
{\rm Im}\, \Tr S^{(0)}({\bf r-r'},\omega+i \varepsilon) \gamma^0
\nonumber \\
&=&\frac{1}{\pi} \int_{\Omega}d^3{\bf r}\, \lim_{\rm r' \to r} 4\omega \frac{
e^{-{\rm Im}\, k|{\bf r-r'}|}\sin {\rm Re}\, k|{\bf r-r'}|}{ 4\pi |{\bf r-r'}|}
\nonumber \\
&=& \int_\Omega d^3{\bf r}\, \frac{4 \omega {\rm Re}\, k}{4\pi^2}=
  V\frac{4\omega\sqrt{\omega^2 -m^2}}{4\pi^2}.
\end{eqnarray}
To obtain the density of states in terms of the momentum of the
fermion the substitution $\omega \to k$ is performed
\be
  \rho(\omega)d\omega=\rho(\omega(k))\frac{k}{\omega}dk=2\frac{Vk^2}{2\pi^2}dk.
\ee
This expression is the desired result. The factor two taken out in front represents
the degeneracy of spin states for a non--interacting field.
\index{density of states!for a free field|)}

\section{\label{sec:mre}Expansion of the Green's function}
\begin{figure} 
	\begin{center}
		\includegraphics[width=\textwidth]{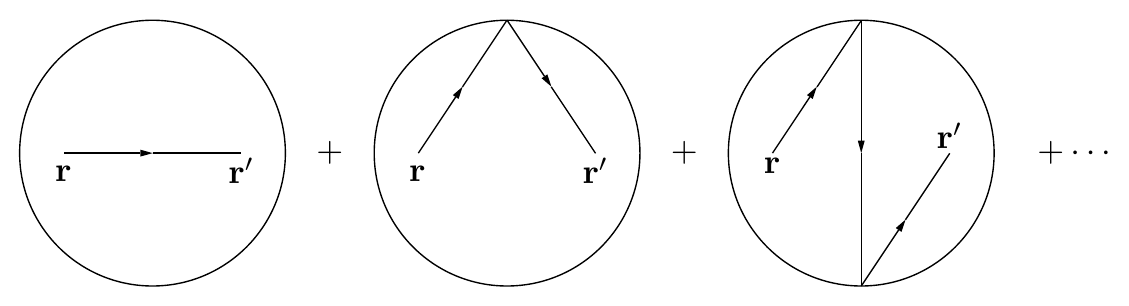}
		\caption{\label{mrefig} The first three terms in the multiple reflection expansion for the propagator.}
	\end{center}
\end{figure}

\index{multiple reflection expansion!of the Green's function|(}
The basic idea of the multiple reflection expansion, is that the
propagation of a particle in $\Omega$ can occur either directly
(as described by the free space propagator), or via one or more
reflections at the surface. Schematically the propagator can be
written as in Fig.~\ref{mrefig}, where each line represents a free space
propagator.
The multiple reflection expansion for the time independent
Green's function has the following form
\begin{eqnarray} \label{mre.general}
  S{(\bf r,r')}&=&S^{(0)}{(\bf r,r')}+\oint_{\dom}d\sigma_\alpha\,
S^{(0)}({\bf r},\bsa)\,K(\bsa)
S^{(0)}(\bsa,{\bf r'})
\nonumber \\
  &+&\oint_{\dom}d\sigma_\alpha d\sigma_\beta\,
S^{(0)}({\bf r},\bsa)\,K(\bsa)
S^{(0)}(\bsa,\bsb)
\,K(\bsb)S^{(0)}(\bsb,{\bf r'})
\nonumber \\
&+& \cdots
\end{eqnarray}
Here $K$ is a reflection kernel describing the ``interaction''
with the surface due to the confining boundary conditions. In the
following I will denote spatial vectors in $\Omega$ by bold face Roman
symbols (e.g. ${\bf r,r'}$), and spatial vectors on $\dom$ by boldface Greek
letters $(\bsa,\bsb,\bsg,\ldots)$. The
multiple reflection expansion was first studied by Balian and
Bloch \cite{Balian70a,Balian74a}. They considered a scalar field with different
boundary conditions, and later also a vector field in a cavity
with conducting walls. The multiple reflection expansion for
scalar, spinor and vector fields with the boundary conditions of
the MIT bag was treated by Hansson and Jaffe \cite{Hansson83a,Hansson83b} and later
also Jaffe and Vintro \cite{Jaffe84a}. The following derivation of the
multiple reflection expansion follows \cite{Hansson83a,Hansson83b}
, but the results
are generalized to a massive spinor field. For a static bag, the
differential equation and boundary condition satisfied by the
time independent Green's function are
\index{Green's function!for quark in MIT bag|(}
\be
\begin{array}{rclrl}
  \left( \gamma^0 \omega +i\bsg \cdot \nabla -m
  \right)S({\bf r,r'},\omega)&=&\delta({\bf r-r'})& \qquad {\bf
  r,r'} &\in \Omega \nonumber \\
  & & &\\
  \left( i\bsg \cdot {\bf n}_\alpha +1
  \right) S(\bsa,{\bf r},\omega)&=&0& 
  \bsa &\in \dom
\end{array}
\ee
The problem of solving the above equations, can be transformed
into an integral equation for $S({\bf r,r'},\omega)$. This is a
well known technique in potential theory. For the classical
Dirichlet and Neumann Problems the integral equations can be
derived directly from the differential equation and boundary
condition, using Green's theorem\index{Green's theorem}
(Gauss' law in 3 dimensions)
for converting a volume integral into a surface integral
\cite{Stakgold}.  Because of the more
complicated nature of the problem at hand a direct derivation is
not so simple. Therefore the solution will be arrived at by an
analogy with electrostatics. The Green's function is split up
into a free space part, and a boundary term
\be
  S({\bf r,r'})=S^{(0)}({\bf r,r'}) + \tilde{S}({\bf r,r'}).
\ee
I have chosen to suppress the $\omega$-variable in the above
expression, and will continue to do so in the following.
The idea is now to write the surface term as a kind of potential
arising from a surface distribution at the boundary
\be \label{Ssurf}
  \tilde{S}({\bf r,r'})=\oint_{\dom}d\sigma_\alpha\, S^{(0)}({\bf r},
  \bsa)\eta(\bsa,{\bf r'}).
\ee
In electrostatics the potential arising from a surface
distribution, $\varsigma$, of charges (a so called simple layer)
is given by \cite{Jackson}
\be \label{phi}
  \varphi ({\bf r})=\int_{\dom} \frac{\varsigma (\bsa 
  ) }{4\pi |{\bf r}- \bsa|} \,
  d\sigma_\alpha .
\ee
This potential has the well known property that it is
continuous across $\dom$, whereas the electric field has a
discontinuous normal component, the discontinuity being
proportional to the surface charge density
\be
  -{\bf n\cdot \nabla}\varphi |_{(+)} +{\bf n\cdot \nabla}\varphi |_{(-)}
  =\varsigma,
\ee
where ${\bf n}$ is the normal to $\dom$, and the indices $+$ and $-$ denote the 
limiting values when approaching the surface from the inside and outside
respectively. (In this case $\dom$ need not be the boundary of a
finite region of
space, so the ``outside'' is defined as the side to which the normal points).
This result can be
obtained from Eq.~(\ref{phi}), or by a direct application of Gauss' law.
The potential due to a surface distribution of dipoles (a dipole
layer) has the form \cite{Jackson}
\be
  \varphi({\bf r})=\int_{\dom} D(\bsa)\,{\bf
  n_\alpha \cdot \nabla_\alpha}\frac{1}{4\pi |{\bf r} - 
  \bsa|} \, d\sigma_\alpha,
\ee
where $D(\bsa)$ is the dipole density.
Unlike a simple layer, a dipole layer causes a discontinuity in
the potential itself
\be
  \varphi_{(-)}(\bsa)- \varphi_{(+)}(\bsa)=D(\bsa).
\ee
Here $\varphi_{(+)}(\bsa)$ is the limiting
value of the potential when approaching the surface from the
inside, and $\varphi_{(-)}(\bsa)$ from the
outside.
Substituting the Green's function for the
Laplace equation\index{Laplace equation} with
the scalar Green's function gives quite analogous results. Thus
the potential generated by a simple layer is continuous in the scalar
case, while its normal derivative is discontinuous at the
surface. The gradient in the connection \ref{spinor:scalar} 
between the spinor and
scalar Green's functions generates a dipole layer, causing the
potential of a ``simple'' spinor layer to be discontinuous. In
addition to not being simple in the sense that it produces a
discontinuity it may also have a matrix structure, complicating
things further. The discontinuity is found by a detailed analysis
to be
\be
  \tilde{S}_{(+)}(\bsb,{\bf r'}) -
  \tilde{S}(\bsb,{\bf r'}) = -\frac{i}{2}
  \bsg \cdot {\bf n}_\beta \eta(\bsb,{\bf r'}).
\ee
The factor $1/2$ comes from the fact that the discontinuity is
from the inside to a point {\em on} the boundary. If the boundary
had been crossed it would produce double the above amount.
However, since we wish to describe quarks confined to a bag,
crossing the boundary is not allowed. The ``source'', $\eta$, is
to be determined by the boundary condition to be satisfied by
$ {S}(\bsa,{\bf r})$. As the free space
part of the Green's function is obviously continuous across $\dom$
the total Green's function has the same discontinuity as the
boundary term
\be
  S_{(+)}(\bsa,{\bf r}) =
  S(\bsa,{\bf r}) -\frac{i}{2} \bsg \cdot {\bf n}_\alpha \eta(\bsa,{\bf r}).
\ee
Since $S({\bf r,r'})$ is continuous within $\Omega$ it is the
limit of $S({\bf r,r'})$ as ${\bf r} \to \bsa$
from the inside that is required to fulfill the
boundary condition.
\begin{eqnarray}
  &&\left( i \bsg \cdot {\bf n}_\alpha +1
  \right) S_{(+)}(\bsa,{\bf r})=0
  \nonumber \\
  &\Updownarrow & \nonumber \\
  &&0=\left( i \bsg \cdot {\bf n}_\alpha +1
  \right) S(\bsa,{\bf r})
  -\left( i \bsg \cdot {\bf n}_\alpha +1
  \right)
   \frac{i}{2} \bsg \cdot {\bf n}_\alpha 
  \eta (\bsa,{\bf r})
  \nonumber \\
  &&\;\;\;=\left( 1+ i \bsg \cdot {\bf n}_\alpha
  \right) S(\bsa,{\bf r}) -\frac{1}{2}
  \left( 1+ i \bsg \cdot {\bf n}_\alpha
  \right) \eta (\bsa,{\bf r}) .
\end{eqnarray}
The factor multiplying $\eta$ is the projection operator,
\be
  P_\alpha=\frac{1}{2}( 1+ i \bsg \cdot {\bf
  n}_\alpha),
\ee
which satisfies $P_\alpha^2=P_\alpha$. With this notation the
equation for $\eta$ is
\begin{eqnarray}
  P_\alpha \eta (\bsa,{\bf r}) &=&
  2P_\alpha S (\bsa,{\bf r}) \nonumber \\
  &=& 2P_\alpha S^{(0)} (\bsa,{\bf r}) 
  +2\oint_{\dom} d\sigma_\beta \, 
  P_\alpha S^{(0)} (\bsa,\bsb) 
  \eta (\bsb,{\bf r}).
\end{eqnarray}
This is an integral equation for $\eta$, involving only the known
free space Green's function $S^{(0)}$. Technically an equation of
this form is known as a Fredholm equation\index{Fredholm equation}
of the second kind.  Hansson and Jaffe show that the form
\be
  \eta (\bsa,{\bf r})=
  P_\alpha \mu (\bsa,{\bf r}),
\ee
is a possible solution,
and then proceed to iterate the equation for $\eta$, giving
\begin{eqnarray}
  \eta( \bsa,{\bf r}) &=&
  2P_\alpha S^{(0)}( \bsa,{\bf r})
  +2^2\oint_{\dom} d\sigma_\beta\, P_\alpha S^{(0)}(\bsa,\bsb) \nonumber \\
  &+&2^3 \oint_{\dom} d\sigma_\beta d\sigma_\gamma \, P_\alpha S^{(0)}(
  \bsa,\bsb) \, P_\beta S^{(0)}(\bsb,\bsg)
  \, P_\gamma S^{(0)}(\bsg,{\bf r})
  \nonumber \\
  &+& \cdots .
\end{eqnarray}
Inserting this in Eq.~(\ref{Ssurf}) gives the multiple reflection
expansion for the Green's function as
\begin{eqnarray}
  S({\bf r,r'})&=&S^{(0)}({\bf r,r'})+2\oint_{\dom} d\sigma_\alpha\,
  S^{(0)}({\bf r},\bsa) P_\alpha 
  S^{(0)}(\bsa,{\bf r'}) \nonumber \\
  &+&2^2\oint_{\dom} d\sigma_\alpha d\sigma_\beta\, 
  S^{(0)}({\bf r},\bsa) P_\alpha 
  S^{(0)}(\bsa,\bsb)\, P_\beta
  S^{(0)}(\bsb,{\bf r'}) \nonumber \\
  &+& \cdots ,
\end{eqnarray}
which is the result advertised in the beginning of this section.
It is not clear, however, that this expansion is suited for the
purpose of doing a perturbation calculation. Such an approach
demands that the successive terms become rapidly smaller. This is
only the case if the integral kernel $P_\alpha S^{(0)}(\bsa,\bsb)$
is sufficiently small. The multiple reflection
expansion is not unique, for example Hansson and Jaffe show that
it may equally well be written
\begin{eqnarray}
  S({\bf r,r'})&=&S^{(0)}({\bf r,r'})+2\oint_{\dom} d\sigma_\alpha\,
  S^{(0)}({\bf r},\bsa) 
  S^{(0)}(\bsa,{\bf r'}) \nonumber \\
  &+&2^2\oint_{\dom} d\sigma_\alpha d\sigma_\beta\, 
  S^{(0)}({\bf r},\bsa)
  S^{(0)}(\bsa,\bsb) \,
  S^{(0)}(\bsb,{\bf r'}) \nonumber \\
  &+& \cdots .
\end{eqnarray}
The two expansions do not correspond term by term, which means
that the notion of the $n$th reflection is not well defined. This
should not concern us too much, since we are looking for a
geometrical expansion of the density of states. Each term in this
expansion may receive contributions from several reflection terms
yielding a result independent of the particular form of the
expansion used. The expansion should thus be chosen in such a way
as to make the calculations as easy as possible.
\index{Green's function!for quark in MIT bag|)}
\index{multiple reflection expansion!of the Green's function|)}

\section{\label{sec:dos}Expansion of the Density of States}
\index{multiple reflection expansion!of the density of states|(}
Using the trace formula, (\ref{trace}), and the general form of the
multiple reflection expansion for the Green's function,
(\ref{mre.general}), yields the following expansion for the density of
states
\begin{eqnarray} \label{DOS.mre}
  \rho (\omega)&=&\mp \frac{1}{\pi} {\rm Im}\,\tr S^{(0)}(\omega\pm
  i\varepsilon)\gamma^0 \nonumber \\
  &\mp&\frac{1}{\pi} \int_\Omega d^3{\bf r} \oint_{\dom}d\sigma_\alpha\,
  \lim_{\bf r' \to r} {\rm Im} \,S^{(0)}({\bf r},\bsa
  ,\omega\pm i\varepsilon)K(\bsa)
  S^{(0)}(\bsa,{\bf r'},\omega\pm i \varepsilon)\gamma^0 \nonumber \\
  &+& \ldots 
\end{eqnarray}
The first term was shown in section \ref{sec:vol} to give the volume
contribution to the density of states. In order to obtain the
geometrical expansion Eq.~(\ref{eq:DOS}), terms proportional to the surface area
$\oint_{\dom} d\sigma$ and the extrinsic curvature $\oint_{\dom} d\sigma\,
(1/R_1+1/R_2)$ must be extracted from the above expansion. For a
specific choice of the reflection kernel, $K(\bsa)$,
this involves the following steps. The integrals over $\dom$ are evaluated in
a series of approximations, where the lowest corresponds to
approximating $\dom$ with a plane surface (its tangent plane), in the next a
second order surface, and so on, at each point $\bsa,\bsb,\ldots$.
This gives corrections of
increasing order in the principal radii of curvature. Within each
approximation the terms of Eq.~(\ref{DOS.mre}) contributing to a given order
$\oint_{\dom} (1/R_1+1/R_2)^n \, d\sigma$, $n=0,1,\ldots$ should be
identified and summed. Finally the integral expressions must be
evaluated, i.e. the Im and tr operations must be performed. This gives
an expansion of the form Eq.~(\ref{eq:DOS}).
\index{multiple reflection expansion!of the density of states|)}
\index{multiple reflection expansion|)}

\clearemptydoublepage
\chapter{Fermi-Dirac and Bose-Einstein Integrals\label{app:integrals}}
In this appendix I give some results for a special class of the
Fermi-Dirac and Bose-Einstein integrals
(massless particles), that are used throughout this
thesis. The work by Elze \cite{Elze86c} for the case of complex
chemical potential
needed in  Chapter \ref{chap:projection}, is reproduced in some detail,
since it has not been published.
\section{Fermions}
I have performed calculations of the Fermi integrals
\be
	I^{+}_n(\eta) \equiv  \int_0^\infty dx\,
		\frac{x^n}{e^{x-\eta} + 1} ,
\ee
for both positive and negative values of $\eta$ and the results agree
with those of Dingle \cite{Dingle57a} and Elze \cite{Elze86c}.
My own calculation is rather lengthy, and if only the combination
\be
S^{+}_n(\eta) \equiv I^{+}_n(\eta) + (-1)^{n+1}I^{+}_n(-\eta) ,
\ee
is needed (as is indeed the
case here) the elegant derivation due to Elze \cite{Elze86c} is much to
be preferred. The form of the result is slightly different, and my own
result as well as that of Dingle \cite{Dingle57a} is
\be \label{eq:dinglefermi}
	S^{+}_n(\eta) =
		2\sum_{k=0}^{\lfloor \frac{n+1}{2} \rfloor}
		\frac{n!\zeta(2k)}{(n+1-2k)!}\left(1-2^{1-2k}\right)
		\eta^{n+1-2k} ,
\ee
where $\zeta(x)$ is the \Index{Riemann zeta function} \cite{Abramowitz},
and $\lfloor x \rfloor$ is the floor of $x$, \ie , the largest integer
smaller than or equal to $x$.
The first few polynomials are
\begin{eqnarray}
	\label{eq:sp0}
	S^{+}_0(\eta) &=& \eta \\
	\label{eq:sp1}
	S^{+}_1(\eta) &=& \frac{\pi^2}{6} +\frac{1}{2}\eta^2 \\
	\label{eq:sp2}
	S^{+}_2(\eta) &=& \frac{\pi^2}{3}\eta +\frac{1}{3}\eta^3\\
	\label{eq:sp3}
	S^{+}_3(\eta) &=& \frac{7\pi^4}{60} + \frac{\pi^2}{2}\eta^2
		+ \frac{1}{4}\eta^4
\end{eqnarray}

\section{Bosons}
For the Bose integral
\be
	I^{-}_n(\eta) \equiv  \int_0^\infty dx\,
		\frac{x^n}{e^{x-\eta} - 1} ,
\ee
the results of Dingle \cite{Dingle57b} can be used to evaluate the
combination 
\be
	S^{-}_n(\eta) \equiv I^{-}_n(\eta) + (-1)^{n+1}I^{-}_n(-\eta),
\ee
which gives
\be
	S^{-}_n(\eta) =
                2\sum_{k=0}^{\lfloor \frac{n+1}{2} \rfloor}
		\frac{n!\zeta(2k)}{(n+1-2k)!}
		\eta^{n+1-2k} .
\ee
The first few polynomials are
\begin{eqnarray}
	\label{eq:sm0Dingle}
	S^{-}_0(\eta) &=& -\eta \\
	\label{eq:sm1Dingle}
	S^{-}_1(\eta) &=& \frac{\pi^2}{3} -\frac{1}{2}\eta^2 \\
	\label{eq:sm2Dingle}
	S^{-}_2(\eta) &=& \frac{2\pi^2}{3}\eta -\frac{1}{3}\eta^3\\
	\label{eq:sm3Dingle}
	S^{-}_3(\eta) &=& \frac{2\pi^4}{15} + \pi^2\eta^2
		- \frac{1}{4}\eta^4
\end{eqnarray}
The Fermi and Bose results can combined to the expression
\be
	S^{\pm}_n(\eta) =
		2\sum_{k=0}^{\lfloor \frac{n+1}{2} \rfloor}
		\frac{n!\zeta(2k)}{(n+1-2k)!}\left[1-(1\pm 1)2^{-2k}\right]
		\eta^{n+1-2k}
\ee

\section{Complex Chemical Potential}
In this section I will reproduce a derivation of the integrals
\be
	S^{\pm}_n(z) \equiv 
	\int_0^\infty dx\, x^n \left( \frac{1}{e^{x -z}\pm 1}
	        +(-1)^{n+1} \frac{1}{e^{x +z}\pm 1} \right) ,
	\qquad n=0,1,2,\ldots
\ee
with complex $z$, which is due to Elze \cite{Elze86c}.

The integrals are periodic in $z$ along the imaginary axis
\be
	S^{\pm}_n(z+2m\pi i) = S^{\pm}_n(z)	, \quad m\in\mathbb{Z},
\ee
with period $2\pi$. The Fermi and Bose cases are connected through
\be
	S^{-}_n(z) = S^{+}_n(z+(2m+1)\pi i), \quad m\in\mathbb{Z}.
\ee
Thus the result for bosons can be obtained from that for fermions, so 
from now on consider just the integral $S^{+}_n(z)$, with $z$ lying in
the strip
\be
	-\pi < \mathrm{Im}\, z < \pi .
\ee
If $z$ is in another strip the result can be found by using the
periodicity. Note that the poles of the integrand are distributed along
the two axes $\mathrm{Re}\, x= \pm \mathrm{Re}\, z$, with period $2\pi$.

The substitution $y=x\pm z$ is performed shifting the integration into
the complex plane, 
\be \label{eq:contourint}
	S^{+}_n(z) = \int_{-z}^{\infty-z} dy\,
		\frac{(y+z)^n}{e^y+1} + (-1)^{n+1}
		\int_z^{\infty+z} dy\,\frac{(y-z)^n}{e^y+1} ,
\ee
and shifting the poles onto the imaginary axis.

\begin{figure}
  \centering
  \includegraphics{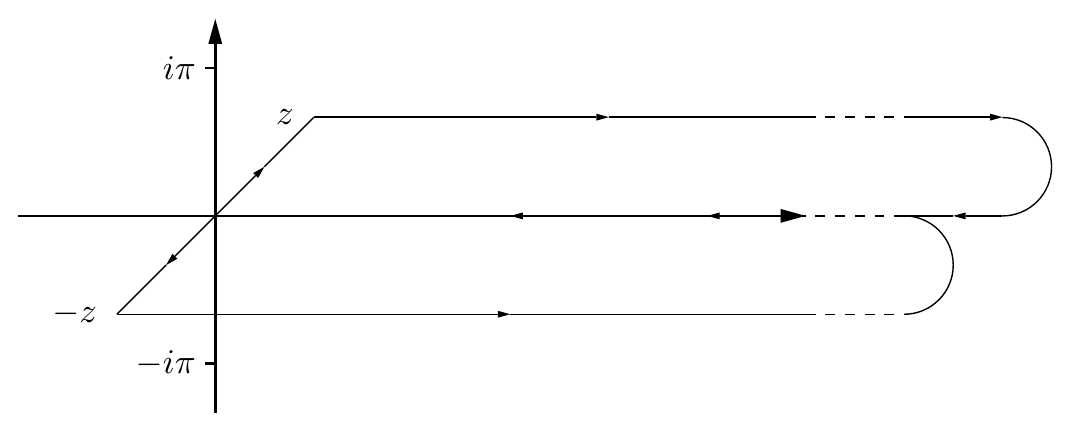} 
  \caption{\label{fig:contour}
	The contours used to evaluate the integrals in
	\Eq{eq:contourint}.
	}
	
\end{figure}

The integration contours are closed as shown in \Fig{fig:contour},
and an application of Cauchy's theorem on the resulting integrals, after
doing a binomial expansion of the factors $(y\pm z)^n$, gives
\begin{eqnarray}
	S^{+}_n(z) &=& \frac{z^{n+1}}{n+1}  \\
		&+& \sum_{k=0}^n {n\choose k} z^k \left[1+(-1)^{n+k+1}\right]
		\left(1-2^{-n+k}\right) \Gamma(n-k+1)\zeta(n-k+1) ,
		\nonumber 
\end{eqnarray}
where the integral representation \cite{Abramowitz}
\be
	\zeta(t) = \frac{1}{(1-2^{1-t})\Gamma(t)}
		\int_0^\infty dx\, \frac{x^{t-1}}{e^x+1} ,
\ee
of the \Index{Riemann zeta function} has been used. $\Gamma(z)$ is the
\Index{gamma function} \cite{Abramowitz}.

Elze expresses this result in terms of \Index{Bernoulli polynomials},
but note that the first term can be included in the sum as the
$(n+1)$th term, since
$\zeta(0)=1/2$. Using the identity $\Gamma(n+1) = n!$ the result of Elze
can be rewritten
\comment{Remember: ${n\choose k} = \frac{n!}{(n-k)!k!}$}
\be
	S^{+}_n(z) =
		\sum_{k=0}^{n+1} \frac{n!}{k!}z^k\left[ 1+(-1)^{n+1+k}\right]
		\left(1-2^{k-n}\right)\zeta(n-k+1) .
\ee

It may not be obvious that this result is identical to
Eq.~(\ref{eq:dinglefermi}), but an evaluation of the low order
polynomials using the above expression, gives results identical to
Eqs.~(\ref{eq:sp0})--(\ref{eq:sp3}), for real $z$.

The Bose integral $S^{-}_n(z)$, as evaluated here, is defined in the
strip $0< \mrm{Im}\, z < 2\pi$, and therefore, as pointed out by Elze
\cite{Elze86c,ElzePrivate}, the expression cannot be extended to the real axis.
The reason for this is that for $\mrm{Im}\, z=0$ there is a pole on the
contour of integration,  resulting in an ill-defined integral.

In the strip $0< \mrm{Im}\, z < 2\pi$ the polynomials of lowest order
are
\begin{eqnarray}
	\label{eq:sm0Elze}
	S^{-}_0(z) &=& i\pi - z ,\\
	\label{eq:sm1Elze}
	S^{-}_1(z) &=& \frac{\pi^2}{3} + i\pi z - \frac{1}{2}z^2 ,\\
	\label{eq:sm2Elze}
	S^{-}_2(z) &=& \frac{2\pi^2}{3} z + i\pi z^2 -\frac{1}{3} z^3 ,\\
	\label{eq:sm3Elze}
	S^{-}_3(z) &=& \frac{2\pi^4}{15} + \pi^2 z^2 + i\pi z^3
		-\frac{1}{4} z^4 .
\end{eqnarray}
Comparing this to Eqs.~(\ref{eq:sm0Dingle})--(\ref{eq:sm3Dingle}) shows
that there are extra terms relative to the result of Dingle
\cite{Dingle57b} valid on the real axis.

As mentioned above, Elze relates $S^\pm_n(z)$ to Bernoulli polynomials
$B_n(x)$ \cite{Abramowitz}.
His result is
\be
	\begin{array}{r@{\;=\;}lr@{\;\in\;}l}
	\displaystyle
	S^+_n(z) & \displaystyle \frac{(2\pi i)^{n+1}}{n+1}
		B_{n+1}\left(\frac{\pi-iz}{2\pi}\right)
		\qquad &\mrm{Im}\,z &  (-\pi,\pi) \\[15pt]
	\displaystyle
	S^-_n(z) & \displaystyle -\frac{(2\pi i)^{n+1}}{n+1}
		B_{n+1}\left(\frac{-iz}{2\pi}\right)
		&\mrm{Im}\,z  & (0,2\pi) .
	\end{array}
\ee

\clearemptydoublepage
\chapter{The \texttt{SHELL} Program\label{app:shell}}
\index{shell model|(}
This appendix contains a short Fortran90 program that implements the
shell model calculation of the energy for a spherical strangelet at zero
temperature. In addition to the routines shown here, the program has to
be linked against the routines \texttt{solver}, \texttt{minimizer}, and
\texttt{sphbes}, used for root finding, function minimization, and the
calculation of spherical Bessel functions\index{Bessel function}
respectively. Such
routines are part of any good repository of numerical techniques (see
\textit{e.g.} Ref.~\cite{NumericalRecipes}).

\section{The Main Loop}

The main routine of the program is merely a loop over baryon number
\texttt{A}.
For every \texttt{A}, the minimum energy is found and the result is
written to standard output.

{\footnotesize
\begin{verbatim}
Program shell
   implicit none
   integer, parameter :: STDOUT = 6
   integer, parameter :: MaxA=100
   integer A
   real E
   real,     external :: MinimumEnergy

   Do A=1,MaxA
      E =  MinimumEnergy(A)
      write(STDOUT,*) A, E/real(A)
   End Do
End ! Program shell
\end{verbatim}
}

\section{Minimizing the Energy}

The function \texttt{MinimumEnergy} calculates the minimum energy of a
strangelet with baryon number \texttt{A} by varying the radius.
For a fixed radius \texttt{R} the function \texttt{Energy} calculates the
energy for a strangelet of optimal composition. This is done in the
routine \texttt{Find\_composition} by trying the various possible
distributions of quarks among the available flavors.

\subsection{\texttt{MinimumEnergy}}

This routine finds the minimum energy as a function of radius \texttt{R}
for a given baryon number \texttt{A}.
The number of levels included in the calculation is determined
by the chemical potentials . These are found
from a liquid drop calculation in the function `\texttt{Energy}'.
We still need a maximum number of allowable levels \texttt{MaxLevels}
to dimension the arrays in the other routines.

{\footnotesize
\begin{verbatim}
Real Function MinimumEnergy(A)
   implicit none
   integer, intent(in)  :: A

   real Rmin, R_guess, R1, R2, R3
   real, parameter :: acc=1.E-3
   real, parameter :: MeVfm = 197.3     ! conversion factor
   real, external  :: Energy, minimizer
   integer MaxLevels

   MaxLevels = 51 !must be odd!

!    Make a guess for initial bracketing triplet
   R_guess = real(A)**(1/3.) / MeVfm
   R1 = R_guess / 5.
   R2 = R_guess
   R3 = R_guess * 5.
   MinimumEnergy = minimizer(R1, R2, R3, Energy, acc, Rmin, &
                             A, MaxLevels)
End ! MinimumEnergy
\end{verbatim}
}

\subsection{\texttt{Energy}}

This routine calculates the energy of a strangelet of optimal
composition. It calls \texttt{Find\_levels} to get the eigenvalue
spectrum, which is then used by the routine \texttt{Find\_composition}
to deduce the
distribution of quarks among flavors. The energy due to the bag constant
is also added here.

In order to get an estimate of how many levels it is necessary to
include in the calculation, the chemical potential is found from a
liquid drop model calculation in the routine \texttt{Find\_chempot}.

{\footnotesize
\begin{verbatim}
Real Function Energy(R, A, MaxLevels)
   implicit none
   real,    intent(in) :: R
   integer, intent(in) :: A,MaxLevels

   real,    parameter :: BBag = 145.**4
   real,    parameter :: PI   = 3.141592654
   integer, parameter :: Nflavor   = 3
   real,    dimension(Nflavor,MaxLevels) :: energies
   integer, dimension(Nflavor,MaxLevels) :: degen
   integer, dimension(Nflavor)           :: NumLevels
   real, dimension(1:3), parameter       :: mass = (/ 0.0, 0.0, 150.0 /)
   real, dimension(1:3)                  :: chemical
   real E

   call Find_chempot(R, A, mass, chemical, Nflavor);
   call Find_levels(R, mass, chemical, energies, degen, &
        Nflavor, MaxLevels, NumLevels)
   call Find_composition(A, E, energies, degen, Nflavor, &
        MaxLevels, NumLevels)

   Energy = E + 4./3.*PI*R**3 * BBag
End ! Energy
\end{verbatim}
}

\subsection{\texttt{Find\_chempot}}

A liquid drop model calculation of the chemical potential \texttt{mu},
assuming only one flavor is done here, using the expressions in Chapter
\ref{chap:mit}. The precision is given by \texttt{tol}.
This is used in \texttt{Get\_eigenvalues} to 
find all single particle energies less than the chemical potential.

{\footnotesize
\begin{verbatim}
Subroutine Find_chempot(R, A, mass, chemical, Nflavor)
   implicit none
   real,                        intent(in)  :: R
   integer,                     intent(in)  :: A, Nflavor
   real, dimension(1:3),        intent(in)  :: mass
   real, dimension(1:3),        intent(out) :: chemical

   real, external               :: solver, Number
   real, parameter              :: tol = 1.E-3
   real                         :: m, mu, mumin, mumax
   integer                      :: i

   Do i = 1,Nflavor
      mumin = mass(i)   ! Lower bound mu = m
      mumax = 1500.     ! Upper bound on mu_i (in MeV)
      m = mass(i)
      mu = solver(Number, mumin, mumax, tol, A, m, R)
!       I add 30% to the found value, since this is only an estimate.
      chemical(i) = 1.3 * mu
   End Do
End ! Find_chempot
\end{verbatim}
}

\subsection{\texttt{Number}}

The function \texttt{Number} is solved for the chemical potential
\texttt{mu} that satisfies the condition that the total baryon number of
the system is $\mathtt{A} = \frac{1}{3} \mathtt{Number}$, where
\texttt{Number} is given in terms of volume, surface, and curvature
number densities
(\Eq{eq:nummasslessV}, \Eq{eq:nummasslessC},
and Eqs.~(\ref{eq:numvolume})--(\ref{eq:numcurvature}))
obtained from the liquid drop model. If the quark mass is less than
\texttt{eps} the expressions for massless quarks are used.
{\footnotesize
\begin{verbatim}
Real Function Number(mu, A, mass, R)
   implicit none
   real,        intent(in) :: mu, mass, R
   integer,     intent(in) :: A

   real, parameter         :: PI = 3.141592654, eps = 1.E-3
   real                    :: V, S, C, NV, NS, NC, x

   V = 4./3.*PI*R**3
   S = 4.*PI*R**2
   C = 8.*PI*R
   If (mass < eps) Then
      NV = mu**3/PI**2
      NS = 0.0
      NC = -4.*mu/PI**2
   Else
      x = mu/mass
      NV = mass**3/PI**2 * (x**2-1)**(3./2.)
      NS = -3.*mass**2/4./PI * ( 0.5*(x**2-1)   &
           - 1/PI*( x**2*atan(sqrt(x**2-1))     &
           - sqrt(x**2-1) ) )
      NC = 3.*mass/8./PI**2 * (-0.5*PI*(x**2-1) &
           + 1/3.*sqrt(x**2-1)                  &
           + x**2 * atan(sqrt(x**2-1)) )
   End If
   Number = NV*V + NS*S + NC*C - 3*A
End ! Number
\end{verbatim}
}

\section{The Optimal Composition}

\subsection{\texttt{Find\_composition}}

This routine finds the composition of a strangelet, \ie , the
distribution of quarks among the levels, that minimizes the
total energy, subject to the constraint that the baryon number
equals \texttt{A}. On return \texttt{levels} and \texttt{deg}
have been sorted (by \texttt{Sort\_levels}) in ascending order.
This procedure is not as difficult as it looks. The idea is to try all
possible combinations of flavors $\mathtt{n1} + \mathtt{n2} + \mathtt{n3}
= 3\mathtt{A}$, and for each
flavor fill the levels from the bottom up. Allowing for from 1 to 
3 flavors makes this look more complicated than it really is.

{\footnotesize
\begin{verbatim}
Subroutine Find_composition(A, E, levels, deg, Nflavor, &
                            MaxLevels, NumLevels)
   implicit none
   integer,intent(in) :: A, Nflavor, MaxLevels
   real,                                   intent(out)  :: E
   real,   dimension(Nflavor, MaxLevels), intent(inout) :: levels
   integer,dimension(Nflavor, MaxLevels), intent(inout) :: deg
   integer,dimension(Nflavor),            intent(in)    :: NumLevels

   real, parameter :: HUGE = 1.E30
   integer n1, n2, n3, cur_lev, n_cur_lev, cur_fla, n_cur_fla
   real E_min, E1
   external Sort_levels

!    Sort the energy levels according to energy, in ascending
!    order,  and rearrange the `deg' array to reflect this.
   call Sort_levels(levels, deg, Nflavor, MaxLevels, NumLevels)

!    Now find the optimal composition of the strangelet.
   E_min = HUGE
   If (Nflavor == 1) Then
      n1 = 3*A
      E = 0.0
      cur_lev = 1;   n_cur_lev = 0
      cur_fla = 1;   n_cur_fla = 0
q11:   Do
         If (n_cur_fla == n1) Exit q11
         E = E + levels(cur_fla,cur_lev)
         n_cur_fla = n_cur_fla + 1
         n_cur_lev = n_cur_lev + 1
         If (n_cur_lev == deg(cur_fla,cur_lev)) Then
            cur_lev = cur_lev + 1
            n_cur_lev = 0
         End If
      End Do q11
      E_min = E
   Else
      Do n1 = 0,3*A
         E = 0.0
         cur_lev = 1;   n_cur_lev = 0
         cur_fla = 1;   n_cur_fla = 0
q12:      Do
            If (n_cur_fla == n1) Exit q12
            E = E + levels(cur_fla,cur_lev)
            n_cur_fla = n_cur_fla + 1
            n_cur_lev = n_cur_lev + 1
            If (n_cur_lev == deg(cur_fla,cur_lev)) Then
               cur_lev = cur_lev + 1
               n_cur_lev = 0
            End If
         End Do q12
         If (Nflavor == 2) Then
            n2 = 3*A-n1
            cur_lev = 1;   n_cur_lev = 0
            cur_fla = 2;   n_cur_fla = 0
q21:        Do
               If (n_cur_fla == n2) Exit q21
               E = E + levels(cur_fla,cur_lev)
               n_cur_fla = n_cur_fla + 1
               n_cur_lev = n_cur_lev + 1
               If (n_cur_lev == deg(cur_fla,cur_lev)) Then
                  cur_lev = cur_lev + 1
                  n_cur_lev = 0
               End If
            End Do q21
            If (E < E_min)   E_min = E
         Else
            E1 = E
            Do n2 = 0,3*A-n1
               E = E1
               cur_lev = 1;   n_cur_lev = 0
               cur_fla = 2;   n_cur_fla = 0
q22:           Do
                  If (n_cur_fla == n2) Exit q22
                  E = E + levels(cur_fla,cur_lev)
                  n_cur_fla = n_cur_fla + 1
                  n_cur_lev = n_cur_lev + 1
                  If (n_cur_lev == deg(cur_fla,cur_lev)) Then
                     cur_lev = cur_lev + 1
                     n_cur_lev = 0
                  End If
               End Do q22
               n3 = 3*A-n1-n2
               cur_lev = 1;  n_cur_lev = 0
               cur_fla = 3;  n_cur_fla = 0
q3:            Do
                  If (n_cur_fla == n3) Exit q3
                  E = E + levels(cur_fla,cur_lev)
                  n_cur_fla = n_cur_fla + 1
                  n_cur_lev = n_cur_lev + 1
                  If (n_cur_lev == deg(cur_fla,cur_lev)) Then
                     cur_lev = cur_lev + 1
                     n_cur_lev = 0
                  End If
               End Do q3
               If (E < E_min)   E_min = E
            End Do   ! n2
         End If
      End Do   ! n1
   End If
   E = E_min
End ! Find_composition
\end{verbatim}
}

\subsection{\texttt{Sort\_levels}}

This subroutine performs a comb sort \cite{Lacey91a}
, on the array \texttt{levels},
sorting it in ascending order, and rearranges the array \texttt{deg}
accordingly.

{\footnotesize
\begin{verbatim}
Subroutine Sort_levels(levels, deg, Nflavor, MaxLevels, &
                       NumLevels)
   implicit none
   integer,intent(in)  :: Nflavor, MaxLevels
   real,   dimension(Nflavor, MaxLevels), intent(inout) :: levels
   integer,dimension(Nflavor, MaxLevels), intent(inout) :: deg
   integer,dimension(Nflavor),            intent(in)    :: NumLevels

   real, parameter :: SHRINK = 1.3
   integer q, switches, i_hold, i, j, top, gap
   real r_hold

   Do q = 1, Nflavor
       gap = NumLevels(q)
loop:  Do
          gap = int(real(gap)/SHRINK)
          Select Case (gap)
          Case (0)
             gap = 1
          Case (9:10)
             gap = 11
          Case Default
             gap = gap
          End Select
          switches = 0
          top = NumLevels(q) - gap
          Do i=1,top
             j = i + gap
             If (levels(q,i) > levels(q,j)) Then
                r_hold = levels(q,i)
                levels(q,i) = levels(q,j)
                levels(q,j) = r_hold
                i_hold = deg(q,i)
                deg(q,i) = deg(q,j)
                deg(q,j) = i_hold
                switches = switches + 1
             End If
          End Do
          If ( (switches == 0) .AND. (gap == 1) ) Exit loop
       End Do loop
   End Do 
End ! Sort_levels
\end{verbatim}
}

\section{The Eigenvalues}

\subsection{\texttt{Find\_levels}}

This routine does the quantum number book-keeping, while leaving the
actual calculation of single particle levels to the routine
\texttt{Get\_levels}. 
\comment{Elaborate on this, with reference to earlier Chapters
        (when written).}

{\footnotesize
\begin{verbatim}
Subroutine Find_levels(R, mass, chem, levels, degen, &
                       Nflavor, Maxlevels, NumLevels)
   implicit none
   real,                    intent(in)  :: R
   real,    dimension(1:3), intent(in)  :: mass, chem
   integer, intent(in)                  :: Nflavor, MaxLevels
   real,    dimension(Nflavor, MaxLevels), intent(out) :: levels
   integer, dimension(Nflavor, MaxLevels), intent(out) :: degen
   integer, dimension(Nflavor),         intent(out)    :: NumLevels

   integer, parameter :: STDERR = 0
   integer l, lmax, kappa, q, nu, numax
   real, dimension(MaxLevels) :: eps
   external Get_eigenvalue

!   lmax is the number of l-values that it takes to accomodate
!   the number of quarks of a single flavor that can fill
!   MaxLevels levels if only the nu=1 levels are used. This gives
!   an upper limit to l.

!    The maximum number of l values (MaxLevels is odd)
   lmax = (MaxLevels - 1)/2

   Do q = 1, Nflavor        ! Loop over quark flavors
!       Initialize NumLevels to zero
      NumLevels(q) = 0
!       First of all the ground state (l=0, j=1/2, kappa=-1) is
!       found. It has degeneracy 3*(2j+1) = 6
      kappa = -1
!       Here we find all eps < mu_q
      call Get_eigenvalues(eps, chem(q), mass(q), kappa, &
                           numax, R, MaxLevels)
      If (numax == 0) Exit
      Do nu=1,numax
         NumLevels(q) = NumLevels(q) + 1
         levels(q,NumLevels(q)) = eps(nu)
         degen(q,NumLevels(q))  = 6
      End Do
      ! Now for the remaining levels
      Do l = 1, lmax ! Loop over orbital angular momentum
         kappa = l   ! First possible value of kappa (j=l-1/2)
         call Get_eigenvalues(eps, chem(q), mass(q), kappa, &
                              numax, R, MaxLevels)
         If (numax == 0) Exit
         Do nu=1,numax
            NumLevels(q) = NumLevels(q) + 1
            levels(q ,NumLevels(q))   = eps(nu)
            degen(q,NumLevels(q))     = 6*l 
         End Do
         kappa = -(l+1) !Second possible value of kappa (j=l+1/2)
         call Get_eigenvalues(eps, chem(q), mass(q), kappa, &
                              numax, R, MaxLevels)
         If (numax == 0) Exit
         Do nu=1,numax
            NumLevels(q) = NumLevels(q) + 1
            levels(q ,NumLevels(q)) = eps(nu)
            degen(q,NumLevels(q))   = 6*(l+1)
         End Do
      End Do
      If (NumLevels(q) >= MaxLevels) Then
         Write(STDERR,*) 'MaxLevels exceeded in "Find_levels"'
         Exit
      End If
   End Do
End ! Find_levels
\end{verbatim}
}

\subsection{\texttt{Get\_eigenvalues}}

For a fixed value of \texttt{kappa}
the routine \texttt{Get\_eigenvalues} finds \texttt{numax} roots of
the Dirac equation\index{Dirac equation!numerical solution},
with precision \texttt{eps}, 
so that the \texttt{numax} roots have energies less than the chemical
potential \texttt{mu}. It returns the energies \texttt{eps} and \texttt{numax}.

{\footnotesize
\begin{verbatim}
Subroutine Get_eigenvalues(eps, mu, mass, kappa, numax, &
                           R, MaxLevels)
   implicit none
   integer,                    intent (in) :: MaxLevels
   real, dimension(MaxLevels), intent(out) :: eps
   real,                       intent (in) :: mu, mass
   integer,                    intent (in) :: kappa
   integer,                    intent(out) :: numax
   real,                       intent (in) :: R

   real, parameter :: acc  = 1.E-3
   real, parameter :: dx   = 0.1
   real x, xmin, xmax, fx, fxmin, energy
   real, external :: Dirac, solver
   integer nu

!    Search for an xvalue for which Dirac(x) changes sign and
!    use  that as xmax. The first root (nu = 1) is x=2.04..
!    for kappa = -1.
   x = 1.0
   Do nu=1,MaxLevels
      xmin = x + dx
      fxmin = Dirac(xmin, kappa, mass, R)
      x = xmin + dx
      Do
        fx = Dirac(x, kappa, mass, R)
        If (fx*fxmin > 0.0) Then
              xmin = x
              x = x + dx
           Else
              xmax = x
              Exit
           End If
      End Do
              
!    Now find the root between xmin and xmax
      x =  solver(Dirac, xmin, xmax, acc, kappa, mass, R)
      energy = sqrt( (x/R)**2 + mass**2 )
      if (energy > mu) Exit
      eps(nu) = energy
   End Do
   numax = nu-1 ! The last radial quantum number
End ! Get_eigenvalues
\end{verbatim}
}

\subsection{\texttt{Dirac}}\index{Dirac equation!numerical solution}

The equation $\mathtt{Dirac}=0$ is the Dirac equation for a single quark
with quantum number \texttt{kappa} in an MIT bag of radius \texttt{R}.
It has several solutions corresponding to different values of the radial
quantum number.

{\footnotesize
\begin{verbatim}
Real Function Dirac(x, kappa, mass, R)
   implicit none
   real,    intent (in) :: x, mass, R
   integer, intent (in) :: kappa

   real mR
   integer kapm1
   real, external :: fkx

   mR = mass*R
   kapm1 = kappa - 1
   Dirac = fkx(kappa, x)                                 & 
           + x * fkx(kapm1, x) / (sqrt(x**2+mR**2) + mR)
End ! Dirac
\end{verbatim}
}

\subsection{\texttt{fkx}}

The function \texttt{fkx} ($f_\kappa(x)$ defined in \Eq{eq:fkx})
occurs in \texttt{Dirac}
and it is given in terms of spherical Bessel\index{Bessel function}
functions.

{\footnotesize
\begin{verbatim}
Real Function fkx(k,x)
   implicit none
   integer, intent(in) :: k
   real,    intent(in) :: x

   external sphbes  ! Spherical Bessel function of integer order
   real sj, sy, sjp, syp
   integer k2

   If (k >= 0) Then
      call sphbes(k, x, sj, sy, sjp, syp)
      fkx = sj
   Else
      k2 = -(k+1)
      call sphbes(k2, x, sj, sy, sjp, syp)
      fkx = (-1)**(k+1) * sj
   End If
End ! fkx
\end{verbatim}
}
\index{shell model|)}

\clearemptydoublepage

\bibliographystyle{prsty}
\bibliography{/usr/users/dmj/artikler/bibliography/new}
\addcontentsline{toc}{chapter}{\bibname}

\clearemptydoublepage
\cleardoublepage
\addcontentsline{toc}{chapter}{\indexname}
\printindex

\end{document}